\documentclass[12pt, a4paper]{article} 

\usepackage{etex} 
\usepackage{setspace}
\usepackage[small, font = onehalfspacing]{caption}
\usepackage[authoryear]{natbib}
\usepackage[english]{babel}
\usepackage[latin1]{inputenc}
\usepackage[T1]{fontenc}
\usepackage{lmodern}
\usepackage{amssymb}
\usepackage{xcolor,graphicx,xspace,caption}
\usepackage{listings,eso-pic}
\usepackage{lscape}
\usepackage{multirow}
\usepackage{enumerate}
\usepackage{amsmath}
\usepackage{amsthm}
\usepackage[flushleft]{threeparttable} 
\usepackage[title]{appendix} 
\usepackage{subcaption}
\usepackage{url}
\usepackage{comment}
\pdfpageattr {/Group << /S /Transparency /I true /CS /DeviceRGB>>}

\usepackage{xr-hyper}
\usepackage[colorlinks=true,urlcolor=blue]{hyperref}
\hypersetup{linkcolor=red,citecolor=blue,filecolor=red}
\makeatletter
\newcommand*{\addFileDependency}[1]{
  \typeout{(#1)}
  \@addtofilelist{#1}
  \IfFileExists{#1}{}{\typeout{No file #1.}}
}
\makeatother

\newcommand*{\myexternaldocument}[1]{
    \externaldocument{#1}
    \addFileDependency{#1.tex}
    \addFileDependency{#1.aux}
}
\myexternaldocument{Online_Supplement}

\AtBeginDocument{
\label{CorrectFirstPageLabel}

}

\usepackage{etex} 
\usepackage[figuresleft]{rotating}
\bibpunct{(}{)}{;}{a}{,}{,}
\usepackage[T1]{fontenc}
\usepackage{lmodern}
\usepackage{xcolor,graphicx,xspace,caption}
\usepackage{mathrsfs}
\usepackage{array}
\pdfpageattr {/Group << /S /Transparency /I true /CS /DeviceRGB>>}

\def\mB{\mathcal{B}}

\newtheorem{theorem}{Theorem}

\definecolor{hint}{RGB}{191,63,0}

\definecolor{hellgelb}{rgb}{1,1,0.8}
\definecolor{colKeys}{rgb}{0,0,1}
\definecolor{colIdentifier}{rgb}{0,0,0}
\definecolor{colComments}{rgb}{1,0,0}
\definecolor{colString}{rgb}{0,0.5,0}

\newcommand{\E}{\mathop{\mbox{\sf E}}}

\newcommand{\F}{\mathcal{F}}

\newcommand{\IF}{\mathbf{I}}

\newcommand{\cov}{\mathop{\hbox{Cov}}}
\newcommand{\Cov}{\mathop{\mbox{Cov}}}

\newcommand{\Var}{\mathop{\mbox{\sf Var}}}

\def\proofname{PROOF}

\newcommand{\N}{\mathop{\mbox{\sf N}}}

\newtheorem{lemma}{Lemma}[section]
\newtheorem{proposition}{Proposition}[section]
\newtheorem{assumption}{Assumption}[section]
\newtheorem{definition}{Definition}[section]

\theoremstyle{definition}

\newcommand{\II}{\mathbf{I}}

\renewcommand{\P}{\operatorname{P}}

\newcommand{\vps}{{\varepsilon}}

\renewenvironment{proof}[1][\proofname]{{\noindent\bfseries #1.}}{\qed}

\renewcommand{\hat}{\widehat}

\newcommand{\RR}{\mathbb{R}}

\renewcommand{\hat}{\widehat}
\renewcommand{\leq}{\leqslant}
\renewcommand{\geq}{\geqslant}

\newcommand{\tpsi}{\tilde{\psi}}
\newcommand{\rit}{\mathbb{R}}
\newcommand{\yit}{\mathbb{Y}}

\newcommand{\Dit}{\mathbb{D}}
\newcommand{\fit}{\mathbb{F}}
\newcommand{\cit}{\mathbb{C}}
\newcommand{\Ait}{\mathbb{A}}

\newcommand{\nit}{\mathbb{N}}
\def\defeq{\stackrel{\mathrm{def}}{=}}  
\newcommand{\Abt}{\mathbf{A}}

\def\defeq{\stackrel{\mathrm{def}}{=}}  

\title{Dynamic Network Quantile Regression Model}
\author{Xiu Xu \footnote{Dongwu Business School, Soochow University, 50 Donghuan Road, Suzhou, Jiangsu 215021, PR China. Email: \href{mailto:xiux@suda.edu.cn}{xiux@suda.edu.cn}.}\; Weining Wang \footnote{Department of Economics and Related Studies, University of York, Heslington, York, YO10 5DD, UK. Email: \href{mailto:weining.wang@york.ac.uk}{weining.wang@york.ac.uk}.}\;   Yongcheol Shin \footnote{Department of Economics and Related Studies, University of York, Heslington, York, YO10 5DD, UK. Email: \href{mailto:yongcheol.shin@york.ac.uk}{yongcheol.shin@york.ac.uk}.}  \;Chaowen Zheng\footnote{Department of Economics and Related Studies, University of York, Heslington, York, YO10 5DD, UK. Email: \href{mailto:cz1113@york.ac.uk}{cz1113@york.ac.uk}.}}

\date{}

\usepackage[top = 1in, bottom = 1in, left = 1in, right = 1in]{geometry}

\begin{document}

\maketitle
\begin{abstract}

We propose a dynamic network quantile regression model to investigate the quantile connectedness using a predetermined network information. We extend the existing network quantile autoregression model of \cite{zhu2019} by explicitly allowing the contemporaneous network effects and controlling for the common factors across quantiles.
To cope with the endogeneity issue due to simultaneous network spillovers, we adopt the instrumental variable quantile regression (IVQR) estimation and derive the consistency and asymptotic normality of the IVQR estimator using the near epoch dependence property of the network process.
Via Monte Carlo simulations, we confirm the satisfactory performance of the IVQR estimator across different quantiles under the different network structures. Finally, we demonstrate the usefulness of our proposed approach with an application to the dataset on the stocks traded in NYSE and NASDAQ in 2016.

\end{abstract}

\bigskip


\newpage

\section{Introduction}
\label{Introduction}


The topology of financial networks is central to the study of financial contagion and systemic risk, see \cite{fafchamps2007}, \cite{acemoglu2015}, \cite{hautsch2015} among others. Given the relevance of tail dependence for financial supervision and risk management \citep{betz2016}, a joint analysis of network effect and tail dependence becomes more important because the implications derived from network models evaluated by conventional conditional mean estimators cannot necessarily be generalized to the tails. \cite{ando2021connect} show that major adverse events are associated with an increase in average connectedness but that their effects on the  tails significantly differ.

Quantile regression (QR) has been a powerful tool for characterizing the heterogeneous policy impacts and measuring tail-event driven risk (e.g. \cite{hardle2016}). Following a seminal work by \citet{bassett1978}, QR can be used to evaluate the entire range of the conditional distribution. Recently, the literature on quantile time series regression has been rapidly growing. \cite{koenker2006quantile} propose a quantile autoregressive model while \cite{galvao2013} study QR in an autoregressive dynamic framework with exogenous stationary covariates. Following the analysis of quantile cointegration in \cite{xiao2009quantile}, \cite{cho2015quantile} bring QR to the autoregressive distributed lag (ARDL) model literature. The quantile ARDL process captures both the long-run and short-run relationships at any desired location in the conditional distribution. 
\cite{engle2004} advance a conditional autoregressive value at risk model whilst \cite{white2015} propose a vector autoregressive (VAR) model for analyzing quantile dynamics.

However, in most financial systems, multiple entities are often intertwined and interacted with each other, which can be represented as networks \citep{hautsch2015, hardle2016, chen2019}. In this context, \cite{zhu2017} develop a network autoregression (NAR) model, which has gained great popularity in social network analysis. A number of extensions have been developed. \citet{zhu2020multivariate} consider a multivariate spatial autoregression model. \citet{zhu2019portal} investigate a screening method to select influential nodes. \cite{zhu2020nonconcave} studies nonconvex penalized estimation methods in high-dimensional VAR models while \cite{zhu2018grouped} extend the network VAR model to allow for group-specific parameters.

In particular, \citet{zhu2019} extend the NAR model by \cite{zhu2017} 
and propose a network quantile autoregression (NQAR) model in order to analyze tail dependency in a dynamic network with a large number of nodes. The NQAR model consists of a system of equations, in which a continuous response is related to its lagged connected nodes, the response of the same node in the previous time period as well as node specific characteristics in a quantile autoregression process. However, main weakness of the NQAR model lies in that it does not accommodate the contemporaneous impact of connected nodes, even though the simultaneous network/peer effects are pervasive in empirical studies  \citep{liu2014,forni2010}.   
If they are statistically significant, the estimation of the NQAR model would become inconsistent and misleading.

In this paper, as a main contribution, we extend the NQAR model and propose a general dynamic network quantile regression model (DNQR) by explicitly incorporating contemporaneous and lagged network effects of connected nodes as well as the impacts of node-sepcific variables and observed common effects. Notice, however, that the simultaneous network effects are inherently endogenous to the system, which leads to inconsistent estimates at conditional mean regression as well as in QR, see \cite{wuthrich2019, wuthrich2020, chernozhukov2020network}. To cope with the endogeneity issue in the different contexts, many studies have attempted to apply the instrumental variable quantile regression (IVQR) estimation advanced by \cite{Chernozhukov2006}, e.g. \cite{frolich2013}, \cite{su2016} and \cite{machado2019}.

To deal with this challenging issue we adopt the IVQR approach. The social network data are similar to the spatial data, in the sense that observations from connected users are correlated. This makes the spatial autoregressive model good candidate for network data analysis. However, our work can be regarded as a nontrivial extension of \cite{Su2011}, who apply the IVQR approach to analyzing the cross-section data using a linear spatial autoregressive model, to a dynamic network quantile model with nodal heterogeneity and common factors, which can shed further lights on uncovering the complex tail dependency in dynamic networks with a large number of nodes and time periods. Our study is also closely related to the growing literature on the tail comovements in a complex financial system, see \cite{diebold2014}, \cite{hautsch2014}, \cite{white2015} 
and \cite{ando2021connect}.

More importantly, we follow \cite{Jenish2012} and \cite{xu2015maximum}, and develop the general asymptotic theory for the IVQR estimator by applying the spatial near epoch dependence (NED) of the underlying network processes. We derive the detailed conditions on the network processes in order to establish the consistency and asymptotic normality of the IVQR estimator. Via Monte Carlo simulations, we confirm that the finite sample performance of the IVQR estimator is satisfactory across quantiles and the different error distributions under the different network structures. 

Finally, we demonstrate the utility of our approach with an application to the dataset on the stocks traded in NYSE and NASDAQ in 2016, through the common shareholder network constructed using the information on the common mutual fund ownership, e.g. \cite{Anton2014}. We find that the contemporary network effects (measured by returns of connected stocks in the same period) are positive and significant and dominate all the other effects across all quantiles. Furthermore, they are stronger at the lower tails than at the upper tails,
suggesting that the contemporaneous network effects should be explicitly and carefully analysed in the dynamic network quantile model.

This paper proceeds as follows. In Section \ref{model} we outline the DNQR model and derive the stationarity condition of the underlying network process. 
Section \ref{estimation} introduces the IVQR estimation and develops its asymptotic properties using the spatial NED approach. In Section \ref{simulation} we provide simulation results, showing that the IVQR estimation performs satisfactory.
In Section \ref{application} the DNQR model is applied to the US financial market data.
Section \ref{conclusion} provides concluding remarks. The mathematical proofs and the additional simulation and empirical results are presented in the Online Appendix. The replication code can be found \href{https://github.com/ChaowenZheng/Dynamic-Network-Quantile-Model}{here} on GitHub.

 \textbf{Notations:}
 For a vector $v = (v_1, \ldots, v_m)^\top \in \RR^m$, we denote $|v|_k = (\sum_{i=1}^m |v_i|^k)^{1/k}$, $\|v\|_k= (\sum_{i=1}^m\E|v_{i}|^k)^{1/k}$,  
 and $|v|_\infty = \max_{i \le m} |v_i|$, where $k$ is a positive integer, and $\E$ is the expectation operator. For any $n \times m$ matrix $A = (a_{i j})_{1\le i\le n, 1\le j \le m}$, we define the two norm and the max norm by $|A|_2 = \sup_{\{v\in \RR^m ,|v|_2 = 1\}} |A v|_2$ and $|A|_{\max} = \max_{i,j}|a_{ij}|$, respectively.  Define the column-sum and the row-sum by $\|A\|_1 = \displaystyle \max_{1\leq j\leq m} \displaystyle \sum^{n}_{i=1} |a_{ij}|$ and $\|A\|_{\infty} = \displaystyle \max_{1\leq i\leq n} \displaystyle \sum^{m}_{j=1} |a_{ij}|$.  We write $a_n=O(b_n)$ or $a_n\lesssim b_n$ if there exists a positive constant $C$ such that $a_n/b_n\leq C$ for all large $n$, and denote $a_n=o(b_n)$ (resp. $a_n\sim b_n$), if $a_n/b_n\rightarrow 0$ (resp. $a_n/b_n \rightarrow c$ for a positive constant $c$). For two sequences of random variables $(X_n)$ and $(Y_n)$, we write $X_n=o_p( Y_n)$ if $X_n/Y_n \rightarrow 0$ in probability. Let $I_{N}$ be an $N \times N$ identity matrix, $\II(\cdot)$ the indicator function, $\mathbf{1}_{N}$ a $N \times 1$ vector with each element as one, and $\nit$ the integer set.

\section{The Model} \label{model}

Consider the large scale network time series data with $N$ nodes for $1 \leq i \leq N$, and $T$ time periods for $1\leq t\leq T$, which is observationally equivalent to a regular panel data. To describe their relationship, we construct an adjacency matrix, $A = (a_{ij}) \in \mathbb{R}^{N \times N}$, where $a_{ij} = 1$ if the node $i$ follows the node $j$, and $a_{ij} = 0$ otherwise. We do not allow the self-following relation, $a_{ii} = 0$. Define the row-normalised network matrix as $W = (w_{ij}) \in \mathbb{R}^{N \times N}$, where $w_{ij} = n_i^{-1} a_{ij}$ and $n_i = \sum_{j=1}^N a_{ij}$. 
Let $\mathbb{Y}_{t}=(Y_{1t}, \cdots, Y_{Nt})^{\top} \in \mathbb{R}^{N}$ be the continuous response (e.g., tweet length) collected at time $t$, and $U_{it}$ be a sequence of $i.i.d.$ uniform random variables on the set $[0,1]$.

We then consider the following DNQR model:
\begin{align}\label{model1}
Y_{it} =&  \;  \gamma_{0}^0(U_{it}) + \displaystyle \sum_{l=1}^{q} \alpha^0_{l}(U_{it}) Z_{il} + \gamma_{1}^0(U_{it}) \displaystyle \sum_{j=1}^{N} w_{ij}Y_{jt}  \\ \nonumber
& + \gamma_{2}^0(U_{it}) \displaystyle \sum_{j=1}^{N} w_{ij} Y_{j,t-1} 
+ \gamma_{3}^0(U_{it})Y_{i,t-1}
+ \displaystyle \sum_{k=0}^{p} F^{\top}_{t-k}\beta_{k}^0(U_{it}),
\end{align}
for $i = 1, \cdots, N$ and $t = 1, \cdots, T$, where $\gamma_{j}^0(\cdot)$ for $j=0, 1, 2, 3$, $\alpha_{l}^0$ for $l=1, \cdots, q$, and each elements in $\beta_{k}^0 \in \rit^{m} $ for $k = 0, 1, \cdots, p$ are unknown parameter functions from $[0,1]$ to $\rit$, and the superscript $0$ is used to denote the true value of parameters. $Z_{i} = (Z_{i1}, \cdots, Z_{iq})^{\top} \in \rit^{q}$ is a $q \times 1$ vector of time-invariant node-specific covariates, and $F_{t} = (F_{t1}, \cdots, F_{tm})^{\top} \in \rit^{m}$ is an $m \times 1$ vector of time-varying common covariates that capture the systematic influences on response variable, $Y_{it}$.

If the right hand side of the DNQR model \eqref{model1} is monotonically increasing in $U_{it}$, then we can write the $\tau$-th conditional quantile function of $Y_{it}$ as
\begin{align}\label{modelquantile}
Q_{Y_{it}}(\tau|\F_{t})  = & \; \gamma_{0}^0(\tau) + \displaystyle \sum_{l=1}^{q} \alpha_{l}^0(\tau) Z_{il} + \gamma_{1}^0(\tau) \displaystyle \sum_{j=1}^{N} w_{ij}Y_{jt}  \\ \nonumber
& + \gamma_{2}^0(\tau) \displaystyle \sum_{j=1}^{N} w_{ij} Y_{j,t-1} + \gamma_{3}^0(\tau)Y_{i,t-1}
+ \displaystyle \sum_{k=0}^{p} F^{\top}_{t-k} \beta_{k}^0(\tau),
\end{align}
where $\F_{t} = \{Z_{1}, \cdots, Z_{N}, \yit_{t-1}, \yit_{t}, F_{t}, F_{t-1},\cdots, F_{t-p}\}$ is the information set. 
The first component, $\gamma_{0}^0(\tau) + \sum_{l=1}^{q}  \alpha _{l}^0(\tau ) Z_{il}$ is the quantile-specific nodal impact of the node $i$, where $\gamma_{0}^0(\tau)$ is the baseline function and $Z_{il}$s are assumed to be independent from $U_{it}s$. 
Next, network interactions between nodes are captured via both contemporaneous and lagged network variables, $\sum_{j=1}^{N}w_{ij}Y_{jt}$ and $\sum_{j=1}^{N}w_{ij}Y_{j,t-1}$, with $\gamma_{1}^0(\tau)$ capturing the quantile-specific simultaneous network effects and $\gamma_{2}^0(\tau)$ measuring the lagged diffusion network effects. $\gamma_{3}^0(\tau)$ is the quantile-specific momentum function, capturing the temporal dynamics for the same node. Furthermore, we control for the dynamic impacts of the (observed) common macroeconomic and financial factors, $F_{t}$, which can mitigate any remaining {common shock effect} in the data.

Let 
$\mathbb{F}_{t} =(F_{t}^{\top},\cdots, F_{t-p}^{\top })^{\top} \in \mathbb{R}^{(p+1)m}$. Define $\mathbf{A}_{0t} = (\gamma_{0}^0(U_{it}) + \sum_{l=1}^{q} \alpha^0_{l}(U_{it}) Z_{il}, 1\leq i\leq N)^{\top }\in \mathbb{R}^{N}$, $\mathbf{A}
_{1t}=diag\{\gamma _{1}^0(U_{it}),1\leq i\leq N\} 
\in \mathbb{R}^{N\times N}$, $\mathbf{A}_{2t}=diag\{\gamma^0 _{2}(U_{it}),1\leq i\leq
N\} 
\in \mathbb{R}^{N\times N}$, $\mathbf{A}_{3t}=diag\{\gamma^0
_{3}(U_{it}),1\leq i\leq N\} 
\in \mathbb{R}^{N \times N}$, and $\mathbf{B}_{t} = ((\beta_{0}^{0\top}(U_{it}), \cdots ,\beta _{p}^{0\top }(U_{it}))^{\top}, 1\leq i\leq N )^{\top} \in 
\mathbb{R}^{N\times (p+1)m}$.
The DNQR model, \eqref{model1} can be expressed compactly in a matrix form:
\begin{equation} 
\mathbb{Y}_{t}=\Gamma +\mathbf{A}_{1t}W\mathbb{Y}_{t}+\mathbf{H}_{t}\mathbb{Y%
}_{t-1}+\mathbf{B}_{t}\mathbb{F}_{t}+V_{t},  \label{modelmatrix}
\end{equation}%
where $\mathbf{H}_{t}=\mathbf{A}_{2t}W+\mathbf{A}
_{3t}\in \mathbb{R}^{N\times N}$, $\Gamma =\E\left( \mathbf{A}_{0t}\right)$,  and $V_{t}=\mathbf{A}_{0t}-\Gamma \in 
\mathbb{R}^{N}$ is $i.i.d.$ over $t$ with mean $\mathbf{0}$ and variance-covariance matrix, $\Sigma_{V} = \sigma_{V}^{2} I_{N} \in \mathbb{R}^{N \times N}$. 

Notice that the DNQR model can be regarded as a substantial extension of \cite{koenker2006quantile}, who provide a classic framework for the analysis of the  random-coefficient model in the quantile autoregression. 
Moreover, the DNQR model encompasses the NAR model by \cite{zhu2017} and the NQAR model by \citet{zhu2019}, through jointly incorporating contemporaneous and lagged network effects of connected nodes as well as exogenous common effects.

We show that the DNQR model is subject to the endogeneity issue due to contemporaneous network spillovers across nodes. Consider a simple two-equation system: 
\begin{eqnarray}
Y_{1t}& =& \gamma_0^0(U_{1t}) + \gamma_1^0(U_{1t})  a_{12} Y_{2t},\label{equIVQRexmp1}\\
Y_{2t}& =& \gamma_0^0(U_{2t}) + \gamma_1^0(U_{2t})  a_{21}Y_{1t}.
\end{eqnarray}
Assuming that $1 - a_{21}a_{12} \gamma_1^0(U_{1t}) \gamma_1^0(U_{2t}) \neq 0$, we obtain the following solutions:
\begin{eqnarray}
Y_{1t}& =&( \gamma_0^0(U_{1t})+\gamma_0^0(U_{2t})\gamma_1^0(U_{1t})a_{12})/  (1 - a_{21}a_{12}\gamma_1^0(U_{1t})\gamma_1^0(U_{2t})) ,\\
Y_{2t}& =&(\gamma_0^0(U_{2t}) +\gamma_0^0(U_{1t})\gamma_1^0(U_{2t})a_{21})/(1 - a_{21}a_{12}\gamma_1^0(U_{1t})\gamma_1^0(U_{2t})).
\end{eqnarray}
As $Y_{1t}$ is a function of $U_{1t}$ and $U_{2t}$, the monotone argument cannot be applied because
\begin{equation}
\P \left(Y_{it} \leq \gamma_{0}^0(\tau) + \gamma _{1}^0(\tau) \overline{Y}_{it}|\overline{Y}_{it}\right) \neq \tau \quad a.s. \text{ for } i= 1,2,  \label{equIVQRexmp}
\end{equation}%
where $\overline{Y}_{1t} = a_{12} Y_{2t}$ and $\overline{Y}_{2t} = a_{21} Y_{1t}$. This shows that the endogeneity is caused by the contemporaneous network term, $\overline{Y}_{it}$.

The simultaneous network spillover would cause inconsistency. Consider the simple mean regression, 
$Y_{it} = \lambda \sum_{j \neq i} w_{ij} Y_{jt} + \varepsilon_{it}$. Let $w_{i} = (w_{i1}, ..., w_{ij}, ..., w_{iN})^{\top} \in \mathbb{R}^{N} $, and $\tilde{w}_{ij}$ as the $(i,j)$ element of the matrix $(I_{N} - \lambda W)^{-1}$. Assuming that $\mathop{\mbox{\sf E}} (\varepsilon_{it} \varepsilon_{jt})= 0$ if $i\neq j$ and $\mathop{\mbox{\sf E}}(\varepsilon_{it}^2) = \sigma_i^{2}$, then the bias term (the average correlation between the endogenous variable and the error term) will be of the order, $\mbox{lim}_{N,T \to \infty} (NT)^{-1}\sum_i\sum_t \mathop{\mbox{\sf E}}  (w_i^{\top}  \mathbb{Y}_{t}\varepsilon_{it} ) =\mbox{lim}_{N \to\infty} N^{-1} \sum_i\sum_{j\neq i} w_{ij}  \tilde{w}_{ji} \sigma_i^2 \approx c$, where $c$ is a constant. This is not equal to zero unless $\mbox{lim}_{N \to \infty} N^{-1}\sum_i\sum_{j\neq i} w_{ij} \tilde{w}_{ji} \sigma_i^2 = o(1)$. Thus, the estimation is likely to be biased unless the link of the network is very weak. 
In the quantile case, the leading bias term will be of the order: 
\begin{equation}
 \mbox{lim}_{N,T \to\infty} (NT)^{-1}\sum_i\sum_t\E ( (\tau-\II(\varepsilon_{it,\tau}\leq 0)) w_i^{\top}  \mathbb{Y}_{t}), 
\end{equation}
which does not tend to zero, where $\varepsilon_{it,\tau}=Y_{it}- Q_{Y_{it}}(\tau|\mathbb{Y}_{t})$ is the $\tau$-th QR error.

The nontrivial estimation issue for the DNQR model lies in that the endogeneity caused by contemporaneous network spillovers renders the ordinary QR estimator to be inconsistent. 
\citet{Chernozhukov2005,Chernozhukov2006, Chernozhukov2008} propose the IVQR approach to estimating quantile treatment effects and develop the robust inference. \cite{chernozhukov2020network} develop a novel technique to constructing simultaneous confidence bands for quantile functions and quantile effects in nonlinear network panels.

We follow the IVQR approach to cope with the simultaneous network endogeneity. Notice that \cite{Su2011} apply the IVQR approach to an analysis of the cross-section data using a linear spatial autoregressive model. However, our work can be regarded as a nontrivial extension of \cite{Su2011} to a dynamic network quantile model with nodal heterogeneity and common factors, which can shed further lights on uncovering the complex tail dependency in dynamic networks  with a large number of nodes and time periods. More importantly, we derive a general asymptotic theory by using the spatial NED property of the network process in Section \ref{asy_dis}.

\subsection{Stationarity}\label{sec_stationary}

In this section we derive the stationarity conditions for $\mathbb{Y}_{t}$ in \eqref{modelmatrix}, and its asymptotic distribution. 
Notice that the DNQR model can easily produce  predictions of quantiles, $\widehat{Q}_{Y_{it}}(\tau|\F_{t})$, given the network structure and the data history, by plugging the estimated parameters into \eqref{modelquantile}. To this end, it is important to derive the conditions under which the network process is stationary. Further, stationarity may be required to identify some parameters. For example, if we wish to uncover the variance structure of series of interest, it would be crucial to check whether it changes over time or not. 

Define $S_{t} = I_N -\mathbf{A}_{1t}W$. 
Then, we make the following assumptions:

\begin{assumption}\label{Ass_stationray} (1) Let $\Upsilon = \max_i |\gamma_{1}^0(U_{it})| \leq c_{1}<1$ and $|W|_{2}\leq 1$ 
, where $W$ is a row normalized network matrix, and $c_{1}$ is a positive constant. Assume that $U_{it}$ and $Z_{i}$ are $i.i.d.$ over $i$ and $t$, and $F_{t}$ are $i.i.d.$ The $k$th moments of $F_t$ and $Z_i$ are finite, $k>2$.

(2) $\max_i |\gamma_{2}^0(U_{it})| + \max_i |\gamma_{3}^0(U_{it})| \leq c_{23}<1$ and $c_1 + c_{23} < 1$, where $c_{23}$ is a positive constant.

(3) $\max_{i}|\gamma^0 _{0}(U_{it})|+\max_{i}\sum_{l=1}^{q}|\alpha^0_{l}(U_{it})||Z_{il}|\leq d_{z}$, and $|\mathbf{B}_{t}|_{\infty }|\mathbb{F}_{t}|_{1}\leq d_{f}$, where $d_{z}$ and $d_{f}$ are random variables with bounded moments. Let 
$\mathbb{D}_{t}=S_{t}^{-1}(\mathbf{B}_{t}\mathbb{F}_{t}+\mathbf{A}_{0t})$ with
$\mathbb{D}=\mathop{\mbox{\sf E}}\mathbb{D}_{t}$ and the elementwise maximum value $\mathbb{D}_{\max} < \infty$. Then, $\max_t|\mathop{\mbox{\sf E}}\{\Vec{(}\mathbb{D}_{t-l_{1}}%
\mathbb{D}_{t-l_{1}}^{\top })\}|_{\infty }\leq \sigma _{d \max} < \infty$, where $l_{1} = (0, 1, \cdots, t-1)$.

(4) The right hand side of the model \eqref{model1} is monotonically increasing in $U_{it}$.
\end{assumption}

Assumption \ref{Ass_stationray}(1) assures the invertibility of 
$S_{t}$. The model \eqref{modelmatrix} has a unique solution if and only if every principal minor of $I_N -\mathbf{A}_{1t}W$ is positive, which is met by Assumption \ref{Ass_stationray}(1), though it is only a sufficient condition. Assumption \ref{Ass_stationray}(2) is necessary to obtain the strict stationarity of $\{\mathbb{Y}_{t}\}_t$. Under Assumptions \ref{Ass_stationray}(2) and (3), the covariance stationarity can be achieved. 
Then, we have the following lemma. 

\begin{lemma} \label{lemma_stationray} Let $\mathcal{C}_z \defeq \sigma(Z_{1},\cdots ,Z_{N})$, where $\sigma(\cdot)$ denotes a sigma field. Then, under Assumption \ref{Ass_stationray} and conditional on $\mathcal{C}_z $, the process 
$\{\mathbb{Y}_{t}\}_t$ is strictly stationary as well as covariance stationary. 
 \end{lemma}
 
We introduce the NED concept in Section \ref{asy_dis} to ensure that the dependency of the processes is decaying appropriately, which is the key in proving the consistency and asymptotic normality of the proposed IVQR estimator. In sum, stationarity is required for moment estimation and forecasting while the NED property is utilized to prove the parameter consistency and asymptotic normality.

Once $\mathbb{Y}_{t}$ is shown to be strictly stationary, 
$\mathbb{Y}_{t}$ is covariance stationary if $\mathop{\mbox{\sf Var}}(\mathbb{Y}_{t})$ and $\Gamma_{l} = \mathop{\hbox{Cov}}(\mathbb{Y}_{t},\mathbb{Y}_{t-l})$ exist. Rewrite the model \eqref{modelmatrix} as
\begin{equation} \label{modelreducedform}
 \mathbb{Y}_{t} = S_{t}^{-1}\mathbf{H}_{t}\mathbb{Y}_{t-1}+S_{t}^{-1}\mathbf{B}_{t}\mathbb{F}_{t}+S_{t}^{-1} \mathbf{A}_{0t}
\end{equation}
where $S_{t}=I_{N}-\mathbf{A}_{1t}W$. Then, we have the following covariance stationary solution: 
\begin{equation}
\mathbb{Y}_{t}=\sum_{l=0}^{\infty }\Pi _{l}\mathbb{D}_{t-l}=\sum_{l=0}^{%
\infty }\Pi _{l}S_{t-l}^{-1}\mathbf{B}_{t-l}\mathbb{F}_{t-l}+\sum_{l=0}^{\infty }\Pi
_{l}S_{t-l}^{-1}\mathbf{A}_{0t} , \label{model_solution}
\end{equation}%
where $\mathbb{D}%
_{t}=S_{t}^{-1}(\mathbf{B}_{t}\mathbb{F}_{t}+\mathbf{A}_{0t})$, $M_{t}=S_{t}^{-1}\mathbf{H}_{t}$ and $\Pi_{l}=M_{t}\times \cdots \times M_{t-l+1}$ for $l>1$ with $\Pi_{0}=I_N$ and $\Pi_{1}=M_{t}$. In the Online Appendix \ref{proof_lemma_stationray} we prove that the covariance of $\mathbb{Y}_{t}$ exist under Assumption \ref{Ass_stationray}.

\subsection{Asymptotic Stationary Distribution} \label{sec_stationary_dis}

Define any vector $a\in \mathbb{R}^{N}$ with $|a|_{2}=1$ and fixed $d$ number of non zero elements. Let $\tilde{\mathbb{Y}}_{t} =\mathbb{Y}_{t}-\mu _{\mathbb{Y}}$, $L_{T}=\sum_{t=1}^{T}a^{\top}\tilde{\mathbb{Y}}_{t}$, and $L_{t}=L_{\lfloor t \rfloor } + (t - \lfloor t \rfloor)a^{\top} \tilde{\mathbb{Y}}_{\lfloor t \rfloor + 1}, t \geq 1$, where $\mu _{\mathbb{Y}} = \sf E(\mathbb{Y}_{t})$ and $\lfloor t \rfloor = \max \{k\in \mathbb{Z}:k\geq t\}$ is the floor function. We then show that the average response is asymptotically normally distributed.

\begin{theorem}
\label{thrm_stationary}Consider any vector $a\in \mathbb{R}^{N}$ with $|a|_{2}=1$ and fixed $d< N$ number of nonzero elements. Under Assumption \ref{Ass_stationray} and conditional on $\mathcal{C}_z $, then
\begin{equation}
\frac{L_{Tu}}{\sqrt{T}} \Rightarrow \sigma_{a\mathbb{Y}} \mathfrak{B}(u), \quad
0 \leq u \leq 1  \label{inv}
\end{equation}%
where $\sigma_{a\mathbb{Y}}^{2} \defeq \sum_{l\geq 0}a^{\top} \Gamma_{l}a$ is the
long run variance of $a^{\top} \tilde{\mathbb{Y}}_{t}$ and $\mathfrak{B}(u)\;(0\leq u\leq 1)$ is a Brownian motion.
\end{theorem}

\textbf{Remark} For $u =1$, Theorem \ref{thrm_stationary} implies: 
\begin{equation}
\sqrt{T}(a^{\top }(\overline{\mathbb{Y}}-\mu _{\mathbb{Y}}))%
\mathrel{\mathop{\longrightarrow}\limits_{}^{\mathcal{L}}}%
\mathop{\mbox{\sf
N}}(0,\sigma _{a\mathbb{Y}}^{2}),\mbox{ ~~~as~~$T\rightarrow\infty$.}
\label{stationary}
\end{equation}%
where $\overline{\mathbb{Y}}=T^{-1}\sum_{t=1}^{T}\mathbb{Y}_{t}$. Thus, the mean of $\mathbb{Y}_{t}$ converges in law to a normal distribution.

\section{The IVQR Estimation} \label{estimation}

We first introduce the estimation algorithms of the IVQR approach. We then discuss the underlying assumptions and develop the asymptotic theory.

\subsection{IVQR Estimator}
\label{IVQRestimation}

Suppose that there exists an $N \times \ell$ matrix of instrumental variables (IV), denoted $\mathbf{R}_{t}=(R_{1t}, \cdots ,R_{Nt})^{\top} \in \mathbb{R}^{N\times \ell}$, which is assumed to be independent of $U_{it}$. Then, we have the following quantile conditions: 
\begin{equation}
\P\left( Y_{it}\leq \gamma _{1} ^{0}(\tau )\overline{Y}_{it}+X_{it}^{\top
}\phi ^{0}(\tau )|X_{it},R_{it}\right) =\tau \;a.s.  \label{equIVQR1}
\end{equation}
where $\overline{Y}_{it} = \sum_{j=1}^{N} w_{ij} Y_{jt}$ and $X_{it}=\left( 1,Z_{i}^{\top}, \overline{Y}_{i,t-1}, Y_{i,t-1}, F_{t}^{\top}, \cdots, F_{t-p}^{\top}\right)^{\top}$ with $\phi^{0}(\tau) = [\gamma^{0}_{0}(\tau), \alpha_{1}^{0}(\tau), ...,\alpha_{q}^{0}(\tau), \gamma^{0}_{2}(\tau), \gamma^{0}_{3}(\tau), \beta_{0}^{0 \top}(\tau),..., \beta_{p}^{0 \top}(\tau)]^{\top} \in \mathbb{R}^{3+q+(p+1)m}$. The above conditional probability is a measurable function of $(X_{it},R_{it})$. 

In general, the valid IVs should satisfy the quantile conditions in \eqref{equIVQR1}, and do not lead to collinearity among $R_{it}$ and $X_{it}$. See Theorem \ref{theorem_estimation} for the asymptotic formula of the variance matrix of the IVQR estimator. The estimation efficiency will be improved by choosing $R_{it}$ appropriately.
Following the literature, we may choose $R_{it}$ to be the higher network orders of lagged dependent variables such as $e_i^{\top}W^2\mathbb{Y}_{t-1}$,  $[e_i^{\top}W^2\mathbb{Y}_{t-1}, e_i^{\top}W\mathbb{Y}_{t-2}]$ and so on, where $e_i$ is a vector with unity on the $i$-th element and zeros otherwise. Based on the satisfactory simulation evidence reported in Section \ref{simulation}, we suggest using $[e_i^{\top}W^2\mathbb{Y}_{t-1}, e_i^{\top}W^3\mathbb{Y}_{t-1}]$ as IVs.

To solve \eqref{equIVQR1} we need to find the unknown true parameters $(\gamma^{0}_{1}(\tau ),\phi^{0\top}(\tau ))^{\top}$ such that $\boldsymbol{0}$ is a solution to the quantile estimation of $Y_{it} - \gamma^{0}_{1}(\tau ) \overline{Y}_{it}-X_{it}^{\top} \phi^{0}(\tau)$ on $(X_{it},R_{it})$: 
\begin{equation}
\boldsymbol{0}\in \arg \min_{g\in \mathscr{G}}\mathop{\mbox{\sf E}}\left[ \rho _{\tau
}\left\{ Y_{it}-\gamma^{0} _{1}(\tau )\overline{Y}_{it}-X_{it}^{\top }\phi^{0} (\tau
)-g\left( X_{it},R_{it}\right) \right\} \right] , \label{equIVQRsol}
\end{equation}%
where $\mathscr{G}$ is the class of measurable functions of $(X_{it},R_{it})$ and $\rho_{\tau }(u)=u\{\tau - \II(u<0)\}$ is the check function with $\II(\cdot)$ the indicator function. 
We then restrict $\mathscr{G}$ to the class of linear-in-parameter functions:
\begin{equation}
\mathscr{G}=\{g\left( X_{it},R_{it}\right) =  R_{it}^{\top }  \lambda (\tau
):\lambda \in \Lambda \},  \label{syg}
\end{equation}
where $\Lambda $ is a compact set in $\mathbb{R}^{\ell }$. Alternatively, we may construct the transformed IVs by the least squares projection of $\overline{Y}_{it}$ on $(X_{it},R_{it})$ as in \cite{Chernozhukov2005,Chernozhukov2006}. Then, we obtain the sample analogue of the objective function: 
\begin{equation}
 Q(\gamma _{1}(\tau ),\phi (\tau ),\lambda (\tau ))=\displaystyle%
\sum_{i=1}^{N}\displaystyle\sum_{t=1}^{T}\left[ \rho _{\tau }\left\{
Y_{it}-\gamma _{1}(\tau )\overline{Y}_{it}-X_{it}^{\top }\phi (\tau )-{R }_{it}^{\top }\lambda (\tau )\right\} \right].  \label{equIVQRobj}
\end{equation}

Let $\theta(\tau ) =(\gamma_{1}(\tau ),\phi^{\top}(\tau ))^{\top}$ and $\eta(\tau)=(\phi^{\top} (\tau),\lambda^{\top} (\tau))^{\top}$. The IVQR estimator, $(\hat{\gamma}_{1}(\tau),\widehat{\phi}^{\top}(\tau),\widehat{\lambda}^{\top}(\tau))^{\top}$, obtained by minimizing \eqref{equIVQRobj}, is expected to converge to the true parameters, $(\gamma^{0}_{1}(\tau), \phi^{0\top}(\tau), \mathbf{0}^{\top})^{\top}$. 
For a given value of endogenous parameter, $\tilde{\gamma}_{1}(\tau)$, over a grid set of the interval $(-1,1)$, we first run the ordinary QR of $Y_{it}-\tilde{\gamma} _{1}(\tau )\overline{Y}_{it}$ on $%
(X_{it},R _{it})$ and obtain the corresponding estimator, denoted $\widehat{\eta}(\tilde{\gamma}_{1}(\tau),\tau) = \left[ \widehat{\phi}^{\top}(\tilde{\gamma}_{1}(\tau),\tau), \widehat{\lambda}^{\top} (\tilde{\gamma}_{1}(\tau),\tau)\right]^{\top}$.
Next, we select $\tilde{\gamma}_{1}(\tau)$ which minimizes $|\widehat{\lambda} (\tilde{\gamma}_{1}(\tau),\tau) |_2^2$ over the interval $(-1,1)$, denoted as $\widehat{\gamma}_{1}(\tau)$. The IVQR estimator of $\theta (\tau)$ is then obtained by $\hat{\theta}(\tau) = (\widehat{\gamma}_{1}(\tau), \widehat{\phi}^{\top}(\widehat{\gamma}_{1}(\tau),\tau))^{\top}$. For a given quantile index $\tau $, the IVQR estimation can proceed as follows:

Step 1. For a given value of $\tilde{\gamma}_{1}(\tau)$, run the QR of $Y_{it} - \tilde{\gamma}_{1}(\tau) \overline{Y}_{it}$ on $(X_{it},R_{it})$ and obtain:
\begin{equation}
\widehat{\eta}(\tilde{\gamma}_{1}(\tau ),\tau) = \displaystyle \arg \min_{(\phi,\lambda)} Q(\tilde{\gamma}_{1}(\tau),\phi(\tau),\lambda (\tau)).
\label{equIVQRstep1}
\end{equation}

Step 2. Minimize a weighted norm of $\widehat{\lambda}(\tilde{\gamma}_{1}(\tau),\tau)$ over $\tilde{\gamma}_{1}(\tau)$ to obtain the IVQR estimator of $\gamma_{1}(\tau)$: 
\begin{equation}
\widehat{\gamma}_{1}(\tau) = \displaystyle \arg \min _{\tilde{\gamma}_{1} \in \left(-1,1\right)} \widehat{\lambda }^{\top}(\tilde{\gamma}_{1}(\tau),\tau) \;\mathrm{A}\;\widehat{\lambda}(\tilde{\gamma}_{1}(\tau),\tau),  \label{equIVQRstep2}
\end{equation}  
where $\mathrm{A}$ is some positive definite matrix. Without loss of generality we set $\mathrm{A} = I$ throughout the paper.

Step 3. Run the QR of  $Y_{it}-\widehat{\gamma}_{1}(\tau) \overline{Y}_{it}$ on $X_{it}$, and obtain the estimator of $\phi(\tau)$, denoted $\widehat{\phi}(\tau) = \widehat{\phi}(\widehat{\gamma}_{1}(\tau),\tau)$. Finally, we obtain the IVQR estimator by 
\begin{equation}
\widehat{\theta }(\tau )=(\widehat{\gamma }_{1}(\tau ),\widehat{\phi }^{\top}(\tau
))^{\top}=(\widehat{\gamma }_{1}(\tau ),\widehat{\phi }^{\top}(\widehat{\gamma }_{1}(\tau
),\tau ))^{\top}.  \label{equIVQRstep3}
\end{equation}

\subsection{Asymptotic Theory} \label{asy_dis}

To develop the asymptotic theory for the IVQR estimator, we need to deal with some topological properties of $\mathbb{Y}_{t}$ that are spatially and temporally dependent. We follow \citet{Jenish2009,Jenish2012} and utilize NED to address the spatial dependence of the statistics. The derivation of the asymptotic property follows from the standard M-estimation, including the quantile loss function as a special case. First, conditional on the common factors, we show in Section \ref{sec_NED} that the elements of $\{\mathbb{Y}_{t}\}_t$ is an NED process. Then, in Section \ref{sec_asydis}, we derive the asymptotic distribution of the IVQR estimator under certain regularity conditions. As we aim to apply the DNQR model to a network dataset with the large number of nodes and time periods, we mainly employ the large $N$ and large $T$ asymptotics, though the asymptotic theory can be equally developed for {large $T$ and fixed $N$ or fixed $N$ and large $T$ (as pointed out by an anonymous referee.)}

\subsubsection{NED Properties of the Network Processes}\label{sec_NED}

\citet{Jenish2009,Jenish2012} extend the notion of NED processes used in the time series to random fields. This class of NED processes can accommodate a wide range of models with a spatial dependence. They derive the central limit theorem and the law of large numbers for NED random fields. We first review the definition and some properties of NED random fields in \cite{Jenish2012}. 

The observations for each node can be modeled as a realization of a dependent heterogeneous process indexed by a point in $\mathbb{R}^{d} $ with  $d \geq 1$. We consider a random field $D \subseteq \mathbb{R}^{d}$. 
The space $\mathbb{R}^{d}$ is endowed with the metric $\rho(j,j') = \max_{1 \leq l \leq d} |j_{l}-j'_{l}|$ with the corresponding norm, $|j| = \max_{1 \leq l \leq d}|j_{l}|$, where $j_{l}$ is the $l$-th element of $j$. The distance between any subsets $U,V\subseteq D$ is defined as $\rho (U,V) = \inf \{\rho (j,j'): j\in U \text{ and } j' \in V\}$. Let $|U|$ denote the cardinality of a finite subset, $U$. In the two dimensional case with $d =2 $ and $j = j(i,t)$, we have: $\rho((i,t), (i^{\prime}, t^{\prime })) = \max(|i-i^{\prime}|, |t-t^{\prime}|)$.

\begin{assumption}
\label{Ass1} Let the lattice $D_{NT} \subseteq D\subseteq \mathbb{R}^{d}$ with $d \geq 1$, be countably infinite, where the cardinality of $D_{NT}$ satisfies $\displaystyle \lim_{N , T \rightarrow \infty} |D_{NT}| \rightarrow \infty$. Then, $\rho(j,j') \geq  \rho_{0}, \forall j,j'\in D$, where $\rho_0$ is a constant. We set $\rho_{0}=1$ w.l.o.g.
\end{assumption}

The minimum distance assumption in Assumption \ref{Ass1} is used for increasing domain asymptotics. It ensures the growth of the sample size as the sample regions $D_{NT}$ expands. The setting is introduced in \cite{Jenish2012} for spatial mixing and NED processes. Note that the space $D$ can be a space of socio-economomic characteristics or geographical space, and the metric is not restricted to physical distance.

\begin{definition}[NED]
\label{DefNED} Let $\mathcal {Z} = \{\mathcal {Z}_{it} (i,t) \in D_{NT},NT \geq 1\}$ and $\zeta = \{\zeta_{it},(i,t) \in D_{NT},NT \geq 1\}$ be random fields with $\Vert \mathcal {Z}_{it} \Vert_{p'} < \infty$ for $p' \geq 1$, where $D_{NT} \subseteq D$ with its cardinality given by $|D_{NT}|=NT$. Let $\{d_{it},(i,t) \in D_{NT},NT \geq 1\}$ be an array of positive constants. Then, the random field $\mathcal {Z}$ is $L_{p'}$-NED on the random field $\zeta$ if 
\begin{equation*}
\Vert \mathcal {Z}_{it}-\mathop{\mbox{\sf E}}(\mathcal {Z}_{it}|\mathscr{F}_{it}(s))\Vert
_{p'}<d_{it}\varphi (s),
\end{equation*}%
for some sequence $\varphi(s) \geq 0$ with $\displaystyle \lim_{s \rightarrow
\infty} \varphi(s)=0$, where $\varphi(s)$ is the NED coefficient and $d_{it}$ is the NED scaling factor. $\mathscr{F}_{it}(s) = \sigma(\zeta_{i't'}: (i',t')\in D_{NT}, \rho((i',t'),(i,t) )\leq s)$ is the $\sigma$-field generated by $\zeta_{i't'}$ within distance $s$ from $(i,t)$. If $\displaystyle \sup_{N,T} \sup_{(i,t) \in D_{NT}} d_{it} < \infty$, then $\mathcal {Z}$ is uniformly $L_{p'}$-NED on $\zeta $.
\end{definition}


Next, we present the $L_{2}$-NED properties of random field $\mathcal {Z}$ on some $\alpha$-mixing random field $\zeta$. The $\alpha$-mixing coefficient is defined below.

\begin{definition}\label{Def2}
	Let $\mathscr{A}$ and $\mathscr{B}$ be two $\sigma$-algebras of $\mathscr{F}$. Define:
	\begin{equation*}
	\alpha(\mathscr{A}, \mathscr{B}) = \sup(|P(AB) - P(A)P(B)|, A \in \mathscr{A}, B \in \mathscr{B})
	\end{equation*}
	For $U \subseteq D_{NT}$ and $V \subseteq D_{NT}$, let $\sigma_{NT}(U) = \sigma(\zeta_{it}, (i,t) \in U)$ and $\alpha_{NT}(U, V) = \alpha(\sigma_{NT}(U), \sigma_{NT}(V))$. For $u, v, h \in \mathbb{N}$, the $\alpha$-mixing coefficient for the random field $\zeta$ is defined as
	\begin{equation*}
	\overline{\alpha}(u, v, h) = \displaystyle \sup_{N,T} \displaystyle \sup_{U,V}(\alpha_{NT}(U, V), |U| \leq u, |V| \leq v, \rho(U, V)\geq h).
	\end{equation*}
\end{definition}

Unlike the standard mixing time-series processes, the mixing coefficients for random fields depend not only on the distance between two sets but also on their sizes. We further assume that $\overline{\alpha }(u,v,h)\leq
\varphi (u,v)\widehat{\alpha }(h)$, where the function $\varphi (u,v)$ is nondecreasing with $u$ and $v$, and $\widehat{\alpha }(h)\rightarrow 0$ as $h\rightarrow \infty $. This implies the two different sources of dependence: (i) the decay of dependence with the distance, and (ii) the accumulation of dependence as the sample region expands. In the random field literature, $\varphi
(u,v)=(u+v)^{a},a\geq 0$ or $\varphi (u,v)=\min (u,v)$, can be commonly selected, see \citet{Jenish2012}.

Define $\mathcal{C}_f \defeq \sigma ( \fit_{t}, \cdots, \fit_{t-p}) $, $\mathcal{C}_z \defeq \sigma ( Z_{1}, \cdots, Z_{N}  )$, where $\sigma(\cdot)$ denotes a sigma field, and  $\mathcal{C} = \mathcal{C}_f \cup \mathcal{C}_z $. We now discuss the NED properties of $\{Y_{it}\}_{i,t}$ on the base $\{U_{it}\}_{i,t}$; $\mathscr{F}_{it}(s) = \sigma(U_{i't'}, \mathcal{C}: (i',t')\in D_{NT}, \rho((i',t'),(i,t) )\leq s)$ is the $\sigma$-field generated by random vectors, $ U_{i't'}$ located within distance $s$ from $(i,t)$.

Notice that the innovation $U_{it}$ is assumed to be $i.i.d.$ over $i$ and $t$ in Assumption \ref{Ass_stationray}(1), though it is well-known that $i.i.d.$ is a special case of $\alpha$-mixing. 
The above condition implies that  $\{U_{it}\}_{i,t}$ is an $\alpha$-mixing random field with an $\alpha$-mixing coefficient, $\overline{\alpha}(u,v,h) \leq (u+v)^{\varsigma} \widehat{\alpha}(h)$, $\varsigma \geq 0$, where $\widehat{\alpha}(h)$  satisfies  $\displaystyle \sum_{h=1}^{\infty} h^{d(\varsigma_{0}+1) - 1} \widehat{\alpha}^{\delta /(4+2\delta)}(h) <\infty$ with $\varsigma_{0}=\delta \varsigma
/(2+\delta)$ and some constant $\delta >0$.

Following \cite{xu2015maximum}, we outline some conditions on NED properties of $\mathbb{Y}_{t}$.

\begin{assumption}
\label{Ass3}The network matrix, $W$ is non-stochastic with zero diagonals and uniformly bounded for all $N$ with absolute row and column sums such that the matrix $S_{t}=I_{N}-\mathbf{A}_{1t}W$ is nonsingular. We consider two cases for $w_{ij}\geq 0$ for any $i, j$.

(1) Case 1: $|w_{ij}|\leq \pi _{0}\rho (i,j)^{-c_{w}}$ with constants $\pi
_{0}\geq 0$ and $c_{w}>d$. In addition, there exists at most  $K(\geq 1)$
number of columns in $W$, with $\min_{u}|\gamma^0
_{1}(u)|\sum_{i=1}^{n}|w_{ij}|>\Upsilon ,$ where the positive constant $\Upsilon $ is defined in Assumption \ref{Ass_stationray} (1).

(2) Case 2: Two nodes influence each other only if they are located sufficiently close; namely, $w_{ij}\not=0$ if $\rho (i,j) \leq \bar\rho _{0}$ and $w_{ij}=0$ otherwise, where we set $\bar{\rho}_{0} > 1$ w.l.o.g.
\end{assumption}

Assumption \ref{Ass3} is mainly used to restrict the NED coefficients, $\varphi (s)\rightarrow 0$ as $s\rightarrow \infty $. Assumption \ref{Ass3} (1) allows two individuals to have direct interaction even though their locations are far away from each other, with the requirement that the strength of interaction $w_{ij}$ declines with the distance $\rho (i,j)$ in the power of $c_{w}$. This assumption is in line with \cite{xu2015maximum}. By excluding a limited number of nodes $K(\geq 1)$, the total effects on other units from each node should be bounded, i.e., we assume that  ${\sup }|\gamma_{1}^0(u)|\sup_{j}\sum_{i=1}^{N}|w_{ij}|<\Upsilon $ or $\mbox{sup}|\gamma_{1}^0(u)| \sup_{j}\sum_{i=1}^{N}|w_{ij}|<1$ w.l.o.g. This corresponds to the existence of a narrow number of units with large aggregate effects on others even as the total number of nodes rises. Assumption \ref{Ass3}(2) allows two individuals to have direct interaction only if they are located within a specific distance. Notice that this assumption does not allow  star nodes, but one can see \citet{pesaran2020econometric,pesaran2021estimation} and \citet{kapetanios2021detection} for some extensions.

Let $u_{it} = u_{it}(\gamma_{1},\phi ,\lambda ,\tau ) = Y_{it} -\gamma_{1}(\tau) \overline{Y}_{it} - X_{it}^{\top}\phi(\tau) -{R}_{it}^{\top}\lambda (\tau)$ with the check function, $\rho_{\tau}(u)=(\tau -\II(u \leq 0))u$ and $\psi_{\tau}(u) =\tau -\II(u \leq 0)$ (the directional derivative of $\rho_{\tau}(u)$). Proposition \ref{Pro1} provides the NED properties of $\{Y_{it}\}_{i,t}$, and its transformations $\{\rho_{\tau}(u_{it})\}_{i,t}$,  $\{\psi_{\tau}(u_{it})\}_{i,t}$ on the base $\{U_{it}\}_{i,t}$.

\begin{proposition}
\label{Pro1} (1) Under Assumptions \ref{Ass_stationray}(1), \ref{Ass1} and \ref{Ass3}(1), and conditional on $\mathcal{C}$, $\{Y_{it}\}_{i,t}$ is geometrically $L_{2}$-NED on $\{U_{it}\}_{i,t}$ such that $\Vert Y_{it} -\mathop{\mbox{\sf E}}(Y_{it}|\mathscr{F}_{it}(s))\Vert_{2} < Cs^{-(c_{w}-d)}$ for $c_{w}>d$ and some $C >0$ that does not depend on $i$ and $t$. Similarly for $\{u_{it}\}_{i,t}$. The transformations $\{\psi _{\tau}(u_{it})\}_{i,t}$ and $\{\rho _{\tau }(u_{it})\}_{i,t}$ are also $L_{2}$-NED on $\{U_{it}\}_{i,t}$.

(2) Under Assumptions \ref{Ass_stationray}(1), \ref{Ass1} and \ref{Ass3}(2), and conditional on $\mathcal{C}$, $\{Y_{it}\}_{i,t}$ is geometrically $L_{2}$-NED on $\{U_{it}\}_{it}
$ such that $\Vert Y_{it} -\mathop{\mbox{\sf E}}(Y_{it}|\mathscr{F}_{it}(s)) \Vert_{2} <C \Upsilon^{s/\bar{\rho}_{0}}\;(\Upsilon <1)$ for some $C>0$ that does not depend on $i$ and $t$. Similarly for $\{u_{it}\}_{i,t}$. The transformations, $\{\psi_{\tau}(u_{it})\}_{i,t}$
and $\{\rho_{\tau }(u_{it})\}_{i,t}$ are also $L_{2}$-NED on $\{U_{it}\}_{i,t}$.
\end{proposition}

Define $s_{it}(\gamma _{1}^{0},\eta ^{0}(\gamma _{1}^{0},\tau ),\tau )=\psi
_{\tau }\left\{ Y_{it}-\gamma _{1}^{0}(\tau )\overline{Y}_{it}-\Psi
_{it}^{\top }\eta ^{0}(\gamma _{1}^{0},\tau )\right\} \Psi _{it}$ where $\Psi_{it} = (R^{\top}_{it}, X^{\top}_{it})^{\top}$,  $\check{s}_{it}=%
\check{s}_{it}(\gamma _{1}^{0},\eta ^{0}(\gamma _{1}^{0},\tau ),\tau
)=s_{it}(\gamma_{1}^{0},\eta ^{0}(\gamma_{1}^{0},\tau),\tau ) - \mathop{\mbox{\sf E}}s_{it}(\gamma_{1}^{0},\eta ^{0}(\gamma _{1}^{0},\tau ),\tau )$ and $\mathop{\mbox{\sf E}}s_{it}(\gamma_{1}^{0},\eta ^{0}(\gamma _{1}^{0},\tau ),\tau ) = 0$ . Conditioning on $\mathcal{C}$, it is easily seen that the process $\{\check{s}_{it}\}_{i,t}$ is NED. 
Define an array of positive constants by $\{c_{it},(i,t) \in D_{NT},NT\geq 1\}$. To derive the central limit theorem for $G_{NT}^{0} = \frac{1}{\sqrt{NT}}
\sum_{i=1}^{N}\sum_{t=1}^{T}\check{s}_{it}  = 
\frac{1}{\sqrt{NT}}\sum_{i=1}^{N}\sum_{t=1}^{T}[s_{it}(\gamma _{1}^{0},\eta^{0}(\gamma _{1}^{0},\tau ),\tau )-\mathop{\mbox{\sf E}}s_{it}(\gamma_{1}^{0},\eta ^{0}(\gamma _{1}^{0},\tau ),\tau )]$, where the variance of $G_{NT}^{0}$ is given by $\displaystyle \Omega_{0} = \tau(1- \tau)  \lim_{N, T\to\infty} (NT)^{-1}\sum_{i,t}\E (\Psi_{it} \Psi_{it}^{\top}|\mathcal{C}) $, we make the following assumptions:

\begin{assumption}
\label{Ass5} 
$\rho _{\tau }(u_{it})/c_{it}$ is uniformly $L_{p'}$-bounded for $p'>1$, i.e.,\newline
$\sup_{N,T}\sup_{(i,t) \in D_{NT}} \mathop{\mbox{\sf E}}|\rho _{\tau
}(u_{it})/c_{it}|^{p'}<\infty$, where $p'$ is an integer.

\end{assumption}

\begin{assumption} \label{Ass6} 
(Uniform $L_{2+\delta}$ Integrability)

(1) $\displaystyle\lim_{k\rightarrow \infty
}\sup_{N,T}\sup_{(i,t)\in D_{NT}}\mathop{\mbox{\sf E}}\{|\check{s}%
_{it}/c_{it}|^{2+\delta } \II(|\check{s}_{it}/c_{it}|>k)\}=0$ for $\delta >0$.

(2) $\displaystyle \inf_{N,T} |D_{NT}|^{-1}M_{NT}^{-2}\Omega_{0} > 0$ (implying that the matrix is positive definite), 
where $\displaystyle M_{NT} = \max_{(i,t) \in D_{NT}} c_{it}. $

(3) NED coefficients satisfy: $\sum_{h=1}^{\infty }h^{d-1}\varphi
(h)<\infty $, and NED scaling factors satisfy: \newline
$\displaystyle\sup_{N,T}\sup_{(i,t)\in D_{NT}}c_{it}^{-1}d_{it}\leqslant
C<\infty $,  where $C$ is a finite constant.
\end{assumption}

Assumption \ref{Ass5} imposes the moment conditions of $\rho_{\tau}(u_{it})$ while Assumption \ref{Ass6} sets the mixing coefficients of the underlying mixing fields. Since $U_{it}$ is assumed to be $i.i.d.$, these conditions are automatically satisfied.

\subsubsection{Asymptotic Distribution of the IVQR Estimator}
\label{sec_asydis}

\begin{assumption}[Conditions for identification and estimation]  \label{Ass_id}
(1) (Compactness and Convexity) For all $\tau$, $(\gamma_1(\tau), \phi(\tau)) \in \mathcal{A} \times \mathcal{B}$, where $\mathcal{A} \times \mathcal{B}$ is compact and convex. 
	
\noindent (2) (Full Rank and Continuity) $\yit_{t}$ has bounded conditional density, a.s. $\sup_{\yit_{t} \in \rit^{N}} f_{\yit_{t} | \F_{t}}(y) < K$, where $\F_{t}=\{Z_{1},\cdots,Z_N,  \yit_{t-1}, \yit_{t}, F_{t}, F_{t-1}, \cdots, F_{t-p}\}$ 
is the information set. Define
	\begin{eqnarray}\label{equ_s}
	S_{NT}(\pi, \tau)  
	&=& \frac{1}{NT}\sum_{i=1}^{N} \sum^T_{t=1} \left[\psi_{\tau}\left\{Y_{it} - \gamma_{1}(\tau)\overline{Y}_{it} - X^{\top}_{it}\phi(\tau) - R^{\top}_{it} \lambda(\tau) \right\} \Psi_{it} \right] ,\\	
	{S}_{\infty}(\pi, \tau) &=&
	\lim_{N, T\to\infty}\E\left[S_{NT}(\pi, \tau) |\mathcal{C} \right] ,\quad {S}^*_{\infty}(\pi, \tau) =
	\lim_{N, T\to\infty}\E\left[S_{NT}(\pi, \tau) \right], \\
	S_{NT}(\theta, \tau) &=& \frac{1}{NT}\sum_{i=1}^{N} \sum^T_{t=1} \left[\psi_{\tau}\left \{Y_{it} - \gamma_{1}(\tau)\overline{Y}_{it} - X^{\top}_{it}\phi(\tau)  \right\} \Psi_{it} \right], \\	
	{S}_{\infty}(\theta, \tau) &=&
	\lim_{N, T\to\infty}\E\left[S_{NT}(\theta, \tau)  |\mathcal{C} \right] ,\quad {S}^*_{\infty}(\theta, \tau) =
	\lim_{N, T\to\infty} \E\left[S_{NT}(\theta, \tau)  \right], 
	\end{eqnarray}
	where $\pi \equiv (\gamma_1, \phi^{\top}, \lambda^{\top})^{\top}$, $\theta \equiv (\gamma_1, \phi^{\top})^{\top}$,  $\psi_{\tau}(u) = \tau - \IF(u<0)$, and $\Psi_{it} = (R^{\top}_{it}, X^{\top}_{it})^{\top}$. Then, the Jacobian matrices, $\dfrac{\partial S_{\infty}(\theta, \tau)}{\partial (\gamma_1, \phi^{\top}) }$  and $\dfrac{\partial S_{\infty}(\pi, \tau)}{\partial (\phi^{\top}, \lambda^{\top}) }$ are continuous and have full rank, uniformly over $\mathcal{A} \times \mathcal{B}\times \mathcal{G}\times \mathcal{T}$, where $\mathcal{G}$ is a compact set with $\lambda(\tau) \in \mathcal{G}$, $\mathcal{T}$ is a compact set with $\tau \in \mathcal{T}$, and the image of $\mathcal{A} \times \mathcal{B}$ under the mapping $\theta \equiv (\gamma_1, \phi^{\top})^{\top} \mapsto {S}_{\infty}(\theta, \tau) $ is simply connected.
	
\noindent (3) For a fixed $\tau \in \mathcal{T}$, the unknown true parameter, $\theta^{0}(\tau) = (\gamma_1^{0}(\tau), \phi^{0\top}(\tau))^{\top}$ uniquely solves ${S}_{\infty}(\theta, \tau) = 0$ over $\mathcal{A} \times \mathcal{B}$.
	\label{assumption_estimation}
\end{assumption}

The compactness of the parameter space in Assumption \ref{Ass_id}(1) is needed for $\gamma_{1}(\tau)$ due to the non-convex objective function. Assumption \ref{Ass_id}(2) implies the global identification of the parameters while the continuity condition is required for deriving the asymptotic normality. Assumption \ref{Ass_id}(3) requires that $\theta^{0}(\tau) = (\gamma_1^{0}(\tau), \phi^{0\top}(\tau))^{\top}$ to be the unique solution to ${S}_{\infty}(\theta, \tau) =0$, which is necessary for consistency of the estimator.

Let $\hat{\theta}(\tau) = (\hat{\gamma}_{1}(\tau),\hat{\phi}^{\top}(\tau))^{\top}$ be the IVQR estimator of $\theta^0 (\tau) =(\gamma_{1}^0(\tau ),\phi^{0\top}(\tau ))^{\top}$, where $\hat{\phi}(\tau) = \hat{\phi}(\hat{\gamma}_{1}(\tau),\tau)$. 
Define the $(4+q+(p+1)m)\times (4+q+(p+1)m)$ matrices:
\begin{eqnarray} \label{jr}
J(\tau) &= & \dfrac{\partial S_{\infty}(\pi, \tau)}{\partial (\gamma_1, \phi^{\top})} \Big|_{ \gamma_1 = \gamma^0_1,\phi = \phi^0,\lambda = \mathbf{0}}, \quad J^*(\tau)=  \dfrac{\partial S^*_{\infty}(\pi, \tau)}{\partial (\gamma_1, \phi^{\top})} \Big|_{ \gamma_1 = \gamma^0_1,\phi = \phi^0,\lambda = \mathbf{0}}.
\end{eqnarray}

\begin{theorem}[Linearization] \label{theorem_linearization}  
Under Assumptions \ref{Ass1}--\ref{Ass5} and \ref{assumption_estimation}, as $\min \{N,T\} \rightarrow \infty$, 
\begin{align} \label{equ_esti}
	\sqrt{NT} \left\{\hat{\theta}(\tau) - \theta^{0}(\tau))\right\} = & - J^{-1}(\tau) G^{0}_{NT} (\theta^{0},\tau) + o_{p}(1).
	\end{align}
\end{theorem}

Under Assumption \ref{Ass6}, the NED process, $\{\check{s}_{it}\}_{i,t}$ satisfies the central limit theorem: $G_{NT}^{0}(\theta^{0},\tau) = \frac{1}{\sqrt{NT}} \sum_{i=1}^{N}\sum_{t=1}^{T} \check{s}_{it}$ follows a zero mean Gaussian process with covariance function, $\displaystyle \Omega_{0} = \tau(1- \tau) \lim_{N, T\to\infty} (NT)^{-1}\sum_{i,t}\E (\Psi_{it} \Psi_{it}^{\top}) $.

\begin{theorem} \label{theorem_estimation}
Under Assumption \ref{Ass1}--\ref{assumption_estimation}, we have $\Omega_0^{-1}\Omega_0^{*}\to_p I$, and $J^{-1}(\tau)J^{*}(\tau)\to_p I$, where 
$\displaystyle \Omega_{0} = \tau(1- \tau) \lim_{N, T\to\infty} (NT)^{-1}\sum_{i,t}\E (\Psi_{it} \Psi_{it}^{\top}|\mathcal{C}) $, $\displaystyle \Omega_{0}^* = \tau(1- \tau) \lim_{N,T \to \infty}(NT)^{-1}\sum_{i,t} \E (\Psi_{it} \Psi_{it}^{\top})$, and $J(\tau),J^{*}(\tau)$ are defined in (\ref{jr}). Then, as $\min \{N,T\} \rightarrow \infty$,
\begin{equation}\label{equ_esti_asym}
\sqrt{NT}\left\{\widehat{\theta}(\tau) - \theta^0(\tau)\right\} \stackrel{d} {\rightarrow}  \N  (0, \Sigma_{\theta}),
\end{equation}
where $\Sigma_{\theta} =  J^{*}(\tau)^{-1} \Omega_{0}^* J^{*}(\tau)^{-1}$.
\end{theorem}

We estimate $\Omega_{0}^*$ and $J^{*}(\tau)$ consistently by
\begin{eqnarray}\label{varcovesti}
\hat\Omega_{0}^* &= & \displaystyle (NT)^{-1}\tau(1-\tau) \sum_{i=1}^{N}\sum_{t=1}^{T} \Psi_{it} \Psi_{it}^{\top}, \\    
\hat{J}^{*}(\tau) & = &\displaystyle (2NTh_{b})^{-1} \sum_{i=1}^{N}\sum_{t=1}^{T} \IF(\hat{u}_{it} \leq h_{b}) \Psi_{it} (\overline{Y}_{it},  X_{it}^{\top}),
\end{eqnarray}
where $\hat{u}_{it} = u_{it}(\widehat{\theta}(\tau), \tau )$, and $h_{b}$ is the bandwidth (see \eqref{h_hs} below for details). Notice that the use of valid IVs do not result in $\hat{J}^{*}(\tau)$ and $\hat\Omega_{0}^*$ having or being closer to singularities, which causes the variance of the estimator $\hat{\theta}(\tau)$ to be unreliably large.

\section{Monte Carlo Simulations} \label{simulation}

In this section, we examine finite sample properties of the IVQR estimator via a Monte Carlo simulation study 
using three different network structures.

\subsection{The Setup}\label{mcsetup}         

We construct the data generating process based on the DNQR model as follows: First, we generate the five nodal covariates, $Z_{i}=(Z_{i1},\cdots ,Z_{i5})^{\top} \in \mathbb{R}^{5}, (q=5)$ from a multivariate normal distribution $\N (\mathbf{0},\Sigma _{z})$, where $\Sigma _{z}=(\sigma _{j_{1}j_{2}})$ and $\sigma
_{j_{1}j_{2}}=0.5^{|j_{1}-j_{2}|}$. Then, we construct the two common covariates, $F_{t}=(F_{1t},F_{2t})^{\top }\in \mathbb{R}^{2} , (m=2)$ from the i.i.d standard normal distribution. Let the true parameters $\gamma ^{0} _{j,it}=\gamma ^{0}_{j}(U_{it})$ for $%
j=0,1,2,3$, $\alpha^{0} _{j,it}=\alpha^{0} _{j}(U_{it})$ for $j=1,...,5$, and $\beta^{0}
_{jk,it}=\beta^{0} _{jk}(U_{it})$ for $j=1,2$ and $k=0,1,$ where we set the lag
of two common covariates to 1 ($p=1)$. We then generate the random coefficients by
\begin{align}
\gamma^{0} _{0,it}=& \mathfrak{u}_{it},~\gamma^{0} _{1,it}=0.1\Phi (\mathfrak{u}_{it}),~\gamma^{0}
_{2,it}=0.4\{1+\exp (\mathfrak{u}_{it})\}^{-1}\exp (\mathfrak{u}_{it}),~\gamma^{0} _{3,it}=0.4\Phi
(\mathfrak{u}_{it}),  \notag \\
\alpha^{0} _{1,it}=& 0.5\Phi (\mathfrak{u}_{it}),~\alpha^{0} _{2,it}=0.3\mathbf{G}%
(\mathfrak{u}_{it},1,2),~\alpha^{0} _{3,it}=0.2\mathbf{G}(\mathfrak{u}_{it},2,2),  \notag \\
\alpha^{0} _{4,it}=& 0.25\mathbf{G}(\mathfrak{u}_{it},3,2),\alpha^{0} _{5,it}=0.2\mathbf{G}%
(\mathfrak{u}_{it},2,1),  \notag \\
\beta^{0} _{10,it}=&0.1\Phi (\mathfrak{u}_{it}),~\beta^{0} _{11,it}=0.3\mathbf{G}%
(\mathfrak{u}_{it},2,2),\beta^{0} _{20,it}=0.2\mathbf{G}(\mathfrak{u}_{it},1,2),~\beta^{0} _{21,it}=0.3%
\mathbf{G}(\mathfrak{u}_{it},2,1),  \notag
\end{align}%
where $\Phi (\cdot )$ is the standard normal distribution function, $\mathbf{G}(\cdot ,a,b)$ is the Gamma distribution function with shape parameter $a$ and scale parameter $b$, and $\mathfrak{u}_{it}$s are i.i.d random variables, generated either from (a) the standard normal distribution or from (b) the $t$-distribution with $5$ degrees of freedom. Notice that $U_{it}$ can be generated by $U_{it}=F(\mathfrak{u}_{it})$, where $F(\cdot)$ is cumulative distribution function of $\mathfrak{u}_{it}$. 
Finally, $\mathbb{Y}_{t}$s are generated by \eqref{model1}.

To check the robustness of the finite sample performance of the IVQR estimator, we consider the following three different adjacency matrices, 
e.g. \cite{zhu2019}.

\textsc{Type 1.} (Dyad Independence Model) \cite{holland1981exponential} introduce this model with a dyad, $D_{ij}=(a_{ij},a_{ji})$ for 
$1\leq i<j\leq N$, where $D_{ij}$s are assumed to be independent. We set the probability of dyad being mutually connected to $\P(D_{ij}=(1,1))=2N^{-1}$
to ensure the network sparsity. Then, we set $\P(D_{ij}=(1,0))=\P(D_{ij}=(0,1))=0.5N^{-0.8}$, which implies that the expected degree for each node is $O(N^{0.2})$. Accordingly, we
have $\P(D_{ij}=(0,0))=1-2N^{-1}-N^{-0.8}$, which tends to 1 as $N \rightarrow \infty $.

\textsc{Type 2.} (Stochastic Block Model) We first consider the Stochastic Block Model 
with an important application in community detection by \cite{zhao2012consistency}. We
follow \cite{nowicki2001estimation} and randomly assign each node a block label index from 1 to $L$, where $L\in \{5,10,20\}$. We then set $%
\P(a_{ij}=1)=0.3N^{-0.3}$ if $i$ and $j$ are in the same block, and $\P(a_{ij}=1)=0.3N^{-1}$ otherwise. Thus, the nodes within the same block have higher probability of connecting with each other than the nodes between blocks.

\textsc{Type 3.} (Power-law Distribution Network) In practice, the majority of nodes in the network have a small number of links while a small
number of nodes have a large number of links, see  \cite{barabasi1999emergence}. In this case the degrees of nodes can be characterized by the power-law distribution. We simulate the adjacency matrix as follows: For each node, we generate the in-degree, $d_{i}=\sum_{j}a_{ji}$ according to the discrete power-law distribution such as $\P(d_{i}=\check{k})=c\check{k}^{-\beta }$, where $c$ is a
normalizing constant and the exponent parameter $\beta $ is set at $2.5$ as in \cite{clauset2009power}. Finally, for the $i$-th node, we randomly select $d_{i}$ nodes as its followers.

To estimate the variance of the IVQR estimator, we follow \cite{koenker2006quantile} and select the bandwidth $h_{b}$ in \eqref{varcovesti} as follows (\cite{hall1988distribution}):
    \begin{equation}\label{h_hs}
        h_{b}=(NT)^{-1/3}\varrho_{\alpha}^{2/3}\left[\frac{1.5 \bar{\varphi}^2(\Phi^{-1}(\tau))}{2(\Phi^{-1}(\tau))^2+1}\right]^{1/3} ,
    \end{equation}
where $\bar{\varphi}(\cdot)$ and $\Phi(\cdot)$ are the probability density and distribution function of standard normal distribution and $\varrho_{\alpha}$ satisfies $\Phi(\varrho_{\alpha})=1-\alpha/2$ for the construction of $1-\alpha$ confidence intervals. We have also considered an alternative selection criterion by \cite{bofinger1975estimation}, and obtained qualitatively similar results that are available upon request.

For the IVQR estimation, we suggest using $\mathbf{R}_{t}=[W^{2}\mathbb{Y}_{t-1},W^{3}\mathbb{Y}_{t-1}]$, where $\mathbf{R}_{t}=(R_{1t},\cdots ,R_{Nt})^{\top}$ and $W$ is the row-sum normalized network matrix. Although we may select the higher network orders such as $[W^{2}
\mathbb{Y}_{t-1},W^{3}\mathbb{Y}_{t-1},W^{2}\mathbb{Y}_{t-2},W^{3}\mathbb{Y}_{t-2},...]$, we find that these two instruments are often the best choice.

\subsection{Simulation Results}\label{mcresults}

Using $1000$ replications, we evaluate the finite sample performance of the IVQR estimator at the different quantiles, $\tau = 0.1,0.5,0.9$ for the $(N,T)$ pairs with $N,T = 100,200,500$.
Table \ref{tab_simu_RMSE_W1} presents the simulation results for \textsc{Type 1} network in terms of RMSE. Overall, RMSEs of all the parameters decrease monotonically as $N$ or $T$ increases across the different quantiles and for the different distributions of $\mathfrak{u}_{it}$, which is in line with the asymptotic theory. But,  RMSEs of $\gamma_{1}$ are larger than those of other parameters, especially in small samples ($N,T =100$), which mainly reflects uncertainty associated with the selection of the IVs. As the sample size grows ($N=500$ or $T=500$), all RMSEs decline sharply. Turning to the simulation results for \textsc{Type 2} and \textsc{Type 3} networks, reported in \ref{tab_simu_RMSE_W2} and \ref{tab_simu_RMSE_W3} in the Online Appendix, we observe qualitatively similar findings. 
In the Online Appendix, we also provide the simulation results for biases, see Tables \ref{tab_simu_Bias_W1}--\ref{tab_simu_Bias_W3}. 
The biases of the IVQR estimators are mostly negligible across the different quantiles, for the different distributions of $\mathfrak{u}_{it}$ and for all the sample pairs of $(N, T)$.

{[Insert  Table \ref{tab_simu_RMSE_W1} here]}

Next, in Table \ref{tab_simu_CP_W1}, we report the coverage probability for the \textsc{Type 1} network  by evaluating if the estimates fall into the 95\% confidence interval at each replication.
Overall, we find that the coverage probabilities of all the parameters are close to the nominal 95\% level across the different quantiles and the different distributions of $\mathfrak{u}_{it}$, and for all the sample pairs $(N,T)$. 
This implies the accuracy of the inference for the IVQR estimator. From the results for \textsc{Type 2} and \textsc{Type 3} networks, reported in Tables \ref{tab_simu_CP_W2} and \ref{tab_simu_CP_W3} in the Online Appendix, we also find that the coverage probabilities of all parameters are close to the nominal 95\% level.

[Insert Table \ref{tab_simu_CP_W1} here]

For comparison, we report the simulation results by applying the ordinary QR estimator in Tables \ref{tab_simu_RMSE_W1_NonIV}--\ref{tab_simu_Bias_W3_NonIV} in the Online Appendix \ref{appendixb_b2}. We observe that RMSEs are larger. More importantly, RMSEs barely decrease with the sample size, especially for $\gamma_{1}$. Furthermore, coverage probabilities are well below the nominal $95\%$ level,  and the biases are large especially for $\gamma_{1}$, which remain substantial as the sample size increases.
This clearly demonstrates the importance of using the IVQR estimator for the DNQR model. 

\section{Application}\label{application}

We explore the financial network quantile connectedness among the stock returns. \citet{Anton2014} find that stock returns tend to display significant comovements due to common active mutual fund owners. In addition, \citet{Pirinsky2006} document strong comovements in the stock returns of firms headquartered in the same geographic area. \citet{Garcia2012} point out that the firms headquartered in the same geographic area have achieved uniformly excessive returns compared to geographically dispersed firms (the return local bias).

We consider the two different financial network structures: the common shareholder based network and the headquarter location based network. Notice that the order of network nodes can be randomly pre-determined and has no effect on the estimation results. For simplicity, we set up $\mathbb{Y}_{t} \in \mathbb{R}^{N}$ in which the row elements are arranged in alphabetical order by the unique trading code. Then the order of nodes in the pre-determined networks are the same as the individual order in $\mathbb{Y}_{t}$, that is alphabetically ordered by the unique stock trading code. The pre-determined networks are constructed by using information on the common mutual fund ownership and the uniform headquarter location. In particular, we let the stocks be connected if they are invested in by at least five common shareholders ($W_{CS5}$) while the companies with headquarters located in the same city are treated as connected ($W_{HQ}$).

We collect the data on all the stocks traded in NYSE and NASDAQ in 2016 from Datastream. The dataset on mutual fund holdings are downloaded from Thomson Reuters whilst the addresses of firms' headquarters are collected from COMPUSTAT. After merging these data from the databases according to the unique trading code and removing the stocks with missing values, we finally obtain 943 stocks ($N = 943$) over the whole time period $T = 252$. We then collect these stock return data from Datastream. We also obtain node-specific covariates such as market capitalization, book value per share, cash flow and price-earning ratio from Datastream, which are then standardized. Finally, we collect VIX from Datastream and the Fama-French three factors (excess market return, SMB, HML) from the website of French's homepage as the common covariates. 

\begin{figure}[!htbp]
\begin{subfigure}{0.5\textwidth}
\centering
\includegraphics[scale=0.29]{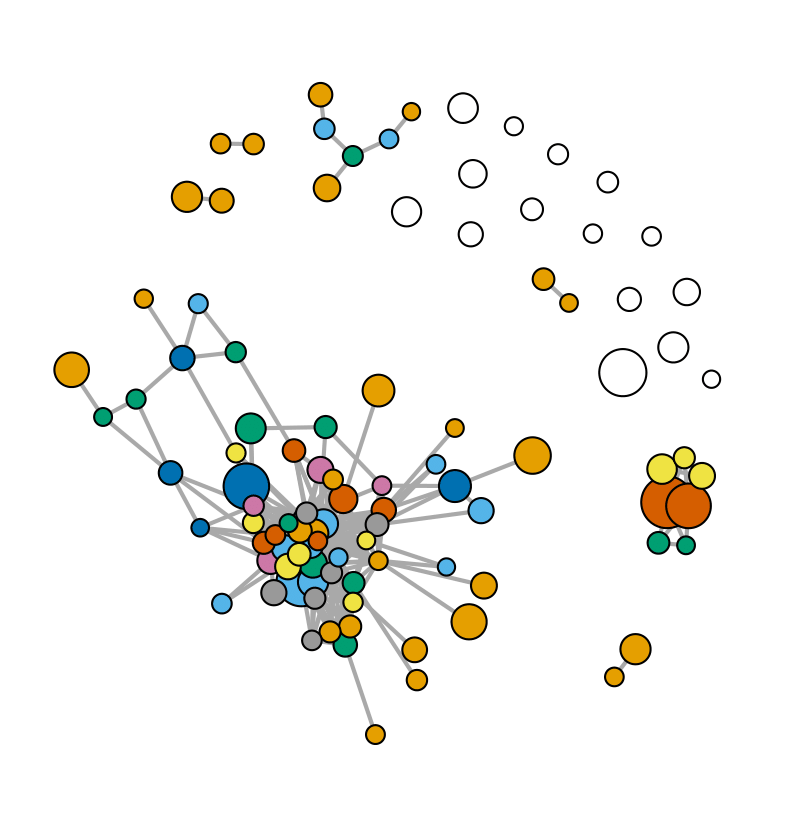}
\caption{The Topology of the Network $W_{CS5}$}\label{NetworkfigK5}
\end{subfigure}
\begin{subfigure}{0.5\textwidth}
\centering
\includegraphics[scale=0.29]{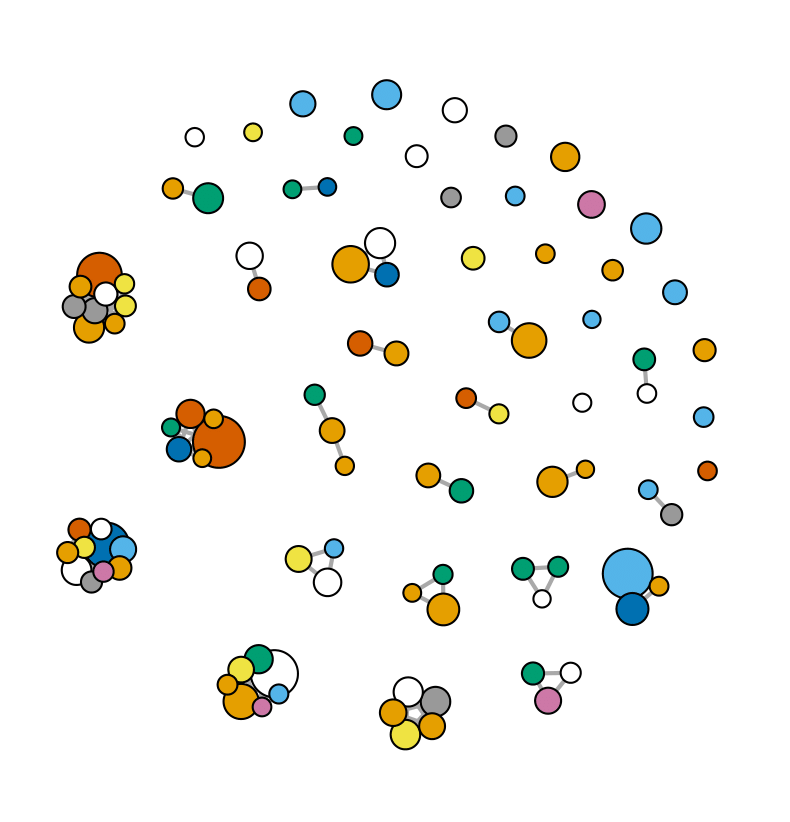}
\caption{The Topology of the Network $W_{HQ}$}\label{NetworkfigHQ}
\end{subfigure}
\caption{\footnotesize We depict the top 100 market value stocks out of 943 firms selected. (a): the common shareholder network $W_{CS5}$, constructed by checking if the stocks are invested in by at least five common shareholders. (b): the uniform headquarter location network $W_{HQ}$, constructed by checking if the headquarters of companies are located in the same city. The larger nodes imply higher market capitalization while the darker nodes present higher connectedness.} \label{Networkfig}
\end{figure}

The network density is 3.24\% for $W_{CS5}$ and 0.63 \% for $W_{HQ}$, respectively. In Figure \ref{Networkfig} we display the topology of two networks for the top 100 market-value stocks only for visualization convenience. The larger nodes imply the higher market capitalization while the darker nodes present the higher connectedness especially for the network with $W_{CS5}$. Here we observe quite different network structures. There is a large connected group in Figure \ref{NetworkfigK5}, showing that the stocks are more centrally connected by common investors. On the contrary Figure \ref{NetworkfigHQ} displays more small groups, implying that the stocks are more locally connected when the network is measured by uniform headquarter locations.

[Insert Figure \ref{Networkfig} here]

Following the simulation results, we select IVs as
$\mathbf{R}_{t}=[W^{2}\mathbb{Y}_{t-1},W^{3}\mathbb{Y}_{t-1}]$. We present the estimation results for the proposed DNQR model together with the two alternative models: (i) the original NQAR model without contemporaneous network effects and common covariates, and (ii) the factor-augmented NQAR model without contemporaneous network effects, denoted NQARF.
To compare the relative performance of the alternative models, we follow \citet{Koenker1999} and evaluate the goodness of fits across the different quantiles. Consider a linear model for the conditional quantile function, 
\begin{align}\label{gfitmodel}
Q_{Y_{it}}(\tau|X_{it})  = & X^{\top}_{1it} \theta_{1}(\tau) + X^{\top}_{2it} \theta_{2}(\tau) 
\end{align}
where $X_{it}=(X^{\top}_{1it},X^{\top}_{2it})^{\top}$. Let $\widehat{\theta}(\tau)=(\widehat{\theta}_{1}^{\top}(\tau),\widehat{\theta}_{2}^{\top}(\tau))^{\top}$ be the unrestricted estimator, which is the minimizer of  $\widehat{V}(\tau)  = \min \displaystyle \sum_{i=1}^{N}
\displaystyle \sum_{t=1}^{T} \rho_{\tau} \left\{Y_{it} - X^{\top}_{it} \theta \right\}$
while $\widetilde{\theta}(\tau) = (\widetilde{\theta}^{\top}_{1}(\tau), \boldsymbol{0}^{\top})^{\top} $ denotes the minimizer for the constrained model, $\widetilde{V}(\tau)  = \min \displaystyle \sum_{i=1}^{N}
\displaystyle \sum_{t=1}^{T} \rho_{\tau} \left\{Y_{it} - X^{\top}_{1it} \theta_{1} \right\}$.
Define the goodness-of-fit criterion as 
\begin{align}
R^2(\tau) = & 1 - \widehat{V}(\tau)/\widetilde{V}(\tau) , 
\end{align}
which measures the overall decreased percentage of the quantile loss function of the unrestricted model with respect to the restricted model.

The estimation results for the network $W_{CS5}$ are presented in Table \ref{TabAppW1}. For convenience we present the coefficients and the standard errors multiplied by $10^{2}$. 
We find that the estimated contemporaneous network effects ($\gamma_1$) by the DNQR model, are significantly positive and dominate all other effects across all quantiles. The lagged network coefficient ($\gamma_2$) and the dynamic coefficient ($\gamma_3$) are also significant across quantiles with relatively smaller magnitudes. 
Furthermore, the goodness of fit, $R^2(\tau)$ reported in the last row, shows that the overall loss function of the DNQR model drops about 7\%--9.5\% relative to the NQAR and 6.9\%--9.5\% relative to the NQARF, respectively, suggesting that the contemporaneous network effects should be explicitly accommodated in the dynamic network quantile model. 
Moreover, the effects of node-specific covariates are significant across quantiles (except Cash) while those of common covariates are all significant across quantiles.

[Insert  Table \ref{TabAppW1} here]

Next, we display the QR coefficients across the different quantiles in Figure \ref{AppParW1}. The dashed line is the QR coefficient while the grey area indicates a kernel density based confidence band advanced by \cite{powell1991}. The contemporaneous network and dynamic coefficients, $\hat{\gamma}_1(\cdot)$ and $\hat{\gamma}_3(\cdot)$, are significant (as their bands exclude the null effect), while the lagged diffusion network coefficient, $\hat{\gamma}_2(\cdot)$ tends to be insignificant only at the middle. $\hat{\gamma}_1(\cdot)$ shows a downward trend with the quantile level, suggesting that the contemporaneous quantile connectedness is stronger at the lower tails (mainly characterised with the market turmoils).
On the other hand, both $\hat{\gamma}_2(\cdot)$ and $\hat{\gamma}_3(\cdot)$ display the $U$-shaped pattern, implying that their effects are stronger at the tails than at the median. 
The QR effects of node-sepcific covariates mostly display a downward trend with the quantile level (except for insignificant Cash), suggesting their effects are stronger at the lower tails than at the upper tails (mainly chracterised with the bulls market). 
Turning to the QR effects of common factors, we observe a mixed finding: the impacts of VIX and the market factor increase with quantiles whilst those of SMB and HML factors decrease with quantiles.

[Insert Figure \ref{AppParW1} here]

Finally, as the robustness check, we provide the two additional estimation results in the Online Appendix. First, we reconstruct the network matrix by changing the number of common shareholders to $CS = 3$ $(W_{CS3})$ and $CS = 7$ $(W_{CS7})$, and the results are reported in Tables \ref{TabAppW3}--\ref{TabAppW7} and Figures \ref{AppParW3}--\ref{AppParW7}. Notice that the network density of $W_{CS3}$ is dense at 25.25\% and relatively sparse at 0.41\% for $W_{CS7}$. 
Overall, we find qualitatively similar results to those reported for $W_{CS5}$. 
One notable observation is that the contemporaneous network effects measured by $\gamma_1$ tend to decrease monotonically as the network becomes more sparse. For example, at $\tau = 0.1$,  $\hat{\gamma}_1$ is estimated at 0.69 for $W_{CS3}$, 0.54 for $W_{CS5}$, and 0.35 $W_{CS7}$, respectively. Still, we find that the patterns of the quantile specific coefficients reported in Figures \ref{AppParW3}--\ref{AppParW7} are qualitatively similar to those displayed in Figure \ref{AppParW1}.

Next, we estimate the models using the headquarter location network $W_{HQ}$, and present the estimation results in Table \ref{TabAppW2} and Figure \ref{AppParW2} in the Online Appendix. Again, we find the qualitatively similar results, highlighting the importance of the contemporaneous network effect, which is also stronger at the lower tails than at the upper tails.

\section{Conclusion} \label{conclusion}

We develop a dynamic network quantile model that accommodates both temporal and cross-sectional dependence. Using the predetermined network information, we analyse the dynamic  quantile connectedness within a network topology. 
The distinguishing feature of the DNQR model lies in that the behavior/response of a given node is not only influenced by its previous behavior/response, but also connected with a weighted average of contemporaneous and lagged behaviors/responses from peers.

The main challenge associated with the DNQR model is the presence of endogeneity stemming from the simultaneous network effect. In this regard, we develop the IVQR estimation, and derive the consistency and asymptotic normality of the IVQR estimator using the NED property of the network process. Monte Carlo exercises confirm the satisfactory performance of the IVQR estimator with the predetermined internal IVs across different quantiles under the different network structures. 

Finally, we demonstrate the usefulness of our proposed approach with an application to the dataset on the stocks traded in NYSE and NASDAQ in 2016. In particular, we find that the contemporary network effects are significant and dominant across all quantiles. Furthermore, their effects display a downward trend with the quantile level, suggesting that the contemporaneous quantile connectedness is stronger at the lower tails. 

\section*{Acknowledgements}
We are mostly grateful for the insightful comments by the editor, the associated editor and three anonymous referees. Xu acknowledges partial financial support of the Natural Science Foundation of China (Grant No. 71803140). Shin and Wang gratefully acknowledge partial financial support from the ESRC (Grant Reference: ES/T01573X/1). Zheng gratefully acknowledges partial financial support from the Royal Economic Society. The usual disclaimer applies.


\begin{table}[!htbp]
\begin{center}
\caption{RMSE ($\times 100$) for \textsc{Type 1} Network}
\label{tab_simu_RMSE_W1}
{\scriptsize 
\begin{tabular}{ccc|ccccccccccccc}
\hline
\hline
&Dist. & $\tau$ &$\gamma_0 $  & $\gamma_1$&  $\gamma_2$ &  $\gamma_3$ & $\alpha_1$ & $\alpha_2$ &$\alpha_3$ & $\alpha_4$ & $\alpha_5$ & $\beta_1$ & $\beta_2$ & $\beta_3$ & $\beta_4$  \\

\hline
&&& \multicolumn{13}{c}{$ N = 100 $} \\
  \multirow{6}*{$T=100$} &$N(0,1)$ &  0.1 & 1.64 & 5.35 & 1.41 & 3.04 & 1.75 & 1.87 & 1.77 & 1.83 & 1.61 & 1.47 & 1.64 & 1.42 & 1.42 \\ 
  &  &  0.5 & 1.49 & 4.75 & 1.19 & 2.66 & 1.38 & 1.58 & 1.46 & 1.51 & 1.31 & 1.17 & 1.33 & 1.14 & 1.13 \\ 
  &  &  0.9 & 1.71 & 5.18 & 1.39 & 2.95 & 1.63 & 1.74 & 1.73 & 1.74 & 1.52 & 1.32 & 1.56 & 1.31 & 1.32 \\ 
  &$t(5)$ &  0.1 & 1.95 & 4.98 & 1.27 & 2.82 & 1.98 & 2.17 & 2.17 & 2.22 & 1.92 & 1.67 & 1.79 & 1.66 & 1.68 \\ 
  &  &  0.5 & 1.55 & 3.81 & 0.94 & 2.13 & 1.37 & 1.68 & 1.59 & 1.53 & 1.41 & 1.16 & 1.28 & 1.14 & 1.15 \\ 
  &  &  0.9 & 2.00 & 4.84 & 1.23 & 2.72 & 1.94 & 2.08 & 2.07 & 2.05 & 1.89 & 1.57 & 1.75 & 1.53 & 1.54 \\ 
  \cline{2-16}
  \multirow{6}*{$T=200$} &$N(0,1)$ &  0.1 & 1.18 & 4.28 & 1.01 & 2.38 & 1.21 & 1.29 & 1.34 & 1.27 & 1.14 & 1.03 & 1.14 & 1.00 & 0.95 \\ 
  &  &  0.5 & 1.06 & 3.62 & 0.87 & 2.00 & 1.00 & 1.14 & 1.06 & 1.06 & 0.92 & 0.81 & 0.95 & 0.75 & 0.76 \\ 
  &  &  0.9 & 1.24 & 4.28 & 1.00 & 2.38 & 1.18 & 1.27 & 1.26 & 1.25 & 1.06 & 0.97 & 1.09 & 0.89 & 0.88 \\ 
  &$t(5)$ &  0.1 & 1.38 & 4.15 & 0.92 & 2.19 & 1.43 & 1.51 & 1.51 & 1.57 & 1.42 & 1.21 & 1.29 & 1.15 & 1.17 \\ 
  &  &  0.5 & 1.07 & 3.09 & 0.64 & 1.55 & 0.98 & 1.14 & 1.10 & 1.06 & 0.97 & 0.80 & 0.93 & 0.80 & 0.81 \\ 
  &  &  0.9 & 1.33 & 4.07 & 0.88 & 2.18 & 1.37 & 1.48 & 1.57 & 1.42 & 1.30 & 1.09 & 1.28 & 1.02 & 1.10 \\ 
  \cline{2-16}
  \multirow{6}*{$T=500$} &$N(0,1)$ &  0.1 & 1.01 & 3.85 & 0.82 & 2.10 & 0.97 & 1.03 & 1.06 & 1.11 & 0.97 & 0.82 & 1.00 & 0.84 & 0.79 \\ 
  &  &  0.5 & 0.89 & 3.24 & 0.70 & 1.76 & 0.83 & 0.93 & 0.86 & 0.83 & 0.74 & 0.63 & 0.77 & 0.62 & 0.68 \\ 
  &  &  0.9 & 1.00 & 3.64 & 0.81 & 2.05 & 0.99 & 1.02 & 1.01 & 0.98 & 0.89 & 0.80 & 0.89 & 0.74 & 0.77 \\ 
  &$t(5)$ &  0.1 & 1.14 & 3.50 & 0.72 & 1.81 & 1.12 & 1.22 & 1.27 & 1.28 & 1.13 & 0.98 & 1.03 & 0.94 & 0.93 \\ 
  &  &  0.5 & 0.91 & 2.56 & 0.56 & 1.29 & 0.82 & 0.94 & 0.90 & 0.93 & 0.78 & 0.67 & 0.77 & 0.61 & 0.63 \\ 
  &  &  0.9 & 1.13 & 3.22 & 0.71 & 1.76 & 1.18 & 1.17 & 1.23 & 1.16 & 1.05 & 0.90 & 0.99 & 0.89 & 0.87 \\ 
\hline
&&& \multicolumn{13}{c}{$ N = 200 $} \\

  \multirow{6}*{$T=100$} &$N(0,1)$ &  0.1 & 1.13 & 4.11 & 0.98 & 2.28 & 1.19 & 1.33 & 1.29 & 1.26 & 1.10 & 1.05 & 1.17 & 1.00 & 0.96 \\ 
  &  &  0.5 & 1.00 & 3.39 & 0.84 & 1.92 & 0.91 & 1.11 & 1.00 & 1.03 & 0.92 & 0.82 & 0.91 & 0.79 & 0.78 \\ 
  &  &  0.9 & 1.14 & 3.88 & 0.96 & 2.27 & 1.11 & 1.22 & 1.18 & 1.20 & 1.07 & 0.95 & 1.08 & 0.91 & 0.93 \\ 
  &$t(5)$ &  0.1 & 1.32 & 3.89 & 0.88 & 1.98 & 1.39 & 1.52 & 1.51 & 1.49 & 1.31 & 1.19 & 1.36 & 1.21 & 1.21 \\ 
  &  &  0.5 & 1.02 & 2.81 & 0.65 & 1.44 & 1.00 & 1.10 & 1.03 & 1.08 & 0.94 & 0.83 & 0.92 & 0.83 & 0.82 \\ 
  &  &  0.9 & 1.34 & 3.64 & 0.89 & 2.00 & 1.30 & 1.48 & 1.41 & 1.40 & 1.27 & 1.12 & 1.28 & 1.14 & 1.07 \\ 
    \cline{2-16}
  \multirow{6}*{$T=200$} &$N(0,1)$ &  0.1 & 0.81 & 3.39 & 0.76 & 1.80 & 0.88 & 0.92 & 0.91 & 0.86 & 0.77 & 0.72 & 0.85 & 0.70 & 0.70 \\ 
  &  &  0.5 & 0.71 & 2.68 & 0.64 & 1.52 & 0.66 & 0.76 & 0.64 & 0.67 & 0.64 & 0.59 & 0.70 & 0.53 & 0.52 \\ 
  &  &  0.9 & 0.85 & 3.30 & 0.68 & 1.74 & 0.90 & 0.88 & 0.86 & 0.81 & 0.77 & 0.71 & 0.83 & 0.65 & 0.62 \\ 
  &$t(5)$ &  0.1 & 0.93 & 2.95 & 0.63 & 1.53 & 1.00 & 1.08 & 1.05 & 1.04 & 0.97 & 0.90 & 0.96 & 0.85 & 0.81 \\ 
  &  &  0.5 & 0.72 & 2.17 & 0.47 & 1.08 & 0.64 & 0.83 & 0.74 & 0.73 & 0.67 & 0.60 & 0.63 & 0.55 & 0.55 \\ 
  &  &  0.9 & 0.92 & 2.92 & 0.65 & 1.59 & 0.97 & 1.06 & 0.99 & 1.03 & 0.95 & 0.77 & 0.90 & 0.79 & 0.76 \\ 
    \cline{2-16}
      \multirow{6}*{$T=500$} &$N(0,1)$ &  0.1 & 0.62 & 2.44 & 0.56 & 1.42 & 0.72 & 0.80 & 0.78 & 0.70 & 0.66 & 0.58 & 0.61 & 0.55 & 0.55 \\ 
&     &  0.5 & 0.60 & 1.97 & 0.50 & 1.17 & 0.49 & 0.69 & 0.58 & 0.56 & 0.52 & 0.47 & 0.56 & 0.43 & 0.40 \\ 
 &     &  0.9 & 0.64 & 2.29 & 0.56 & 1.35 & 0.72 & 0.74 & 0.68 & 0.69 & 0.66 & 0.56 & 0.60 & 0.50 & 0.55 \\ 
  &$t(5)$ &  0.1 & 0.71 & 2.35 & 0.57 & 1.29 & 0.78 & 0.84 & 0.85 & 0.86 & 0.82 & 0.69 & 0.72 & 0.69 & 0.69 \\ 
 &     &  0.5 & 0.61 & 1.49 & 0.38 & 0.79 & 0.55 & 0.65 & 0.63 & 0.62 & 0.51 & 0.47 & 0.49 & 0.43 & 0.45 \\ 
  &     &  0.9 & 0.77 & 2.22 & 0.50 & 1.16 & 0.81 & 0.87 & 0.94 & 0.82 & 0.74 & 0.61 & 0.71 & 0.58 & 0.57 \\ 
\hline
&&& \multicolumn{13}{c}{$ N = 500 $} \\
  \multirow{6}*{$T=100$} &$N(0,1)$ &  0.1 & 0.92 & 3.59 & 0.82 & 2.01 & 0.95 & 1.08 & 1.07 & 0.99 & 0.90 & 0.85 & 0.97 & 0.81 & 0.81 \\ 
  &  &  0.5 & 0.82 & 2.85 & 0.66 & 1.57 & 0.75 & 0.83 & 0.83 & 0.84 & 0.73 & 0.66 & 0.78 & 0.61 & 0.62 \\ 
  &  &  0.9 & 0.94 & 3.35 & 0.82 & 1.84 & 0.97 & 1.02 & 1.02 & 1.00 & 0.88 & 0.76 & 0.86 & 0.75 & 0.74 \\ 
  &$t(5)$ &  0.1 & 1.08 & 3.33 & 0.75 & 1.69 & 1.06 & 1.25 & 1.23 & 1.22 & 1.06 & 0.99 & 1.10 & 0.94 & 0.91 \\ 
  &  &  0.5 & 0.85 & 2.35 & 0.55 & 1.21 & 0.73 & 0.93 & 0.81 & 0.85 & 0.79 & 0.67 & 0.74 & 0.67 & 0.66 \\ 
  &  &  0.9 & 1.05 & 3.17 & 0.72 & 1.69 & 1.07 & 1.11 & 1.15 & 1.17 & 0.99 & 0.89 & 1.01 & 0.90 & 0.93 \\ 
  \cline{2-16}
    \multirow{6}*{$T=200$} &$N(0,1)$ &  0.1 & 0.70 & 2.70 & 0.60 & 1.49 & 0.72 & 0.72 & 0.68 & 0.66 & 0.65 & 0.61 & 0.68 & 0.57 & 0.56 \\ 
&     &  0.5 & 0.60 & 2.24 & 0.49 & 1.19 & 0.58 & 0.65 & 0.61 & 0.64 & 0.53 & 0.46 & 0.60 & 0.46 & 0.46 \\ 
 &     &  0.9 & 0.63 & 2.69 & 0.62 & 1.43 & 0.65 & 0.74 & 0.71 & 0.66 & 0.63 & 0.57 & 0.64 & 0.49 & 0.51 \\ 
  &$t(5)$ &  0.1 & 0.70 & 2.42 & 0.51 & 1.26 & 0.86 & 0.89 & 0.95 & 0.89 & 0.78 & 0.70 & 0.78 & 0.65 & 0.67 \\ 
 &     &  0.5 & 0.61 & 1.65 & 0.39 & 0.84 & 0.53 & 0.62 & 0.61 & 0.59 & 0.54 & 0.46 & 0.56 & 0.48 & 0.45 \\ 
  &     &  0.9 & 0.81 & 2.29 & 0.52 & 1.26 & 0.73 & 0.87 & 0.82 & 0.83 & 0.72 & 0.69 & 0.76 & 0.63 & 0.66 \\ 
        \cline{2-16}
  \multirow{6}*{$T=500$} &$N(0,1)$ &  0.1 & 0.51 & 2.33 & 0.52 & 1.25 & 0.59 & 0.57 & 0.58 & 0.64 & 0.60 & 0.45 & 0.56 & 0.46 & 0.41 \\ 
  &  &  0.5 & 0.45 & 1.63 & 0.43 & 0.91 & 0.40 & 0.65 & 0.47 & 0.45 & 0.47 & 0.40 & 0.42 & 0.31 & 0.34 \\ 
  &  &  0.9 & 0.61 & 2.10 & 0.49 & 1.11 & 0.61 & 0.56 & 0.56 & 0.57 & 0.50 & 0.37 & 0.53 & 0.41 & 0.45 \\ 
  &$t(5)$ &  0.1 & 0.56 & 2.10 & 0.45 & 1.11 & 0.70 & 0.59 & 0.71 & 0.82 & 0.66 & 0.60 & 0.68 & 0.63 & 0.51 \\ 
  &  &  0.5 & 0.52 & 1.30 & 0.35 & 0.67 & 0.46 & 0.60 & 0.49 & 0.43 & 0.43 & 0.39 & 0.40 & 0.36 & 0.35 \\ 
  &  &  0.9 & 0.64 & 2.10 & 0.46 & 1.10 & 0.72 & 0.77 & 0.64 & 0.62 & 0.56 & 0.47 & 0.59 & 0.48 & 0.58 \\ 
\hline
\hline
\end{tabular}
}
\end{center}
{\footnotesize{Notes: The simulation results are based on the DGP in Section \ref{mcsetup} with 1000 replications and reported across the three different quantiles, $\tau = (0.1,0.5,0.9$) for the sample pairs, $(N,T) = 100, 200, 500$, where we generate $\mathfrak{u}_{it}$ from either a standard normal distribution, $N(0,1)$ or a $t$-distribution with 5 degrees of freedom, $t(5)$.}}
\end{table}
\normalsize

\begin{table}[!htbp]
\begin{center}
\caption{Coverage Probability ($\times 100$) for \textsc{Type 1} Network}
\label{tab_simu_CP_W1}
{\scriptsize 
\begin{tabular}{ccc|ccccccccccccc}
\hline
\hline
&Dist. & $\tau$ &$\gamma_0 $  & $\gamma_1$&  $\gamma_2$ &  $\gamma_3$ & $\alpha_1$ & $\alpha_2$ &$\alpha_3$ & $\alpha_4$ & $\alpha_5$ & $\beta_1$ & $\beta_2$ & $\beta_3$ & $\beta_4$  \\
\hline
&&& \multicolumn{13}{c}{$ N = 100 $} \\
  \multirow{6}*{$T=100$} &$N(0,1)$ &  0.1 & 93.5 & 97.8 & 92.9 & 97.3 & 93.1 & 93.2 & 94.9 & 94.6 & 94.8 & 94.7 & 95.3 & 94.5 & 94.7 \\ 
  &  &  0.5 & 93.0 & 97.2 & 93.5 & 95.3 & 94.2 & 93.0 & 95.4 & 95.0 & 94.8 & 94.2 & 96.1 & 94.7 & 94.4 \\ 
  &  &  0.9 & 92.8 & 97.3 & 94.0 & 94.4 & 93.8 & 92.1 & 94.8 & 95.1 & 94.6 & 95.6 & 94.6 & 94.8 & 94.6 \\ 
  &$t(5)$ &  0.1 & 93.3 & 98.6 & 92.7 & 95.0 & 94.5 & 94.6 & 94.0 & 94.5 & 95.0 & 95.2 & 94.4 & 94.8 & 94.4 \\ 
&  &  0.5 & 92.1 & 96.8 & 93.0 & 96.7 & 95.0 & 94.3 & 94.3 & 95.1 & 95.6 & 96.2 & 95.0 & 96.2 & 95.8 \\ 
  &  &  0.9 & 93.0 & 97.2 & 94.2 & 95.1 & 93.9 & 94.9 & 94.5 & 94.8 & 95.8 & 94.5 & 94.2 & 95.5 & 94.8 \\ 
  \cline{2-16}
    \multirow{6}*{$T=200$} &$N(0,1)$ & 0.1 & 93.4 & 96.2 & 95.6 & 95.4 & 95.6 & 92.8 & 94.6 & 97.0 & 94.0 & 95.0 & 95.8 & 94.4 & 94.4 \\ 
  &  &  0.5 & 95.3 & 93.6 & 93.9 & 94.7 & 96.4 & 94.8 & 96.5 & 94.6 & 94.6 & 94.5 & 94.0 & 96.0 & 95.3 \\ 
  &  &  0.9 & 93.1 & 92.9 & 93.9 & 94.9 & 93.7 & 94.8 & 94.6 & 94.8 & 94.9 & 93.9 & 95.4 & 94.6 & 95.8 \\ 
  &$t(5)$ &  0.1 & 94.8 & 98.8 & 93.8 & 97.0 & 96.3 & 94.3 & 96.3 & 96.5 & 96.5 & 96.8 & 96.5 & 97.3 & 96.5 \\ 
   &  &  0.5 & 94.2 & 92.8 & 94.2 & 91.4 & 95.4 & 94.8 & 95.8 & 95.2 & 95.0 & 96.0 & 95.0 & 94.8 & 96.8 \\ 
  &  &  0.9 & 93.8 & 98.3 & 92.8 & 96.5 & 93.5 & 96.3 & 95.8 & 96.8 & 96.0 & 96.0 & 95.8 & 93.0 & 96.5 \\ 
  \cline{2-16}
  \multirow{6}*{$T=500$} &$N(0,1)$ & 0.1 & 92.4 & 95.9 & 92.7 & 94.8 & 94.1 & 96.4 & 94.8 & 94.3 & 94.4 & 96.4 & 94.1 & 93.1 & 94.9 \\ 
  &  &  0.5 & 90.9 & 92.6 & 91.5 & 92.9 & 93.5 & 93.5 & 95.7 & 95.7 & 96.0 & 96.9 & 95.2 & 96.7 & 94.7 \\ 
  &  &  0.9 & 90.9 & 95.3 & 92.9 & 92.7 & 92.9 & 94.7 & 95.1 & 95.5 & 95.2 & 94.1 & 95.5 & 94.1 & 95.9 \\ 
  &$t(5)$ &  0.1 & 93.9 & 93.7 & 94.7 & 94.1 & 95.2 & 95.3 & 95.1 & 94.8 & 95.7 & 94.8 & 95.5 & 94.0 & 95.7 \\ 
  &  &  0.5 & 92.3 & 93.2 & 92.7 & 93.3 & 96.1 & 94.0 & 95.6 & 94.1 & 95.6 & 96.0 & 95.6 & 97.1 & 96.8 \\ 
  &  &  0.9 & 91.5 & 92.9 & 93.7 & 92.7 & 93.3 & 95.5 & 94.0 & 94.8 & 94.0 & 95.7 & 96.5 & 95.6 & 95.6 \\ 
\hline
&&& \multicolumn{13}{c}{$ N = 200 $} \\
 \multirow{6}*{$T=100$} &$N(0,1)$ &  0.1 & 93.3 & 96.5 & 95.5 & 95.8 & 95.5 & 94.3 & 96.0 & 96.8 & 95.5 & 94.3 & 94.8 & 95.0 & 94.8 \\ 
  &  &  0.5 & 91.0 & 94.0 & 93.3 & 97.0 & 96.5 & 92.0 & 94.5 & 95.5 & 94.5 & 95.8 & 94.5 & 96.0 & 95.3 \\ 
   &  &  0.9 & 92.5 & 95.5 & 91.5 & 94.8 & 94.5 & 93.8 & 95.3 & 93.5 & 96.3 & 95.0 & 94.0 & 93.3 & 94.0 \\ 
  &$t(5)$ &  0.1 & 94.3 & 97.6 & 92.9 & 97.2 & 93.9 & 94.1 & 94.7 & 95.4 & 94.9 & 94.8 & 94.1 & 94.1 & 93.7 \\ 
  &  &  0.5 & 90.8 & 93.7 & 91.5 & 94.2 & 94.2 & 93.2 & 94.5 & 94.4 & 94.9 & 95.1 & 94.7 & 94.7 & 94.9 \\ 
  &  &  0.9 & 91.1 & 97.3 & 91.9 & 94.5 & 94.8 & 93.5 & 94.3 & 94.6 & 94.5 & 94.7 & 94.1 & 93.7 & 95.3 \\ 
  \cline{2-16}
  \multirow{6}*{$T=200$} &$N(0,1)$ & 0.1 & 94.0 & 93.5 & 91.0 & 92.0 & 94.5 & 93.5 & 95.5 & 93.5 & 94.5 & 93.0 & 95.5 & 96.0 & 94.0 \\ 
  &  &  0.5 & 90.4 & 93.2 & 94.3 & 92.4 & 96.2 & 95.2 & 97.0 & 95.6 & 95.8 & 95.4 & 93.6 & 95.8 & 96.8 \\ 
  &  &  0.9 & 91.3 & 94.2 & 92.4 & 93.4 & 90.4 & 93.6 & 94.2 & 96.4 & 94.4 & 93.0 & 93.8 & 95.2 & 96.0 \\ 
  &$t(5)$ &  0.1 & 94.6 & 93.0 & 93.4 & 94.0 & 94.2 & 95.4 & 95.6 & 95.4 & 94.2 & 93.6 & 92.4 & 95.6 & 95.4 \\ 
  &  &  0.5 & 94.2 & 92.2 & 91.4 & 93.8 & 97.0 & 92.4 & 94.8 & 96.0 & 95.8 & 95.4 & 96.0 & 96.0 & 96.4 \\ 
   &  &  0.9 & 93.5 & 95.0 & 95.5 & 94.5 & 93.5 & 96.0 & 95.0 & 97.5 & 94.5 & 95.5 & 96.0 & 96.0 & 95.0 \\ 
      \cline{2-16}
        \multirow{6}*{$T=500$} &$N(0,1)$ &  0.1 & 95.2 & 92.8 & 92.8 & 93.6 & 93.2 & 94.4 & 93.6 & 95.6 & 94.0 & 96.4 & 96.4 & 95.6 & 95.6 \\ 
    &  &  0.5 & 93.7 & 92.7 & 96.0 & 94.3 & 95.0 & 94.4 & 95.3 & 94.7 & 93.3 & 94.7 & 94.5 & 95.2 & 95.3 \\ 
&  &  0.9 & 92.7 & 93.1 & 94.6 & 93.3 & 94.6 & 94.0 & 94.7 & 96.0 & 94.7 & 95.3 & 94.7 & 94.7 & 94.9 \\ 
 &$t(5)$ &  0.1 & 94.2 & 95.3 & 96.0 & 94.6 & 95.0 & 94.7 & 94.6 & 94.9 & 95.3 & 95.5 & 94.3 & 95.2 & 95.6 \\ 
   &  &  0.5 & 94.3 & 94.4 & 92.0 & 95.6 & 94.0 & 94.8 & 94.8 & 95.6 & 96.4 & 94.4 & 95.2 & 96.8 & 97.2 \\ 
     &  &  0.9 & 92.6 & 95.3 & 94.6 & 94.2 & 95.7 & 95.1 & 95.3 & 94.6 & 95.6 & 96.3 & 94.00 & 94.3 & 96.0 \\ 
\hline
&&& \multicolumn{13}{c}{$ N = 500 $} \\
  \multirow{6}*{$T=100$} &$N(0,1)$ &  0.1 & 93.2 & 96.4 & 94.0 & 94.2 & 95.6 & 93.6 & 94.8 & 95.6 & 96.0 & 96.0 & 93.8 & 94.6 & 95.4 \\ 
  &  &  0.5 & 94.3 & 94.6 & 93.8 & 94.0 & 95.6 & 95.0 & 95.0 & 95.4 & 95.2 & 97.0 & 96.2 & 96.6 & 96.2 \\ 
 &  &  0.9 & 92.5 & 93.5 & 92.0 & 93.5 & 95.0 & 94.0 & 95.0 & 95.3 & 95.5 & 93.7 & 96.5 & 96.3 & 94.1\\ 
  &$t(5)$ & 0.1 & 93.4 & 96.4 & 94.0 & 96.0 & 96.8 & 93.2 & 95.0 & 95.4 & 95.8 & 94.4 & 95.0 & 96.0 & 96.6 \\ 
   &  &  0.5 & 93.0 & 92.5 & 93.0 & 93.5 & 95.0 & 90.0 & 96.0 & 94.0 & 95.0 & 95.5 & 93.5 & 95.5 & 95.5 \\ 
  &  &  0.9 & 93.2 & 94.0 & 92.4 & 94.0 & 95.8 & 96.8 & 95.8 & 95.6 & 95.4 & 97.0 & 94.4 & 96.2 & 94.4 \\ 
  \cline{2-16}
    \multirow{6}*{$T=200$} &$N(0,1)$ &  0.1 & 92.6 & 92.4 & 93.3 & 93.3 & 92.8 & 94.3 & 96.7 & 95.2 & 94.8 & 94.3 & 93.3 & 94.3 & 95.7 \\ 
   &  &  0.5 & 91.7 & 92.5 & 93.0 & 92.8 & 94.5 & 95.9 & 94.5 & 95.5 & 94.8 & 95.7 & 95.3 & 94.5 & 95.2 \\ 
   &  &  0.9 & 91.0 & 92.5 & 93.2 & 93.3 & 93.3 & 95.2 & 93.8 & 94.3 & 94.8 & 93.3 & 96.2 & 98.1 & 97.1 \\ 
  &$t(5)$ &  0.1 & 94.3 & 92.4 & 93.8 & 93.3 & 93.3 & 94.3 & 92.4 & 93.8 & 93.3 & 95.2 & 95.2 & 94.8 & 94.8 \\ 
   &  &  0.5 & 91.6 & 94.8 & 96.2 & 95.7 & 96.7 & 94.8 & 96.2 & 95.2 & 96.2 & 97.1 & 94.3 & 97.1 & 97.1 \\ 
   &  &  0.9 & 93.5 & 93.9 & 94.3 & 95.4 & 95.2 & 95.2 & 95.2 & 94.8 & 95.2 & 94.3 & 94.8 & 94.3 & 94.3 \\ 
      \cline{2-16}
   \multirow{6}*{$T=500$} &$N(0,1)$ &  0.1 & 92.7 & 93.6 & 93.6 & 95.5 & 93.6 & 91.8 & 95.5 & 96.4 & 94.7 & 93.6 & 93.6 & 95.5 & 95.6 \\ 
   &  &  0.5 & 91.8 & 93.6 & 93.6 & 92.7 & 95.5 & 95.2 & 95.5 & 95.5 & 94.3 & 96.4 & 94.6 & 94.6 & 94.6 \\ 
   &  &  0.9 & 92.9 & 94.1 & 94.6 & 92.3 & 92.7 & 90.9 & 90.9 & 93.6 & 95.2 & 95.5 & 92.7 & 94.6 & 94.8 \\ 
  &$t(5)$ &  0.1 & 93.9 & 93.2 & 93.6 & 92.2 & 92.7 & 94.6 & 92.7 & 99.1 & 94.6 & 96.4 & 93.6 & 97.3 & 96.4 \\ 
   &  &  0.5 & 93.1 & 92.4 & 92.7 & 90.0 & 95.5 & 93.1 & 97.3 & 96.4 & 96.4 & 95.5 & 92.7 & 95.5 & 96.4 \\ 
   &  &  0.9 & 93.9 & 95.4 & 93.8 & 93.6 & 94.8 & 95.1 & 95.2 & 95.3 & 94.3 & 95.6 & 94.4 & 95.7 & 95.5 \\ 
  \hline
\hline
\end{tabular}
}
\end{center}
{\footnotesize{Notes: See the notes to Table \ref{tab_simu_RMSE_W1}.}}
\end{table}
\normalsize

\begin{table}[!htbp]
\begin{center}
\scriptsize 
\caption{Estimation Results for the Network $W_{CS5}$} \label{TabAppW1}
\begin{tabular}{c|ccc|ccc|ccc}
\hline
\hline			
			& \multicolumn{3}{c}{DNQR } & \multicolumn{3}{c}{NQARF} & \multicolumn{3}{c}{NQAR } \\ \hline
& $ \tau=0.1$  & $ \tau=0.5$  & $ \tau=0.9$ & $ \tau=0.1$  & $ \tau=0.5$  & $ \tau=0.9$ & $ \tau=0.1$  & $ \tau=0.5$  & $ \tau=0.9$ \\ 
$ \hat\gamma_0 $  &   $-2.28^{***}$ &  $0.02^{***}$ &  $2.33^{***}$ &  $-2.55^{***}$ &  $0.05^{***}$ &  $2.64^{***}$ &  $-2.55^{***}$ &  $0.04^{***}$ &  $2.65^{***}$ \\ 
&   (0.01)  &  (0.00)  &  (0.01)  &  (0.01)  &  (0.00)  &  (0.01)  &  (0.01)  &  (0.00)  &  (0.01)  \\ 
$ \hat\gamma_1 $  &   $54.07^{***}$ &  $57.55^{***}$ &  $47.09^{***}$ & - & - & -  & - & - & -  \\ 
&   (1.43)  &  (0.81)  &  (1.49)  & && & && \\ 
$ \hat\gamma_2 $  &   $3.39^{***}$ &  $0.55^{***}$ &  $2.56^{***}$ &  $3.96^{***}$ &  $-0.62^{***}$ &  $2.67^{***}$ &  $4.90^{***}$ &  $-0.55^{***}$ &  $2.47^{***}$ \\ 
&   (0.48)  &  (0.33)  &  (0.53)  &  (0.66)  &  (0.21)  &  (0.63)  &  (0.65)  &  (0.20)  &  (0.62)  \\ 
$ \hat\gamma_3 $  &   $-1.29^{***}$ &  $-2.41^{***}$ &  $-2.45^{***}$ &  $-0.41$ &  $-2.10$ &  $-2.17$ &  $-0.09$ &  $-1.59$ &  $-2.19$ \\ 
&   (0.35)  &  (0.27)  &  (0.34)  &  (0.42)  &  (0.13)  &  (0.39)  &  (0.41)  &  (0.13)  &  (0.39)  \\ 
&&&&&&&&& \\ 
SIZE  &   $0.09^{***}$ &  $0.00^{***}$ &  $-0.09^{***}$ &  $0.09^{***}$ &  $0.00^{***}$ &  $-0.09^{***}$ &  $0.10^{***}$ &  $0.00^{***}$ &  $-0.09^{***}$ \\ 
&   (0.00)  &  (0.00)  &  (0.00)  &  (0.00)  &  (0.00)  &  (0.00)  &  (0.00)  &  (0.00)  &  (0.00)  \\ 
BM  &   $0.13^{***}$ &  $0.01^{***}$ &  $-0.11^{***}$ &  $0.14^{***}$ &  $0.02^{***}$ &  $-0.12^{***}$ &  $0.14^{***}$ &  $0.02^{***}$ &  $-0.12^{***}$ \\ 
&   (0.00)  &  (0.00)  &  (0.00)  &  (0.00)  &  (0.00)  &  (0.00)  &  (0.00)  &  (0.00)  &  (0.00)  \\ 
Cash  &   $0.00$ &  $0.00$ &  $0.02$ &  $0.00$ &  $0.01$ &  $0.02$ &  $0.00$ &  $0.01$ &  $0.02$ \\ 
&   (0.01)  &  (0.01)  &  (0.01)  &  (0.01)  &  (0.01)  &  (0.01)  &  (0.01)  &  (0.01)  &  (0.01)  \\ 
PE  &   $0.04^{***}$ &  $0.00^{***}$ &  $-0.02^{***}$ &  $0.03^{***}$ &  $0.01^{***}$ &  $-0.02^{***}$ &  $0.03^{***}$ &  $0.01^{***}$ &  $-0.01^{***}$ \\ 
&   (0.01)  &  (0.00)  &  (0.01)  &  (0.01)  &  (0.00)  &  (0.01)  &  (0.01)  &  (0.00)  &  (0.01)  \\ 
VIX  &   $-0.16^{***}$ &  $-0.04^{***}$ &  $-0.07^{***}$ &  $-0.23^{***}$ &  $-0.08^{***}$ &  $-0.12^{***}$ & - & - & -  \\ 
&   (0.02)  &  (0.01)  &  (0.02)  &  (0.02)  &  (0.01)  &  (0.02)  & && \\ 
Rm - Rf  &   $-0.07^{***}$ &  $0.02^{***}$ &  $0.10^{***}$ &  $-0.14^{***}$ &  $0.03^{***}$ &  $0.15^{***}$ & - & - & -  \\ 
&   (0.02)  &  (0.01)  &  (0.02)  &  (0.02)  &  (0.01)  &  (0.02)  & && \\ 
SMB  &   $0.02^{*}$ &  $0.00^{*}$ &  $-0.01^{*}$ &  $0.00$ &  $0.01$ &  $0.04$ & - & - & -  \\ 
&   (0.01)  &  (0.00)  &  (0.01)  &  (0.01)  &  (0.00)  &  (0.01)  & && \\ 
HML  &   $0.03^{**}$ &  $0.00^{**}$ &  $-0.04^{**}$ &  $0.01$ &  $-0.01$ &  $-0.10$ & - & - & -  \\ 
&   (0.01)  &  (0.00)  &  (0.01)  &  (0.01)  &  (0.00)  &  (0.01)  & && \\ 
\hline &&&&&&&&& \\  
Goodn.fit. &  - & - & -  & 8.68 & 9.45 & 6.88 & 9.39 & 9.50 & 7.00 \\ 
\hline
\hline
\end{tabular}
\end{center}
{\footnotesize Notes: The dataset consists of $N = 943$ stocks with $T = 252$ time periods. The network matrix, $W_{CS5}$ is constructed by checking if the stocks are invested in by at least five common shareholders with the network density, 3.24\%. The estimates ($\times 10^{2}$) are reported across different quantiles $\tau = 0.1, 0.5, 0.9$, and the value in parentheses is the standard error ($\times 10^{2}$). DNQR denotes the proposed model, NQAR is the model without contemporaneous network effects and common factors, and NQARF is the factor-augmented NQAR model. Goodn.fit. ($\times 10^{2}$) represents the goodness of fit of DNQR model with respect to the other models. The 1\%, 5\% and 10\% significance levels are denoted by ***, **, *, respectively.} 
\end{table}

\newpage

\begin{figure}[!htbp]
\centering
\includegraphics[scale=0.7]{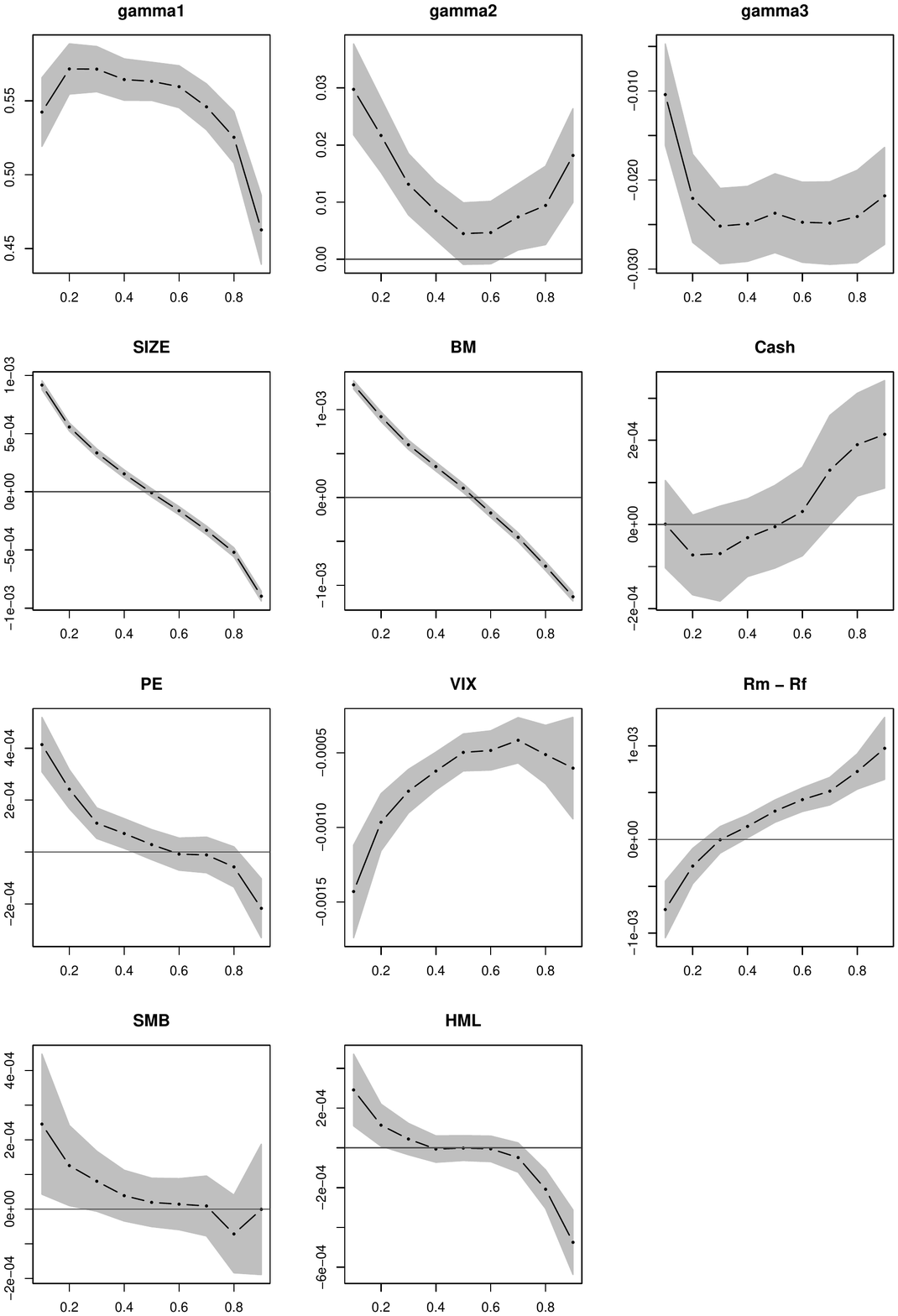}
\caption{Quantile-specific Coefficients for the Network $W_{CS5}$}\label{AppParW1}
\caption*{\footnotesize Notes: The dashed line is the QR coefficient while the grey area indicates a kernel density based confidence band advanced by \cite{powell1991}. They are displayed across quantiles, $\tau = 0.1, 0.2, \cdots, 0.9$.}
\end{figure}

\newpage

\normalsize
\newpage
\bigskip
\begin{center}
	\textbf{ \Large Online Supplement for \\ \medskip Dynamic Network Quantile Regression Model}

\author{Xiu Xu \footnote{Dongwu Business School, Soochow University, 50 Donghuan Road, Suzhou, Jiangsu 215021, PR China. Email: \href{mailto:xiux@suda.edu.cn}{xiux@suda.edu.cn}.}\; Weining Wang \footnote{Department of Economics and Related Studies, University of York, Heslington, York, YO10 5DD, UK. Email: \href{mailto:weining.wang@york.ac.uk}{weining.wang@york.ac.uk}.}\;   Yongcheol Shin \footnote{Department of Economics and Related Studies, University of York, Heslington, York, YO10 5DD, UK. Email: \href{mailto:yongcheol.shin@york.ac.uk}{yongcheol.shin@york.ac.uk}.}  \;Chaowen Zheng\footnote{Department of Economics and Related Studies, University of York, Heslington, York, YO10 5DD, UK. Email: \href{mailto:cz1113@york.ac.uk}{cz1113@york.ac.uk}.}}

\end{center}
\bigskip

	\numberwithin{equation}{section}
	
		Section \ref{appendix} provides the proofs for Theorems \ref{thrm_stationary}--\ref{theorem_estimation}. Section \ref{appendixb} presents the additional simulation results on the performance of the IVQR and the ordinary QR estimators under the different network structures. Section \ref{appendixc} reports the additional empirical results by employing the alternative common shareholder network structures constructed by imposing the different number of common shareholders, and the uniform headquarter location network.

\begin{appendices}
	\section{Proof of Lemmas and Theorems} \label{appendix}
	\subsection{Proof of Lemma \ref{lemma_stationray}} \label{proof_lemma_stationray}
	
	\textbf{(i) Strict Stationarity}
	
	
	Under Assumption \ref{Ass_stationray}(1) and  conditioning on $\mathcal{C}_z $, we have $(I-\mathbf{A}_{1t}W)^{-1} =\sum_{k=0}^{\infty} (\mathbf{A}_{1t}W)^{k}$. Then, we obtain the reduced form of the model \eqref{modelmatrix} by
	\begin{equation*}
		\yit_{t} = M_{t}\mathbb{Y}_{t-1}+\mathbb{C}_{t}, \end{equation*}
	where $M_t = (I - \Abt_{1t}W)^{-1}\mathbf{H}_t$ and $\cit_t =  (I - \Abt_{1t}W)^{-1} (\Gamma + \mathbf{B}_{t}\fit_{t} + V_{t})$. This process belongs to the class of a general autoregressive process with $\{M_{t}, \mathbb{C}_{t}, t\in \mathbb{Z}\}$. By Theorem 1.3 and Lemma 2.1 of \cite{bougerol1992stationarity}, $\mathbb{Y}_{t}$ has a strictly stationary solution, if the sequence of random matrices $\{M_{t}, t\in \mathbb{Z}\}$
	satisfy the following two conditions: \begin{itemize}
		
		\item[a)] {\normalsize $\mathop{\mbox{\sf E}}\log |M_{0}|_2^{+}< \infty$,
			with $\log |M_{0}|_2^{+} = \max(\log |M_{0}|_2, 0) $, }
		
		\item[b)] {\normalsize $\lim_{t \to \infty} |M_{0}M_{-1}\cdots M_{-t}|_2 = 0 
			$ almost surely. }
	\end{itemize}
	
	Recall that $|A|_2 = \sup_{ \{v \in \mathbb{R}^d, v \neq 0\}} |Av|_2/|v|_2$, where $|.|_2$ is the two norm of a vector or matrix. We now prove the two conditions. First, consider:
	\begin{eqnarray*}
		|M_t|_2& =& |(I - \Abt_{1t}W)^{-1} H_{t}|_2\\
		&\leq& |\sum_{k\geq 0} (\Abt_{1t}W)^k H_t|_2 \quad \text{(under {Assumption \ref{Ass_stationray}(1))}}\\
		&\leq &   \sum_{k\geq 0} |(\Abt_{1t}W)^k H_t|_2 \quad \text{(under Minkowski inequality)}\\
		&\leq& \sum_{k\geq 0} |(\Abt_{1t})|_2^k|W|_2^k |H_t|_2  \quad \text{(under {Assumption \ref{Ass_stationray}(1))}}\\
		&\leq & \sum_{k\geq 0} |(\Abt_{1t})|_2^k(|(\Abt_{2t})|_2+|(\Abt_{3t})|_2)  \quad \text{(under {Assumption \ref{Ass_stationray}(1))}}\\
		&=& \sum_{k\geq 0} \{\max_i |\gamma_{1}(U_{it})|\}^k(\max_i |\gamma_{2}(U_{it})|+\max_i |\gamma_{3}(U_{it})|)\\
		&\leq & \sum_{k\geq 0}  c_1^k (\max_i |\gamma_{2}(U_{it})|+\max_i |\gamma_{3}(U_{it})|)\\
		&\leq & \sum_{k\geq 0}  c_1^k c_{23}   \quad \text{(under {Assumption \ref{Ass_stationray}(2))}}\\ 
		&\leq &  c_{23} /(1 - c_1)  .
	\end{eqnarray*}
	Then, $\E\log |M_{0}|_2^{+}\leq \log \E|M_{0}|_2^{+} \leq \max \{\log (c_{23}/(1-c_1)), 0\} < \infty$. Hence, the conditions a) holds. 
	
	Next, the second condition b) can be written as
	\begin{eqnarray*}
		\E|M_{0}M_{-1}\cdot M_{-t}|_2 &\leq & \E \Pi^0_{l=-t}   \{ (\sum_{k\geq 0}c_1^k)(\max_i |\gamma_{2}(U_{il})|+\max_i |\gamma_{3}(U_{il})|) \}\\
		&\leq & (1-c_1)^{-t-1} [\E\{\max_i |\gamma_{2}(U_{il})|+\max_i |\gamma_{3}(U_{il})|\}]^{t+1}\\
		&\leq & (1-c_1)^{-t-1}c_{23}^{t+1}.
	\end{eqnarray*}
	For a small constant, $\vps >0$, we have:
	\begin{eqnarray*}
		&&\sum^{\infty}_{t=1}\P ( |M_{0}M_{-1}\cdot M_{-t}|_2> \vps)\\ 
		&\leq& \sum^{\infty}_{t=1} \frac{\E|M_{0}M_{-1}\cdot M_{-t}|_2}{\vps}     \quad \text{(under Markov's inequality)}\\
		&\leq & \sum^{\infty}_{t=1} (1-c_1)^{-t-1}c_{23}^{t+1} /\vps
		= \dfrac{(c_{23})^2}{(1-c_1)(1-c_1-c_{23})\vps} < \infty.
	\end{eqnarray*}
	Then, by Borel-Cantelli lemma, the condition b) holds. Therefore, any projection of the process in \eqref{modelmatrix} has a strictly stationary solution.

	\noindent{\textbf{(ii) Covariance Stationarity}}
	
	In addition, if $\Var (\yit_t) $ and $\cov (\yit_{t}, \yit_{t-l})$ exits, then $\yit_t$ is covariance stationary.
	The model, \eqref{modelmatrix} can be written as 
	\begin{equation*}
		\yit_t  = \sum^{\infty}_{l = 0}  \Pi_l \Dit_{t-l} =  \sum^{\infty}_{l=0}\Pi_l S^{-1}_{t-l}  \mathbf{B}_{t-l} \fit_{t-l}
		+ \sum^{\infty}_{l=0}\Pi_l S^{-1}_{t-l} \mathbf{A}_{0t}, 
	\end{equation*}
	where $\mathbb{D}
	_{t}=S_{t}^{-1}(\mathbf{B}_{t}\mathbb{F}_{t}+\mathbf{A}_{0t})$, $M_{t}=S_{t}^{-1}\mathbf{H}_{t}$ and $\Pi _{l}=M_{t}\times \cdots \times M_{t-l+1}$ for $l>1$ with $\Pi _{0}=I$ and $\Pi _{1}=M_{t}$. Let $\mathop{\mbox{\sf E}}%
	M_{t}=M$ and $\mathop{\mbox{\sf E}}\mathbb{D}_{t}=\mathbb{D}$. By Assumption \ref{Ass_stationray}(3), $|\mathbb{D}_{t}|_{\infty} \leq (1-c_1)^{-1}(d_f+ d_z )$. Thus, the expected value of $\mathbb{Y}_t$ is given by $\mu_{\mathbb{Y}%
	} = (I -M)^{-1} \mathbb{D}$. Further, we have: $|M_t I_N| \leq_{a}
	c_{23}/(1-c_1) I_N$ for every $t$, where $\leq_{a} $ denotes 'element-wise smaller'. The variance and covariance of $\mathbb{Y}_{t}$ are then given by%
	\begin{align*}
		\Gamma _{0}& =\mathop{\mbox{\sf Var}}(\mathbb{Y}_{t})=\mathop{\mbox{\sf E}}%
		\left\{ \left( \sum_{l=0}^{\infty }\Pi _{l}\mathbb{D}_{t-l}\right) \left(
		\sum_{l=0}^{\infty }\Pi _{l}\mathbb{D}_{t-l}\right) ^{\top }\right\} -\mu _{%
			\mathbb{Y}}\mu _{\mathbb{Y}}^{\top },
		\\
		\Gamma _{l}&=\mathop{\hbox{Cov}}(\mathbb{Y}_{t},\mathbb{Y}_{t-l})=%
		\mathop{\mbox{\sf E}}\left\{ \left( \sum_{j=0}^{\infty }\Pi _{j}\mathbb{D}%
		_{t-j}\right) \left( \sum_{j=0}^{\infty }\Pi _{j}\mathbb{D}_{t-j-l}\right)
		^{\top }\right\} -\mu _{\mathbb{Y}}\mu _{\mathbb{Y}}^{\top}. \notag
	\end{align*}
	
	Consider $\Var(\yit_t) $. Let $c' = (1-c_1)^{-1}c_{23} $, then we have: 
	\begin{equation*}
		e_i^{\top} \E \left \{ \left (\sum_{l\geq 0} \Pi_{l}\Dit_{t-l}\right ) \left (\sum_{l\geq 0} \Pi_{l}\Dit_{t-l} \right )^{\top} \right \} e_j =  I_{1} + I_2 +I_3.    
	\end{equation*}
	First, we show that $ I_1 = \sum_{l_1> l_2}e_{i}^{\top}\otimes e_{j}^{\top}  M^{l_2}\otimes M^{l_2}\E\{I \otimes M_{t-l_2} \cdots I \otimes M_{t-l_1-1} \Vec (\Dit_{t}\Dit^{\top}_{t-l_2})\}  \leq \sum_{l_1> l_2}|\Dit|_{\max} \E(d_z+d_f) \{c'\}^{l_1-l_2} c'^{2l_2} = \sum_{l_1> l_2}|\Dit|_{\max} \E(d_z+d_f) c'^{l_1+l_2} < \infty$. Similarly, $I_2 = \sum_{l_1}e_{i}^{\top}\otimes e_{j}^{\top}  M^{l_1}\otimes M^{l_1}\E\{\Vec (\Dit_{t-l_1}\Dit_{t-l_1}^{\top})\}\leq \sigma_{d\max} \sum_{l_1} c'^{2l_1} $ and $I_3 = \sum_{l_2> l_1}e_{i}^{\top}\otimes e_{j}^{\top}  M^{l_1}\otimes M^{l_1}\E\{ M_{t-l_1} \otimes I \cdots M_{t-l_1-1}  \otimes I\Vec (\Dit_{t-l_1}\Dit_{t}^{\top}) \}\leq \sum_{l_1< l_2}|\Dit|_{\max} \E(d_z+d_f) c'^{l_1+l_2} $. Thus, we have $I_1, I_2, I_3 < \infty$ under Assumption \ref{Ass_stationray}.  Finally, it is easily seen that $\Gamma_{l}$ exists. Thus, $a^{\top}\mathbb{Y}_{t}$ is covariance stationary.

	\subsection{Proof of Theorem \ref{thrm_stationary}}	\label{prf_thrm_stary}

	
	We introduce the functional dependence measure (see \cite{wu2011asymptotic}).
	
	\begin{definition}\label{norm}
		Define $X_{t} =  g(\mathcal{F}_{t})$ with the shift process $\mathcal{F}_{t} = (...,\xi_{t-1},\xi_{t})$, and $X_{t}^\ast = g(\mathcal{F}_{t}^*)$ with $\mathcal{F}_{t}^* = (\ldots, \xi_{-1}, \xi^\ast_0, \xi_{1},\ldots,\xi_{t-1}, \xi_{t})$, where we replace $\xi_{0}$ by an i.i.d. copy of $\xi_{0}^\ast$. For $q \geq 1$, define the functional dependence measures, $\delta_{q,t}(X_t) \defeq \|X_{t}- X_{t}^\ast\|_q$, that measures the dependency of $\xi_{0}$ on $X_{t}$, and $\Delta_{m,q} (X_t)\defeq \sum^{\infty}_{t=m} \delta_{q,t}$, that measures the cumulative effect of $\xi_{0}$ on $X_{t\geq m}$. We aslo define the predictive dependence measure by
		$\mathcal{P}_j X_t = \E(X_t|\mathcal{F}_j) - \E (X_t| \mathcal{F}_{j-1})$.		
	\end{definition}
	
	By Theorem 3 in \cite{wu2011asymptotic} and conditioning on  $\mathcal{C}_z  $, if the condition, $ \sum^{\infty}_{t=0} \|\mathcal{P}_0(a^{\top}\tilde{\yit}_t) \|_q < \infty $ holds, then we can obtain the main result in Theorem \ref{thrm_stationary}. 
	Noting that $\Delta_{0,q} (a^{\top}\yit_t)< \infty$ and $\|\mathcal{P}_0(a^{\top}\tilde{\yit}_t) \|_q $ are equivalent measures, we prove that $\Delta_{0,q} (a^{\top}\yit_t)< \infty$ as follows:
	\begin{eqnarray*}
		\mathcal{P}_0(a^{\top}\tilde{\yit}_t) &=& a^{\top} (\E(\tilde{\yit}_t| \mathcal{F}_{0})- \E(\tilde{\yit}_t| \mathcal{F}_{-1}))\\
		&=&a^{\top}(M^{t}(\Dit_0-\Dit) +\sum_{l\geq t} M^{t-1}(M_0- M)M_{-1}\cdots M_{t-l+1} \Dit_{t-l})\\
		&\leq &  a^{\top} (M^{t}\IF_{N} 2(d_z+d_f) +\sum_{l\geq t} M^{t-1}(M_0- M)M_{-1}\cdots M_{t-l+1} \IF_{N} (d_z+d_f))\\
		&\leq &  a^{\top}(c'^t+ \sum_{l\geq t}2c'^{l})\IF_N  2(d_z+d_f) \leq 4 (d_z+ d_f)c'^{t}/(1-c')
	\end{eqnarray*}
	where $c^{\prime }=(1-c_1)^{-1}c_{23}$. Thus, $\Delta_{0,q} (a^{\top}\mathbb{Y}_t)\lesssim \sum_{t \geq 0} (\|d_z\|_q+ \|d_f\|_q)c^{\prime t} /(1-c^{\prime}) < \infty$. Hence, the condition is satisfied.

	\subsection{Proof of Theorem \ref{theorem_linearization} and Theorem \ref{theorem_estimation}} 
	
	Recall that $\mathcal{C}_f \defeq \sigma ( \fit_{t}, \cdots, \fit_{t-p}) $, $\mathcal{C}_z \defeq \sigma ( Z_{1}, \cdots, Z_{N}  )$, where $\sigma(\cdot)$ denotes a sigma field. $\mathcal{C} = \mathcal{C}_f \cup \mathcal{C}_z $. As the statistic object is involved with spatial/temporal dependence, we should condition on $\mathcal{C}$. Throughout the expectations operator $\E$ is conditional on $\mathcal{C}$.

	\renewcommand*{\thelemma}{A.\arabic{lemma}}
	
	\subsubsection{Lemmas for NED Processes}
	
	Let $\{U_{it}\}_{i,t}$ be the basis of NED processes. Then, we provide a number of Lemmas on the basic properties of NED in random fields, see also \cite{xu2015maximum}. 
	
	\begin{lemma}\label{Lem1}
		If $\{Y_{it}\}$ and $\{Z_{it}\}$ are both uniformly $L_{2+\delta}$ bounded, and uniformly and geometrically $L_{2}$-NED, then $\{Y_{it}Z_{it}\}$ is uniformly and geometrically $L_{2}$-NED.
	\end{lemma}
	
	\begin{lemma} {(Lemma A.1 in \citet{Jenish2009})} \label{Lem2}
		For $h \geq 1$, there exists some $\pi_{1} < \infty$ such that the number of all elements in $D_{NT}$ located within a distance $[h, h+1)$ satisfying $\sum_{j \in D_{NT}: h\leq \rho(i, j) < h + 1} \leq \pi_{1}h^{d-1}$ for any $i \in D_{NT}$. 
	\end{lemma}
	
	\begin{lemma}\label{Lem3}
		Suppose that an $N \times N$ square matrix, $W$ could be decomposed into the sum of two $N \times N$ matrices: $W = A + B$. Denote $|A|_{max} = \max_{ij} |a_{ij}|, i, j = 1, \cdots, N$. Then, for any positive integer $l$, we have $(W^{l} - B^{l})_{ij} \leq |A|_{max} \displaystyle \sum_{k = 0}^{l-1} \| B\|_{\infty}^{k} \| W\|_{1}^{l-1-k}$.
	\end{lemma}
	\begin{proof}\; 
		Let $e_{i} = (0, \cdots, 0, 1, 0, \cdots, 0)^{\top}$ is the unit column vector with one on its $i$-th entry and zeros otherwise.
		By expansion, $W^{l} - B^{l} = \displaystyle \sum_{k = 0}^{l-1} B^{k}AW^{l-1-k} $. Then, $(W^{l} - B^{l})_{ij} = \displaystyle \sum_{k = 0}^{l-1} e_{i}^{\top}B^{k}AW^{l-1-k} e_{j} $. For any matrix $M$ and a vector $e$ of dimension $n$, $\| Me\|_{\infty} \leq |M|_{max} \| e\|_{1}$. Hence, $e_{i}^{\top}B^{k}AW^{l-1-k} e_{j} \leq \| e_{i}^{\top}B^{k}\|_{\infty} \| AW^{l-1-k} e_{j}\|_{1} \leq \| B^{k}\|_{\infty} | A|_{max} \| W^{l-1-k} e_{j}\|_{1} \leq | A|_{max}  \| B^{k}\|_{\infty} \| W\|_{1}^{l-1-k}$ for $k = 0, \cdots, l-1$.
	\end{proof}
	
	\begin{lemma}\label{Lem4}
		For any $\alpha > 0$ and $s \geq 2$, $\sum_{h = [s]}^{\infty} h^{-\alpha-1} <  \frac{2^{\alpha+1}}{\alpha}s^{-\alpha}$, where $[s]$ denotes the largest integer less than or equal to $s$.
	\end{lemma}
	\begin{proof}\;	
		For $h \geq 2$, $h\geq\frac{h+1}{2}$. When $\alpha > 0$, $\sum_{h = [s]}^{\infty} h^{-\alpha-1} \leq \sum_{h = [s]}^{\infty} (\frac{h+1}{2})^{-\alpha-1} \leq 2^{\alpha+1} \int_{s}^{\infty} x^{-\alpha-1} dx = \frac{2^{\alpha+1}}{\alpha}s^{-\alpha}$. Therefore, $\sum_{h = [s]}^{\infty} h^{-\alpha-1} < \frac{2^{\alpha+1}}{\alpha}s^{-\alpha}$.
	\end{proof}
	
	\begin{lemma}\label{Lem5}
		If $\{Y_{it}\}$ and $\{Z_{it}\}$ are both uniformly $L_{2+\delta}$ bounded, and uniformly and geometrically $L_{p}$-NED, then $\{Y_{it} - Z_{it}\}$ and $\{Y_{it} + Z_{it}\}$ are both uniformly and geometrically $L_{p}$-NED.
	\end{lemma}
	\begin{proof}\;
		Define $\Vert Z_{it}-\mathop{\mbox{\sf E}}(Z_{it}|\mathscr{F}_{it}(s))\Vert
		_{p}<d_{it}^{Z}\varphi(s)^{Z}$ and $\Vert Y_{it}-\mathop{\mbox{\sf E}}(Y_{it}|\mathscr{F}_{it}(s))\Vert
		_{p}<d_{it}^{Y}\varphi(s)^{Y}$.	By Minkowski's inequality, $ \| (Y_{it} - Z_{it}) - \E((Y_{it} - Z_{it})|\mathscr{F}_{it}(s)) \|_{p} \leq \| Y_{it} - \E(Y_{it}|\mathscr{F}_{it}(s)) \|_{p} + \| Z_{it} - \E(Z_{it}|\mathscr{F}_{it}(s)) \|_{p} < d_{it}^{Y}\varphi(s)^{Y} + d_{it}^{Z}\varphi(s)^{Z} < d_{it}\varphi(s) $, where $d_{it} = \max(d_{it}^{Y}, d_{it}^{Z})$ and $\varphi(s) = \varphi(s)^{Y} + \varphi(s)^{Z}$. Similarly for $\{Y_{it} + Z_{it}\}$.
	\end{proof}
	
	\begin{lemma}\label{Lem6} (\cite{ibragimov1971independent})
		Let $L_{p}(\mathscr{F}_{1})$ and $L_{p}(\mathscr{F}_{2})$ be the class of $\mathscr{F}_{1}-$measurable and $\mathscr{F}_{2}-$measurable random variables, $x$ with $\| x \|_{p} < \infty $. Let $X\in L_{p}(\mathscr{F}_{1}) $ and $Y \in L_{p}(\mathscr{F}_{2})$. Then, for any $1\leq p, q, r < \infty$ such that $p^{-1} + q^{-1} + r^{-1} = 1$,
		\begin{equation*}
			|Cov(X, Y)| < 4\alpha^{1/r}(\mathscr{F}_{1}, \mathscr{F}_{2}) \| X \|_{p} \| Y \|_{q},
		\end{equation*}
		where $\alpha(\mathscr{F}_{1}, \mathscr{F}_{2}) = \sup_{A \in \mathscr{F}_{1}, B \in \mathscr{F}_{2}} (|P(AB) - P(A)P(B)|)$.
	\end{lemma}

	\begin{lemma}\label{Lemma1}
		Under assumptions \ref{Ass_stationray}(1), \ref{Ass1} - \ref{Ass3}(1),
		
		(1) Define $\Gamma_{u} =\| W\|_{1} < \infty $, and $\Gamma_{w} =\| W\|_{\infty} =1 $.
		
		(2) For any $N$ and positive integer $h$, $\| W^{h}\|_{1} < hK\Gamma_{u}$. 
	\end{lemma}
	\begin{proof}\;
		(1) Using Lemma \ref{Lem2}, $ \| W\|_{1} = \sup_{j} \sum_{i = 1}^{N} |\omega_{ ij}| = \sup_{j} \sum_{h = 1}^{\infty} \sum_{i: h \leq \rho(i, j) < h + 1 } |\omega_{ ij}| \leq \sup_{j} \sum_{h = 1}^{\infty} \sum_{i: h \leq \rho(i, j) < h + 1 } \pi_{0}h^{-c_{w}} \leq \sum_{h = 1}^{\infty} \pi_{1}h^{d-1}\pi_{0}h^{-c_{w}} < \infty$, due to {$c_w>d$ (Assumption \ref{Ass3}(1))}. As $W$ is row-normalized in our model, $\Gamma_{w} =\| W\|_{\infty} =1 $.
		
		(2) Define an index set $V_{N}$ with $\sum_{i = 1}^{N} |\omega_{ij}| > \Gamma_{w}$ if $j \in V_{N}$ and $\sum_{i = 1}^{N} |\omega_{ij}| \leq \Gamma_{w}$ if $j \notin V_{N}$. By Assumption \ref{Ass3}(1), $|V_{N}|\leq K$ for any $N$. Let $e_{k} = (0, \cdots, 0, 1, 0, \cdots, 0)^{\top}$ be a unit column vector with one on the $k$-th entry and zeros on other entries. Note that $e = (1,\cdots, 1)^{\top} = \sum_{k = 1}^{N}e_{k}$ and $I_{N} = \sum_{j = 1}^{N}e_{j}e_{j}^{\top}$. The $k$-th column sum of $W^{h}$ ($e^{\top}W^{h}e_{k}$) can be expressed as
		\begin{align*}
			e^{\top}W^{h}e_{k} & = \displaystyle \sum_{j = 1}^{N} e^{\top} W e_{j}e_{j}^{\top} W^{h-1}e_{k}\\
			& = \displaystyle \sum_{j \in V_{N}} e^{\top} W e_{j}e_{j}^{\top} W^{h-1}e_{k} + \displaystyle \sum_{j \notin V_{N}} e^{\top} W e_{j} e_{j}^{\top} W^{h-1}e_{k}\\
			& \leq K \left(\max_{j \in V_{N}} e^{\top} We_{j}\right) \left(\max_{j \in V_{N}} e_{j}^{\top} W^{h-1}e_{k}\right) + \left(\max_{j \notin V_{N}} e^{\top} W e_{j}\right) \displaystyle \sum_{j \notin V_{N}} e_{j}^{\top} W^{h-1}e_{k} \\
			& \leq K\Gamma_{u} \| W^{h-1}\|_{\infty} + \Gamma_{w}\| W^{h-1}\|_{1} \\
			& \leq K\Gamma_{u} + \| W^{h-1}\|_{1},  \quad \forall k = 1, \cdots,N.
		\end{align*}
		Hence, we have $\| W^{h}\|_{1} \leq K\Gamma_{u} + \| W^{h-1}\|_{1} $. By deduction, we achieve that $\| W^{h}\|_{1} \leq (h-1)K\Gamma_{u} + \Gamma_{u} \leq hK\Gamma_{u}$.
	\end{proof}

	\subsubsection{Proof of Proposition \ref{Pro1}}
	\label{proof_pro1}
	
	\begin{proof}\;
		(1) We first discuss the NED properties of $\{ Y_{it} \}_{i,t}$. Following \citet{Jenish2012}, we can show that the NED property is satisfied if random fields are generated from nonlinear Lipschitz type functions on random field $\{ U_{it}\}$.
		Notice that $\int_{0}^{1}\gamma^0_1(u)du \leq \max_u \gamma^0_{1}(u)$. Define {$\{\mathcal{F}_{it}(s) = \sigma( U_{i',t'}, \mathcal{C}): |i-i'|\leq s, |t-t'|\leq s \}$, $\Pi_{l,i,t}^s \defeq \E(\Pi_l|\mathcal{F}_{it}(s)) $}, $\Dit_{t-l,i,t}^s = \E(\Dit_{t-l}|\mathcal{F}_{it}(s)) $, and $\tilde{\Pi}_{l-s+1} = M_{t-s-1}\cdots M_{t-l+1}$. Conditioning on $\mathcal{C}$,
		\begin{eqnarray*}
			&&||Y_{it}- \E\{ Y_{it}| \mathcal{F}_{it}(s)\}||_{2}\\
			&\leq& ||e_i^{\top}(\sum_{l\geq 0} \Pi_l \Dit_{t-l} -\sum_{l\geq 0} \E (\Pi_l \Dit_{t-l}|\mathcal{F}_{it}(s)))||_{2}\\
			&\leq& || e_i^{\top}\sum_{l\leq s} (|\Pi_l|_a+ |\Pi_{l,i,t}^s|_a) |\Dit_{t-l}- \Dit_{t-l,i,t}^s|_a ||_{2}\\
			&+& || e_i^{\top}\sum_{l> s} (|\Pi_{s+1}|_a+ |\Pi_{s+1,i,t}^s|_a) |\tilde{\Pi}_{l-s+1}\Dit_{t-l}- M^{l-s+1}\mu_{\Dit}|_a ||_{2}\\
			&\leq& T_1+ T_2,
		\end{eqnarray*}
		Then, we handle the first term via a spatial dependency and the second term via a temporal dependency.
		
		Let $\Ait = (\sum_{k\geq 0}c_1^{k} |W|_a^k) (c_2 |W|_a+ c_3I)$. Notice that $|\Pi_l|_a+ |\Pi_{l,i,t}^s|_a \leq 2\Ait^l$, where $|.|_{a}$ is element-wise absolute value. By the row normalization, we have  $(\E |T_2|^q )^{1/q}\lesssim |\mu_{q,\Dit}|_{\infty} (c_{23}/(1-c_1))^{s+1} $, where $|\mu_{q,\Dit}|_{\infty}$ is the maximum element of $ \max_i(\E||\Dit_{t-l}- \mu_{\Dit}|_{a,i}|^q)^{1/q}$.
		Define $\mathcal{B}_{it}(s) = \{(i',t'): |i'-i|\leq s, |t'-t|\leq s\}$. Then, we have: 
		\begin{eqnarray}
			T_1 \leq || \sum_{j\not \in \mathcal{B}_{it}(s)} e_i^{\top} \sum_{l\leq s} \Ait_{.j}^l|(\Dit_{t-l}- \Dit_{t-l,i,t}^s)|_j ||_{2},
		\end{eqnarray}
		as  the term inside $\mathcal{B}_{it}(s)$ cancels for $(\Dit_{t-l}- \Dit_{t-l,i,t}^s)$.
		Note that $||e_i^{\top} \Ait^{l-1}||_{2} \leq (c_{23})^{l-1}/(1-c_1)^{l-1}$ by Assumption \ref{Ass_stationray}.
		For $l> 1$, we have 
		\begin{eqnarray*}
			&&  \sum_{j\not \in \mathcal{B}_{it}(s)}  \sum_{l\leq s} ||e_i^{\top}\Ait^{l}(\Dit_{t-l}- \Dit_{t-l,i,t}^s)||_{2}
			\leq \sum_{j\not \in \mathcal{B}_{it}(s)}  \sum_{l\leq s} ||(|e_i^{\top} \Ait^{l-1}|_1 |\Ait(\Dit_{t-l}- \Dit_{t-l,i,t}^s)|_{\infty})||_{2}\\
			&& \leq \sum_{l\leq s} (c_2+ c_3)^{(l-1)}/(1-c_1)^{(l-1)}(\{|\Dit_{t-l}|_{\infty}\vee |\mu_{\Dit}|_{\infty}\} \sum_{j\not \in \mathcal{B}_{it}(s)} c_{23}g_{ij})\\
			&& \leq  \sum_{l\leq s} (c_{23})^{l}/(1-c_1)^{(l-1)} (1-c)^{-1} \{(d_f+ d_z)\vee |\mu_{\Dit}|_{\infty}\} (\sum_{j\not \in \mathcal{B}_{it}(s)} g_{ij}) \\
			&& \leq C (\sum_{j\not \in \mathcal{B}_{it}(s)} g_{ij}), 
		\end{eqnarray*}
		where $C$ is a constant and $g_{ij} = |(I - |c_{1} W|)^{-1}_{ij}| $. Using Assumption \ref{Ass_stationray}(1)--(3), we obtain the last step.

		Hence, under the condition, $\displaystyle \sup_{i} \sum_{j \not \in \mathcal{B}_{it}(s)} g_{ij} \rightarrow 0$ as $s \rightarrow \infty $. Then, $T_1 \rightarrow 0$. It is straightforward to prove it using the following results: either $T_1 \lesssim c^{-s} |\mu_{q,\Dit}|_{\infty}$ or $T_1 \lesssim s^{-c_w-2}|\mu_{q,\Dit}|_{\infty}$ depending on the assumption. Therefore,
		\begin{equation}
			\| Y_{it} - \E(Y_{it}|\mathscr{F}_{it}(s)) \|_{2} \lesssim \psi(s),
		\end{equation}
		where $\psi(s) \to 0$ as $s \to 0$. 
		
		Next, we discuss condition, $ \displaystyle \sup_{i} \sum_{j\not \in \mathcal{B}_{it}(s)} g_{ij} \rightarrow 0$ as $s \rightarrow \infty $, under Assumption \ref{Ass3}(1) and (2), respectively. 
		
		(i) Under Assumption \ref{Ass3}(1), we use the properties of nilpotent matrix and decompose the matrix, $W$. For any positive integer $h$, we construct two $N \times N$ matrices $A$ and $B$ as follows: $a_{ij} = w_{ij}\II\{\rho(i, j) < N-h+1\}$ and $b_{ij} = w_{ij} \II \{\rho(i, j) \geq N-h+1\}$. Then, $W = A + B$ and $a_{ij}b_{ij} = 0$. Next, we check whether $B$ is a nilpotent matrix, i.e., $B^{h} = 0$. Under Assumption \ref{Ass3}(1), $|w_{ij}| \leq \pi_{0} \rho(i, j)^{-c_{w}} $, and by Lemma \ref{Lem3}, we have:
		\begin{align*}
			(|W|^{h})_{ij} = & (W^{h} - B^{h})_{ij} \leq |A|_{max} \displaystyle \sum_{k = 0}^{h-1} \| B\|_{\infty}^{k} \| W^{h-1-k}\|_{1}  \quad \text{by Lemma \ref{Lem3}} \\
			\leq & \pi_{0} \rho(i, j)^{-c_{w}} \sum_{k = 0}^{h-1} \| W\|_{\infty}^{k} (h-k-1)K\Gamma_{u}  \quad \text{by Assumption \ref{Ass3}(1); Lemma \ref{Lemma1}(2)} \\
			\leq & \pi_{0} \rho(i, j)^{-c_{w}} \sum_{k = 0}^{h-1}  (h-k-1)K\Gamma_{u}  \leq \pi_{0} \rho(i, j)^{-c_{w}} K\Gamma_{u} h^{2}.
		\end{align*}
		Hence, for any $i \neq j$, using $\Upsilon = \displaystyle \sup_{\tau}  |\gamma^{0}_{1}(\tau)|$ (Assumption \ref{Ass_stationray}(1)), then
		\begin{align*}
			g_{ij} = & |(I - |\gamma^{0}_{1} W|)^{-1}_{ij}| = \sum_{h = 0}^{\infty} |\gamma^{0}_{1} W|^{h}_{ij} = \sum_{h = 0}^{\infty} |\gamma^{0}_{1}|^{h} | W|^{h}_{ij} \\
			\leq & \sum_{h = 0}^{\infty} \Upsilon ^{h} \pi_{0} \rho(i, j)^{-c_{w}} K\Gamma_{u}  h^{2} \quad \text{by Assumption \ref{Ass_stationray}(1)} \\
			= &  \pi_{0}K \Upsilon^{2} \Gamma_{u} \rho(i, j)^{-c_{w}}  \sum_{h = 0}^{\infty} \Upsilon^{h-2} h^{2} \\
			\leq &  \pi_{2}\rho(i, j)^{-c_{w}}, \quad \text{for some constant $\pi_{2}>0$.}
		\end{align*}
		For sufficiently large $s$, we have:
		\begin{align*}
			\displaystyle \sup_{i} \sum_{j: \rho(i, j)>s} g_{ij}
			& \leq \displaystyle \sup_{i} \sum_{h = [s]}^{\infty} \sum_{j: h\leq \rho(i, j) < h + 1} \pi_{2}\rho(i, j)^{-c_{w}} \\
			& \leq \sum_{h = [s]}^{\infty} \pi_{1}h^{d-1}  \pi_{2}h^{-c_{w}} =  \sum_{h = [s]}^{\infty} \pi_{1} \pi_{2}h^{-(c_{w} - d) - 1}   \quad \text{by Lemma \ref{Lem2}}  \\
			& \leq \pi_{1} \pi_{2}  \frac{2^{c_{w} - d +1}}{c_{w} - d}s^{-(c_{w} - d)}   \quad \text{by Lemma \ref{Lem4}} \\
			& \leq \pi s^{-(c_{w} - d)}, \quad \text{for some constant $\pi>0$.}
		\end{align*}
		Under Assumption \ref{Ass3}(1), $c_{w} > d$. Hence, as $s \rightarrow \infty$,  $\displaystyle \sup_{i} \sum_{j: \rho(i, j)>s} g_{ij} \leq \pi s^{-(c_{w} - d)} \rightarrow 0$.
		
		(ii) Next, under Assumption \ref{Ass3}(2), we have: $\displaystyle \sup_{i} \sum_{j: \rho(i, j)>s} g_{ij} =  \displaystyle \sup_{i} \sum_{j: \rho(i, j)>s} \sum_{h = 0}^{\infty}  |\gamma^0_{1}|^{h}| W|^{h}_{ij} \\ \leq \displaystyle \sup_{i} \sum_{j: \rho(i, j)>s} \sum_{h = [s/\bar{\rho}_{0}] +1} |\gamma^0_{1}|^{h}| W|^{h}_{ij}
		= \displaystyle \sup_{i} \sum_{h = [s/\bar{\rho}_{0}] +1} \sum_{j: \rho(i, j)>s}  |\gamma^0_{1}|^{h}| W|^{h}_{ij} \\
		\leq \displaystyle \sup_{i} \sum_{h = [s/\bar{\rho}_{0}] +1} \Upsilon^{h}
		\leq (1-\Upsilon)^{-1} \Upsilon^{s/\bar{\rho}_{0}}$.
		Under Assumption \ref{Ass_stationray}(1), $\Upsilon < 1$. Hence, as $s \rightarrow \infty$, we have: $\displaystyle \sup_{i} \sum_{j: \rho(i, j)>s} g_{ij} \leq  (1-\Upsilon)^{-1} \Upsilon^{s/\bar{\rho}_{0}} \rightarrow 0$.

		(2)	Next, we discuss the NED properties of $\{ u_{it} \}_{i,t} $. We first show that $\{\bar{y}_{it}\}_{it}$ is NED, where $\bar{y}_{it} = e_i^{\top} W \yit_t$. Note that $ \| \bar{y}_{it} - \E(\bar{y}_{it}|\mathscr{F}_{it}(s)) \|_{2}
		\leq \displaystyle \sum_{j = 0}^{N} |W_{ij}| \| Y_{j,t} - \E(Y_{j,t}|\mathscr{F}_{it}(s)) \|_{2}$. Under Assumption \ref{Ass3}(1), using the result $ \| Y_{it} - \E(Y_{it}|\mathscr{F}_{it}(s)) \|_{2} < C s^{-(c_{w} - 2)}$ in Proposition \ref{Pro1}(1), we have: $ \| \overline{Y}_{it} - \E(\overline{Y}_{it}|\mathscr{F}_{it}(s)) \|_{2}
		\leq \displaystyle \sum_{j = 0}^{N} |W_{ ij}| C s^{-(c_{w} - d)} \leq \| W\|_{\infty}C s^{d-c_{w}} \leq \pi_{3}s^{d-c_{w}}$ for some positive constant $\pi_{3}$. Hence, $\{ \overline{Y}_{it} \}$ is NED process. Using Proposition \ref{Pro1}(2), similar results can be obtained under Assumption \ref{Ass3}(2). By Lemma \ref{Lem5}, it is easily seen that $\{u_{it} \}_{i,t}$ follow the NED process.
		
		We now prove that this NED property can be transformed. Let $\tilde{u}_{it}$ be a middle point between $u_{it}$ and $0$. Then, for sufficient small $\vps>0$, we have:
		\begin{eqnarray*}
			& &\| \psi_{\tau}( u_{it}) - \E[\psi_{\tau}( u_{it})|\mathscr{F}_{it}(s)] \|_{2} =  \| \II( u_{it} \geq 0) - \E[ \II( u_{it} \geq 0)|\mathscr{F}_{it}(s)] \|_{2} \\
			& \leq &\| \II( u_{it} \geq 0) - 
			\II	\{\E[u_{it}| \mathscr{F}_{it}(s)] \geq 0 \} \|_{2} = \left\{\E\left | \II( u_{it} \geq 0) - \II\{ \E[ u_{it}|\mathscr{F}_{it}(s)] \geq 0 \} \right|^{2} \right \}^{\frac{1}{2}} \\
			& \leq & \P ( u_{it} \geq 0,  \E[u_{it}|\mathscr{F}_{it}(s)] <0) 
			\leq \P (u_{it} \geq \vps,  \E[u_{it}|\mathscr{F}_{it}(s)] <0) + \P(0<u_{it} < \vps) \\
			&\leq & \P (|u_{it} - \E[u_{it} |\mathscr{F}_{it}(s)]| >\vps) + \vps f(\tilde{u}_{it}) 
			\leq \E (|u_{it}- \E[u_{it}|\mathscr{F}_{it}(s)]|^2)/\vps^2 + \vps f(\tilde{u}_{it}) \\
			&\leq& \psi(s)/\vps^2 +\vps c_u.
		\end{eqnarray*}
		Taking $\vps = \psi(s)^{1/3}$ to be sufficiently small, then we achieve the desired result. Hence, conditioning on $\mathcal{C}$, 
		$\{\psi_{\tau}(u_{it})\}_{i,t}$ and $\{\rho_{\tau}(u_{it})\}_{i,t}$ are also $L_{2}$-NED on $\{ U_{it} \}_{i,t}$.
	\end{proof}

	\subsubsection{Proof of Theorem \ref{theorem_linearization} and Theorem \ref{theorem_estimation}}

	We collect some important notations: $\eta(\tau)\equiv(\phi(\tau)^{\top},\lambda(\tau)^{\top})^{\top}$, $\pi(\tau) \equiv (\gamma_1(\tau) , \eta^{\top}(\tau))^{\top} = (\gamma_1(\tau), \phi^{\top}(\tau), \lambda^{\top}(\tau))^{\top}$.  For convenience we denote $\eta \equiv (\phi^{\top},\lambda^{\top})^{\top}$, $\pi \equiv (\gamma_1, \eta^{\top})^{\top} =(\gamma_1, \phi^{\top}, \lambda^{\top})^{\top}$, and $\theta \equiv (\gamma_1, \phi^{\top})^{\top}$. Recall that 
	\begin{align}
		&	Q_{NT}(\gamma_1, \eta, \tau)  \equiv  \frac{1}{NT}\sum_{i=1}^{N} \sum^T_{t=1} \left[\rho_{\tau}\left\{Y_{it} - \gamma_{1}(\tau)\overline{Y}_{it} - X^{\top}_{it}\phi(\tau) - {R}^{\top}_{it} \lambda(\tau) \right\}  \right], \\	
		&	Q_{\infty}(\gamma_1, \eta, \tau)  \equiv  \lim_{N, T\rightarrow\infty}  \E[Q_{NT}(\gamma_1, \eta, \tau)] ,	\\		
		&	\hat\eta(\gamma_1,\tau)  \equiv   (\hat\phi^{\top}(\gamma_1, \tau),\hat\lambda^{\top}(\gamma_1, \tau))^{\top} \equiv \underset{(\phi,\lambda) \in \mathcal{B}\times\mathcal{G} }{\mbox{arg min }} Q_{NT}(\gamma_1, \eta, \tau), \\
		&	\eta^{0}(\gamma_1, \tau)  \equiv  (\phi^{0\top}(\gamma_1, \tau), \lambda^{0\top}(\gamma_1, \tau))^{\top} = \underset{(\phi^0, \lambda^0) \in \mathcal{B}\times\mathcal{G} }{\mbox{arg min }} Q_{\infty}(\gamma_1,\eta, \tau) , \\
		&   s_{it}(\gamma_1, \eta(\gamma_1, \tau), \tau) \equiv \psi_{\tau}\left\{Y_{it} - \gamma_1(\tau)\overline{Y}_{it} - \Psi^{\top}_{it} \eta(\gamma_1,\tau) \right\} \Psi_{it}, \quad \Psi_{it} = (R^{\top}_{it}, {X}^{\top}_{it})^{\top} , \\
		&   \check{s}_{it}(\gamma_1, \eta(\gamma_1, \tau), \tau) \equiv s_{it}(\gamma_1,\eta(\gamma_1,\tau),\tau)-\E s_{it}(\gamma_1,\eta(\gamma_1,\tau),\tau), \\
		& G_{NT}(\gamma_1, \eta(\gamma_1, \tau), \tau)  \equiv  \frac{1}{\sqrt{NT}}\sum_{i=1}^N\sum_{t=1}^T \check{s}_{it}(\gamma_1,\eta(\gamma_1,\tau),\tau) \\ \nonumber
		& \quad \quad \quad \quad \quad \quad \quad \quad = \frac{1}{\sqrt{NT}}\sum_{i=1}^N\sum_{t=1}^T[ s_{it}(\gamma_1,\eta(\gamma_1,\tau),\tau)-\E s_{it}(\gamma_1,\eta(\gamma_1,\tau),\tau)],  \\
		& G^{0}_{NT} \equiv \frac{1}{\sqrt{NT}}\sum_{i=1}^N\sum_{t=1}^T \check{s}_{it}(\gamma_1^{0},\eta^{0}(\gamma_1^{0},\tau),\tau) = \frac{1}{\sqrt{NT}}\sum_{i=1}^N\sum_{t=1}^T[ s_{it}(\gamma_1^{0},\eta^{0}(\gamma_1^{0},\tau),\tau)] .	\end{align}
	and
	\begin{eqnarray}
		\hat\gamma_1(\tau) &\equiv& \underset{\gamma_1 \in \mathcal{A}}{\mbox{arg min }} |\hat\lambda(\gamma_1,\tau)|_2^2,\quad \gamma_1^{*}(\tau) \equiv \underset{\gamma_1 \in \mathcal{A} }{\mbox{arg min }}|\lambda(\gamma_1,\tau)|_2^2 , \\
		\hat\eta(\tau) &\equiv& (\hat\phi^{\top}(\tau),\hat\lambda^{\top}(\tau))^{\top} \equiv \hat\eta(\hat\gamma_1(\tau),\tau), \\
		\eta^{0}(\tau) &\equiv& (\phi^{0\top}(\tau), \mathbf{0}^{\top})^{\top} \equiv \eta^{0}(\gamma_1^{0}(\tau),\tau).	
	\end{eqnarray}
	where $\eta^{0}(\tau)=(\phi^{0\top}(\tau),\mathbf{0}^{\top})^{\top}$ are the true parameters and $\rho_\tau(u)=(\tau-1(u\leq0))u$, where recall that $ \mathcal{A}$ is the parameter space of $\gamma_1(\tau)$. 

	
	Following \citet{Chernozhukov2006}, we fix $\tau$ and prove the Theorems in three steps. 
	
	\textbf{Step 1. (Identification)} By Assumption \ref{assumption_estimation}(3),  $\theta^{0}(\tau)=(\gamma_1^{0}(\tau),\phi^{0}(\tau))$ is the unique solution to ${S}_{\infty}(\theta, \tau) = 0$; namely, it uniquely solves:
	\begin{equation}
		\lim_{N, T\to\infty}\frac{1}{NT}\sum_{i=1}^{N} \sum^T_{t=1} \E\left[\psi_{\tau}\left\{Y_{it} - \gamma_{1}(\tau)\overline{Y}_{it} - X^{\top}_{it}\phi(\tau)  - {R}^{\top}_{it} \mathbf{0} \right\} \Psi_{it} \right].
		\label{exi1} 
	\end{equation}
	
	In view of the global convexity of $Q_{\infty}(\gamma_1, \eta, \tau)$ in $\eta$ for each $\gamma_1$ and $\tau$, if  $\eta^{0}(\gamma_1,\tau)=(\phi^{0}(\gamma_1,\tau),\lambda^{0}(\gamma_1,\tau))$ is in the interior of $\mB\times \mathcal{G}$, then $\eta^{0}(\gamma_1,\tau)$ uniquely solves the first order condition of minimizing $Q_{\infty}(\gamma_1, \eta, \tau)$ over $\eta$:
	\begin{equation}
		\lim_{N\rightarrow\infty,T\rightarrow\infty}\frac{1}{NT}\sum_{i=1}^N\sum_{t=1}^T  \E\left[\psi_{\tau}\left\{Y_{it} - \gamma_{1}(\tau)\overline{Y}_{it} - X^{\top}_{it}\phi(\gamma_1, \tau)  - {R}^{\top}_{it} \lambda(\gamma_1,\tau)  \right\} \Psi_{it} \right] =\mathbf{0}.
		\label{exi2}
	\end{equation}
	We need to find $\gamma_1^{*}(\tau)$ by  minimizing $|\lambda(\gamma_1,\tau)|_2^2$ over $\gamma_1$ subject to the constraint in (\ref{exi2}). By (\ref{exi1}), it is clear that $\gamma_1^{*}(\tau)=\gamma_1^{0}(\tau)$ making $|\lambda(\gamma_1(\tau),\tau)|_2^{2}=0$ and that $\gamma_1^{0}(\tau)$ satisfies (\ref{exi2}). That is, $\gamma_1^{*}(\tau)=\gamma_1^{0}(\tau)\in\mbox{arg min }_{\gamma_1 \in \mathcal{A}}|\lambda(\gamma_1,\tau)|_2^{2}$ subject to the constraint in (\ref{exi2}). It is also the unique solution by (\ref{exi1}). Hence, $\phi(\gamma_1^*(\tau),\tau)=\phi(\gamma_1^{0}(\tau),\tau)=\phi^{0}(\tau)$ by (\ref{exi2}).

	\textbf{Step 2. (Consistency)} In Proposition \ref{Pro1}, we establish that $\{\rho_{\tau}(u_{it})\}_{i,t}$ is $L_{2}$-NED on $\{U_{it} \}_{i,t}$. By Theorem 1 in \citet{Jenish2012} and under Assumption \ref{Ass5}, we have the uniform consistency:  $\sup_{\gamma_1,\phi,\lambda,\tau}|Q_{NT}(\gamma_1,\phi,\lambda, \tau)- \E Q_{NT}(\gamma_1,\phi,\lambda, \tau)|_2^2 = o_p(1)$.
	By the bounded density condition in \ref{assumption_estimation}(2), $Q_{\infty}(\gamma_1, \eta, \tau)$ is continuous over $\mathcal{A} \times (\mathcal{B}\times \mathcal{G})\times \mathcal{T}$. By Lemma B2, $\sup_{(\gamma_1,\eta,\tau) \in \mathcal{A} \times (\mathcal{B}\times \mathcal{G})\times \mathcal{T}}|Q_{NT}(\gamma_1, \eta, \tau)- Q_{\infty}(\gamma_1, \eta, \tau)|_2^2 = o_p(1)$. By Lemma B.1 in \citet{Chernozhukov2006} this implies the uniform convergence: $\sup_{(\gamma_1,\tau) \in \mathcal{A} \times \mathcal{T}}|\hat\eta(\gamma_1,\tau)-\eta^{0}(\gamma_1,\tau)|_2^2 =o_p(1)$ (*). It follows that $\sup_{(\gamma_1,\tau) \in \mathcal{A} \times \mathcal{T}} |\hat\lambda(\gamma_1,\tau)-  \lambda^{0}(\gamma_1,\tau) |_2^2=o_p(1)$, which again implies $\sup_{\tau \in  \mathcal{T}}|\hat\gamma_1(\tau)-\gamma_1^{0}(\tau)|_2^2=o_p(1)$ by Lemma B.1 in \citet{Chernozhukov2006}.  Finally,  by (*) we have: $\sup_{\tau \in  \mathcal{T}}|\hat\phi(\tau)-\phi^{0}(\tau)|_2^2=o_p(1)$ and $\sup_{\tau \in  \mathcal{T}} |\hat\lambda(\hat\gamma_1(\tau),\tau) - \mathbf{0}|_2^2=o_p(1) $.
	
	\textbf{Step 3. (Asymptotics)} Consider a small ball $B_{\epsilon_{NT}}(\gamma_1^{0}(\tau))$ of a radius, $\epsilon_{NT}$ centered at $\gamma_1^{0} \equiv \gamma_1^{0}(\tau)$ for each $\tau$ where $\epsilon_{NT}$ is independent of $\tau$ and $\epsilon_{NT} \rightarrow 0$ slowly enough. Let $\tilde{\gamma}_1 \equiv\tilde{\gamma}_1(\tau)\in B_{\epsilon_{NT}}(\gamma_1^{0}(\tau))$. By the properties of the ordinary quantile regression estimator,  $\hat\eta(\tilde{\gamma}_1, \tau) $ (e.g. Theorem 3.3 in \citet{koenker1978}),
	\begin{equation}
		O(n^{-1/2})=\frac{1}{\sqrt{NT}}\sum_{i=1}^N\sum_{t=1}^T \psi_{\tau}\left\{Y_{it} - \tilde{\gamma}_1(\tau)\overline{Y}_{it} - \Psi^{\top}_{it}\hat\eta(\tilde{\gamma}_1,\tau) \right\} \Psi_{it}.
		\label{exi3}
	\end{equation}
	
	Let $s_{it}(\tilde{\gamma}_1, \hat\eta(\tilde{\gamma}_1, \tau), \tau)=\psi_{\tau}\left\{Y_{it} - \tilde{\gamma}_1(\tau)\overline{Y}_{it} - \Psi^{\top}_{it} \hat\eta(\tilde{\gamma}_1,\tau) \right\} \Psi_{it}$, \\$G_{NT} = \frac{1}{\sqrt{NT}}\sum_{i=1}^N\sum_{t=1}^T[ s_{it}(\tilde{\gamma}_1,\hat\eta(\tilde{\gamma}_1,\tau),\tau)-\E s_{it}(\tilde{\gamma}_1,\hat\eta(\tilde{\gamma}_1,\tau),\tau)]$, and \\ $G^{0}_{NT}= \frac{1}{\sqrt{NT}}\sum_{i=1}^N\sum_{t=1}^T s_{it}(\gamma_1^{0},\eta^{0}(\gamma_1^{0},\tau),\tau)$. Using Lemma \ref{lemmaB1}, for any $\sup_{\tau \in \mathcal{T}}|\tilde{\gamma}_1(\tau)-\gamma_1^{0}(\tau)|_2^2=o_p(1)$, it follows that $\sup_{\tau \in \mathcal{T}}|G_{NT}-G^{0}_{NT}|_2^2=o_p(1)$. Then, the equation \eqref{exi3} can be expressed as
	\begin{align}
		O(n^{-1/2})=&\frac{1}{\sqrt{NT}}\sum_{i=1}^N\sum_{t=1}^T s_{it}(\tilde{\gamma}_1, \hat\eta(\tilde{\gamma}_1, \tau), \tau)  \nonumber\\
		=&\frac{1}{\sqrt{NT}}\sum_{i=1}^N\sum_{t=1}^T\left[ s_{it}(\tilde{\gamma}_1,\hat\eta(\tilde{\gamma}_1,\tau),\tau)- \E s_{it}(\tilde{\gamma}_1,\hat\eta(\tilde{\gamma}_1,\tau),\tau)\right] \\
		& + \frac{1}{\sqrt{NT}}\sum_{i=1}^N\sum_{t=1}^T \E s_{it}(\tilde{\gamma}_1,\hat\eta(\tilde{\gamma}_1,\tau),\tau)  \nonumber\\
		=&G^{0}_{NT} + o_p(1) + \frac{1}{\sqrt{NT}}\sum_{i=1}^N\sum_{t=1}^T  \E s_{it}(\tilde{\gamma}_1,\hat\eta(\tilde{\gamma}_1,\tau)).
		\label{OG}
	\end{align}
	By the mean value theorem and dominated convergence, we have:
	\begin{align}
		&\frac{1}{\sqrt{NT}}\sum_{i=1}^N\sum_{t=1}^T  \E s_{it}(\tilde{\gamma}_1,\hat\eta(\tilde{\gamma}_1,\tau))\nonumber\\
		=&\frac{1}{\sqrt{NT}}\sum_{i=1}^N\sum_{t=1}^T \E \psi_{\tau}\left\{Y_{it} - \tilde{\gamma}_1(\tau)\overline{Y}_{it} - \Psi^{\top}_{it}\hat\eta(\tilde{\gamma}_1,\tau) \right\} \Psi_{it} \nonumber\\
		=&(J_{\gamma_1}(\tau)+o_p(1))\sqrt{NT}(\tilde{\gamma}_1(\tau)-\gamma_1^{0}(\tau)) + (J_{\eta}(\tau)+o_p(1))\sqrt{NT}(\hat\eta(\tilde{\gamma}_1,\tau)-\eta^{0}(\tau)), \label{jj15}
	\end{align}
	where
	\begin{equation*}
		J_{\gamma_1}(\tau) =  \dfrac{\partial S_{\infty}(\theta,\tau)} {\partial \gamma_1} \Big|_{\gamma_1 =\gamma^0_1},   \\ J_{\eta}(\tau) =  \dfrac{\partial S_{\infty}(\pi,\tau)} {\partial(\phi^{\top},\lambda^{\top})} \Big|_{\phi = \phi^0, \lambda = \mathbf{0}}, \\
		J(\tau) = \dfrac{\partial S_{\infty}(\pi,\tau)} {\partial (\gamma_1,\phi^{\top})} \Big|_{\gamma_1 = \gamma^0_1, \phi = \phi^0, \lambda = \mathbf{0}}, 
	\end{equation*}
	with dimensions $(q+4+(p+1)m) \times 1$, $(q+4+(p+1)m) \times (q+4+(p+1)m)$ and $(q+4+(p+1)m) \times (q+4+(p+1)m)$.
	Putting (\ref{OG}) and (\ref{jj15}) together, we have:
	\begin{align}
		O(n^{-1/2}) = & G^{0}_{NT} + o_p(1) +  (J_{\gamma_1}(\tau)+o_p(1))\sqrt{NT}(\tilde{\gamma}_1(\tau)-\gamma_1^{0}(\tau)) \\ \nonumber
		& + (J_{\eta}(\tau)+o_p(1))\sqrt{NT}(\hat\eta(\tilde{\gamma}_1,\tau)-\eta^{0}(\tau)),
	\end{align}
	which implies for any $\sup_{\tau \in \mathcal{T}}|\tilde{\gamma}_1(\tau)-\gamma_1^{0}(\tau)|_2^2=o_p(1)$ that  
	\begin{equation}\label{equ12}
		\sqrt{NT}(\hat\eta(\tilde{\gamma}_1,\tau)-\eta^{0}(\tau))= - J_{\eta}^{-1}(\tau)G^{0}_{NT}- J_{\eta}^{-1}(\tau)J_{\gamma_1}(\tau)[1+o_p(1)]\sqrt{NT}(\tilde{\gamma}_1(\tau) - \gamma_1^{0}(\tau))+o_p(1).
	\end{equation}
	Then,
	\begin{align}
		& \sqrt{NT}(\hat\phi(\tilde{\gamma}_1,\tau)-\phi^{0}(\tau))
		=  \bar J_{\phi}(\tau) G^{0}_{NT}-\bar J_{\phi}(\tau) J_{\gamma_1}(\tau)[1+o_p(1)]\sqrt{NT}(\tilde{\gamma}_1(\tau) - \gamma_1^{0}(\tau))+o_p(1), \\
		& \sqrt{NT}(\hat\lambda(\tilde{\gamma}_1,\tau)-\mathbf{0})= \bar J_{\lambda}(\tau) G^{0}_{NT}-\bar J_{\lambda}(\tau) J_{\gamma_1}(\tau)[1+o_p(1)]\sqrt{NT}(\tilde{\gamma}_1(\tau) - \gamma_1^{0}(\tau))+o_p(1), \label{eq11}
	\end{align}
	where we partition $J^{-1}_{\eta}(\tau)=[\bar J^{\top}_{\phi}(\tau),\bar J^{\top}_{\lambda}(\tau)]^{\top}$. 
	
	By Step 2, with probability approaching one,
	\begin{equation}
		\hat\gamma_1(\tau)=\underset{\tilde{\gamma}_1\in B_{\epsilon_{NT}}(\gamma_1^{0})}{\mbox{ arg min }}|\hat\lambda(\tilde{\gamma}_1,\tau)|_2^2, \text{    for all } \tau \in \mathcal{T}. \label{go20}
	\end{equation}
	As discussed in Section \ref{sec_NED}, the process $\{\check{s}_{it}\}_{i,t}$ is NED process, where \\ $\check{s}_{it} \equiv \check{s}_{it}(\gamma_1^{0},\eta^{0}(\gamma_1^{0},\tau),\tau) = s_{it}(\gamma_1^{0},\eta^{0}(\gamma_1^{0},\tau),\tau)-\E s_{it}(\gamma_1^{0},\eta^{0}(\gamma_1^{0},\tau),\tau)$. By Theorem 2 in \citet{Jenish2012}, and under Assumption \ref{Ass_stationray}(1) and \ref{Ass6}, we have $G^{0}_{NT} \overset{d} {\rightarrow}N(0, \Omega_{0})$, where $\displaystyle \Omega_{0} = \tau(1- \tau) \lim_{N, T\to\infty} (NT)^{-1}\sum_{i,t}\E (\Psi_{it} \Psi_{it}^{\top}|\mathcal{C}) $. Hence, $G^{0}_{NT} = O_p(1)$. Then,
	\begin{equation}
		|\sqrt{NT}\hat\lambda(\tilde{\gamma}_1,\tau)|_2^2=|O_p(1)-\bar J_{\lambda}(\tau) J_{\gamma_1}(\tau)[1+o_p(1)]\sqrt{NT}(\tilde{\gamma}_1(\tau) - \gamma_1^{0}(\tau))|_2^2.\label{go21}
	\end{equation}
	It follows from (\ref{go20}) and (\ref{go21}) that $\sqrt{NT}(\hat\gamma_1(\tau)-\gamma_1^{0}(\tau))=O_p(1)$ by the full rank properties of $\bar J_{\lambda}(\tau) J_{\gamma_1}(\tau)$. Consequently, following Lemma B.1 in \citet{Chernozhukov2006}, and combing \eqref{eq11} and \eqref{go21},
	\begin{align}
		\sqrt{NT}(\hat\gamma_1(\tau)-\gamma_1^{0}(\tau))=&\underset{s\in \mathbb R}{\mbox{ arg min }} | - \bar J_{\lambda}(\tau) G^{0}_{NT} - \bar J_{\lambda}(\tau) J_{\gamma_1}(\tau) s|_2^2+o_p(1)\nonumber\\
		=&[J^{\top}_{\gamma_1}(\tau)\bar J^{\top}_{\lambda}(\tau) \bar J_{\lambda}(\tau) J_{\gamma_1}(\tau)]^{-1}[J^{\top}_{\gamma_1}(\tau) \bar J^{\top}_{\lambda}(\tau)  \bar J_{\lambda}(\tau)] G^{0}_{NT}+o_p(1).
	\end{align}
	Plugging this result into \eqref{equ12}, and after some algebra we can show that
	\begin{align}
		& \sqrt{NT}(\hat\eta(\hat\gamma_1(\tau),\tau)-\eta^{0}(\tau)) \\ \nonumber
		& =J^{-1}_{\eta} (\tau)\left[ I-J_{\gamma_1}(\tau)[J^{\top}_{\gamma_1}(\tau) \bar J^{\top}_{\lambda}(\tau)  \bar J_{\lambda}(\tau) J_{\gamma_1}(\tau)]^{-1}J^{\top}_{\gamma_1}(\tau) \bar J^{\top}_{\lambda}(\tau)  \bar J_{\lambda}(\tau)  \right]G^{0}_{NT}+o_p(1).
	\end{align}
	As $J_{\gamma_1}(\tau)\bar J_{\lambda}(\tau)$ is invertible, we have:
	\begin{align*}
		\sqrt{NT}(\hat\lambda(\hat\gamma_1(\tau),\tau)-\mathbf{0}) = & - \bar J_{\lambda} (\tau)\left[ I-J_{\gamma_1}(\tau)[J^{\top}_{\gamma_1}(\tau) \bar J^{\top}_{\lambda}(\tau)]^{-1}\bar J_{\lambda}(\tau)  \right]G^{0}_{NT}+o_p(1) \\
		= & \mathbf{0} \times O_p(1)  + o_p(1) .
	\end{align*}
	Using the fact that $(\tilde{\gamma}_1(\tau), \hat\eta^{\top}(\tilde{\gamma}_1(\tau),\tau)) = (\hat\gamma_1(\tau), \hat\eta^{\top}(\tau)) = (\hat\gamma_1(\tau), \hat\phi^{\top}(\tau), \mathbf{0} + o_p(\frac{1}{\sqrt{NT}}))$, as $\min(N,T) \rightarrow \infty$, we have:
	\begin{align}
		\sqrt{NT}\left\{\hat{\theta}(\tau) - \theta^{0}(\tau))\right\} = & - J^{-1}(\tau) G^{0}_{NT}(\theta^{0},\tau) + o_{p}(1).
	\end{align}

	Recall that $\displaystyle \Omega_{0} = \tau(1- \tau) \lim_{N, T\to\infty} (NT)^{-1}\sum_{i,t}\E (\Psi_{it} \Psi_{it}^{\top}|\mathcal{C}) $. Using the properties of NED process, $\{\check{s}_{it}\}_{i,t}$, 
	by Theorem 2 in \citet{Jenish2012}, under Assumption  \ref{Ass_stationray} (1) and \ref{Ass6}, and conditioning on $\mathcal{C}$, we have: $\Omega_{0}^{-1/2}G^{0}_{NT} \overset{d} {\rightarrow} N(\mathbf{0}, I)$. Thus,
	{\small
		\begin{align}
			&	\left(
			\begin{array}{c}
				\sqrt{NT}(\hat\gamma_1(\tau)-\gamma_1^{0}(\tau))\\
				\sqrt{NT}(\hat\phi(\tau)-\phi^{0}(\tau))
			\end{array}
			\right) \nonumber \\
			&	= 
			\left(
			\begin{array}{c}
				[J^{\top}_{\gamma_1}(\tau)\bar J^{\top}_{\lambda}(\tau) \bar J_{\lambda}(\tau) J_{\gamma_1}(\tau)]^{-1}[J^{\top}_{\gamma_1}(\tau) \bar J^{\top}_{\lambda}(\tau)  \bar J_{\lambda}(\tau)]\\
				\bar J_\phi (\tau)\left[ I-J_{\gamma_1}(\tau)[J^{\top}_{\gamma_1}(\tau) \bar J^{\top}_{\lambda}(\tau)  \bar J_{\lambda}(\tau) J_{\gamma_1}(\tau)]^{-1}J^{\top}_{\gamma_1}(\tau) \bar J^{\top}_{\lambda}(\tau)  \bar J_{\lambda}(\tau)  \right]
			\end{array}
			\right)G_{NT}^{0} + o_p(1).
	\end{align}}
	Then conditioning on $\mathcal{C}$, we have: $G_{NT}^{0} \overset{d} {\rightarrow} N(\mathbf{0}, \Omega_{0}).$  
	Recall that the conditional version of $J(\tau)$ is $J(\tau)^{*}$ while the unconditional version of $\Omega_0$ is defined by
	$\displaystyle \Omega_{0}^{*} = \tau(1- \tau) \lim_{N, T\to\infty} (NT)^{-1}\sum_{i,t}\E (\Psi_{it} \Psi_{it}^{\top}) $. As we assume that $\Omega_{0}^{-1}\Omega_{0}^{*} \to_p I $ and $J(\tau)^{-1}J(\tau)^{*} \to_p I $, the conclusion follows. $\blacksquare$

	\setcounter{lemma}{0}
	\renewcommand*{\thelemma}{B.\arabic{lemma}}
	
	\subsubsection{Lemmas \ref{lemmaB1} and \ref{lemmaB2}}
	For convenience we collect some important notations: $\pi(\tau) \equiv  (\gamma_1(\tau), \phi^{\top}(\tau) , \lambda^{\top}(\tau) )^{\top} = (\theta^{\top}(\tau) , \lambda^{\top}(\tau) )^{\top} = (\gamma_1(\tau) , \eta^{\top}(\tau) )^{\top} $, with $\theta(\tau) \equiv (\gamma_1(\tau),\phi^{\top}(\tau))^{\top} $ and $\eta(\tau)\equiv(\phi^{\top}(\tau),\lambda^{\top}(\tau))^{\top}$.  For simplicity, we denote  $\pi \equiv (\gamma_1, \phi^{\top}, \lambda^{\top})^{\top} = (\theta^{\top}, \lambda^{\top})^{\top} = (\gamma_1, \eta^{\top})^{\top}$ with $\theta \equiv (\gamma_1, \phi^{\top})^{\top}$ and $\eta\equiv(\phi^{\top},\lambda^{\top})^{\top}$, and the true parameter set, $\pi^{0} \equiv (\gamma_1^{0}, \phi^{0\top}, \lambda^{0\top})^{\top} = (\theta^{0\top}, \mathbf{0}^{\top})^{\top}$ with $\theta^{0} \equiv \theta^{0}(\tau) \equiv (\gamma_1^{0}(\tau), \phi^{0\top}(\tau))^{\top}$ and $\lambda^{0} = \mathbf{0}$. Recall that 
	\begin{align*}
		u_{it} = & Y_{it} - \gamma_1(\tau)\overline{Y}_{it} - \Psi^{\top}_{it} \eta(\gamma_1,\tau),\quad \Psi_{it} = (R^{\top}_{it}, X^{\top}_{it})^{\top},  \\
		u^*_{it} = & Y_{it} - \gamma_1^{0}(\tau)\overline{Y}_{it} - \Psi^{\top}_{it} \eta^{0}(\gamma_1^{0},\tau), \\
		s_{it}(\theta,\lambda,\tau) = &   s_{it}(\gamma_1, \eta(\gamma_1, \tau), \tau)=  \psi_{\tau}\left\{Y_{it} - \gamma_1(\tau)\overline{Y}_{it} - \Psi^{\top}_{it} \eta(\gamma_1,\tau) \right\} \Psi_{it}, \\
		\check{s}_{it}(\theta,\lambda,\tau) = &  \check{s}_{it}(\gamma_1, \eta(\gamma_1, \tau), \tau)=  s_{it}(\gamma_1,\eta(\gamma_1,\tau),\tau)-\E s_{it}(\gamma_1,\eta(\gamma_1,\tau),\tau), \\
		G_{NT} = & \frac{1}{\sqrt{NT}}\sum_{i=1}^N\sum_{t=1}^T \check{s}_{it}(\gamma_1,\eta(\gamma_1,\tau),\tau)\\ \nonumber
		= & \frac{1}{\sqrt{NT}}\sum_{i=1}^N\sum_{t=1}^T[ s_{it}(\gamma_1,\eta(\gamma_1,\tau),\tau)-\E s_{it}(\gamma_1,\eta(\gamma_1,\tau),\tau)] , \\
		G^{0}_{NT} = & \frac{1}{\sqrt{NT}}\sum_{i=1}^N\sum_{t=1}^T \check{s}_{it}(\gamma_1^{0},\eta^{0}(\gamma_1^{0},\tau),\tau) \\ \nonumber
		=  & \frac{1}{\sqrt{NT}}\sum_{i=1}^N\sum_{t=1}^T[ s_{it}(\gamma_1^{0},\eta^{0}(\gamma_1^{0},\tau),\tau)-\E s_{it}(\gamma_1^{0},\eta^{0}(\gamma_1^{0},\tau),\tau)].
	\end{align*}
	
	We need to prove that $\sup_{\tau \in \mathcal{T}}|G_{NT}-G^{0}_{NT}|_{\infty}=o_p(1)$ for any $\sup_{\tau \in \mathcal{T}}|\hat\pi(\tau)-\pi^{0}(\tau)|_{\infty}=o_p(1)$. For any estimator, $\hat\pi(\tau) = (\hat\theta(\tau), \hat\lambda(\tau)) = (\hat\gamma_1(\tau), \hat\phi(\tau) , \hat\lambda(\tau) )$, satisfying
	$|\hat \theta(\tau) - \theta^{0}(\tau)|_a \leq \delta_{1}$ and $|\hat\lambda(\tau) - 0 |_a \leq \delta_{2}$ with a constant vector $\delta = (\delta_1^{\top},\delta_2^{\top})^{\top}$, we define: 
	\begin{equation} \label{a30}
		\tpsi_{it}(\tau,\delta)  =  \check{s}_{it}(\hat\theta , \hat\lambda, \tau) - \check{s}_{it}(\theta^{0} , 0, \tau).
	\end{equation}

	\begin{lemma} \label{lemmaB1}
		\begin{equation}
			\sup_{\tau \in B_{\tau}  }\sup_{ |\delta_1|_1 \leq c_1 /\sqrt{NT}} \sup_{|\delta_2|_1\leq c_2 /\sqrt{NT}} \bigg|\dfrac{1}{\sqrt{NT}}\sum_{i=1}^{N} \sum^T_{t=1} \tpsi_{it}(\tau,\delta) \bigg|_{\infty} = O_p(1).
		\end{equation}
		where $c_1$ and $c_2$ are constants, and $B_{\tau} \in (0,1)$ is a compact set.
	\end{lemma}

	\begin{proof}\;
		Denote $|.|_a$ the element-wise absolute value.
		Let $d= q+4+(p+1)m$ and define the function class with constants $m_1,m_2$:
		\begin{eqnarray*}
			&&\mathcal{V}(m_1,m_2,B_{\tau} )\\
			&\defeq & \{(\theta, \tau, \lambda)\mapsto \{[\tau-\IF(y - \gamma_1 \overline{y} -x^{\top}\phi-R^{\top}\lambda\leq 0)]\Psi \\
			&&- [\tau-\IF(y - \gamma_1^0 \overline{y} -x^{\top}\phi^0-R^{\top}\lambda^0\leq 0)]\Psi \}, |\theta-\theta^0|_1 \leq m_1, |\lambda -\lambda^0|_1\leq m_2,\tau \in B_{\tau} \}.
		\end{eqnarray*}
		$\mathcal{V}$ is a VC subgraph with index $v \geq d+2$, see Lemma 9.12 i) in \cite{kosorok2007introduction} and Lemma A.7 in \cite{andrews1993empirical}. $\mathcal{V}$ has the envelop function $V(\cdot)$.  With probability measure $Q$ and $L_2$ norm, we have: $\|V\|_{Q,2} = (\int |V |_2^2dQ)^{1/2}$. Then, by Theorem 9.3 of \cite{kosorok2007introduction}, we assume that covering numbers of VC-classes of functions are given by $\mathcal{N}(\vps\|V\|_{Q,2}, \mathcal{V}, L_{2}(Q)) \lesssim (\vps)^{-(v-1)}$ .
		
		Let $T_{NT}(f) = G_{NT}(f)-G^{0}_{NT}(f) = (\sqrt{NT})^{-1}\sum_{i}\sum_t\tpsi_{it}(\tau,\delta)$, and\\ $J_{NT}(f) = (NT)^{-1} \sum_{i}\sum_{t} (s_{it}(\theta^0 + \delta_{1},\lambda + \delta_{2} ,\tau)-s_{it}(\theta^0,0,\tau))$. Define $\widetilde{\mathcal V} = \mathcal{V}(\delta_1 \sqrt{NT}^{-1},\delta_2 \sqrt{NT}^{-1},B_{\tau} )$. Then, the rate of the cover of the envelope for $\widetilde{\mathcal V} $ is   $\|\tilde{V}\|_2\lesssim \{NT\}^{-1/4}$. 
		Further, define $A_{\vps}$ as the $\vps\|\tilde{V}\|_{Q,2}$ cover of the functional class $\mathcal{V}$, where for each $f$ in $v$ we define the the closest element to it as $\pi(f)$ and $|\pi(f)- f|_{Q,2} \leq \vps$. Then, it is not straightforward to show that $|\mathcal{A}_{\vps}|\lesssim (\vps)^{-(v-1)}$. 
		For our choice of $\vps$ we assume that $\mathcal{N}(\vps\|V\|_{Q,2}, \mathcal{V}, L_{1,n}(Q)) \lesssim_p \mathcal{N}(\vps\|V\|_{Q,2}, \mathcal{V}, L_{2}(Q) )$. 
		The one step chaining gives:
		\begin{align}\label{decomp}
			&\sup_{\tau \in B_{\tau}  }\sup_{|\delta_1|_1\leq c_1 1/\sqrt{NT}} \sup_{|\delta_2|_1\leq c_2 1/\sqrt{NT}} |(\sqrt{NT})^{-1}\sum_{i}\sum_t\tpsi_{it}(\tau,\delta) |_{\infty}\\
			&\leq \sup_{f\in\widetilde{\mathcal{V}}}|T_{NT}(f)|_{\infty}\notag\\
			&= \sqrt{NT}\sup_{f\in\widetilde{\mathcal V}} \big|\left[J_{NT}(f)-J_{NT}\{\pi(f)\}-\E J_{NT} (f) +\E J_{NT}\{\pi(f)\}\right] \\
			& \quad \quad + [J_{NT}\{\pi(f)\}-\E J_{NT}\{\pi(f)\}]\big|_{\infty}\notag\\
			&\lesssim_p 2\sqrt{NT}\vps (NT)^{-1/4}+ \sqrt{NT}\max_{f\in A_{\vps}}|J_{NT}(f)- \E J_{NT}(f)|_{\infty}\notag \\
			&=2(NT)^{1/4}\vps  + K_{NT},
		\end{align}
		where {$(NT)^{-1/4}$ corresponds to the rate of the envelope.}

		Here, $K_{NT}$ involves partial sums, that are handled via the NED property and the continuity of the function with respect to the parameter (see Lemma \ref{lemmaB2}). By Lemma \ref{lemmaB2}, we have the following rate:
		\begin{equation*}
			\P(\sup_{\tau, \theta,\lambda  \in \Theta_{\vps}}| \frac{1}{\sqrt{NT}} \sum_i \sum_t \tpsi_{it}(\tau,\delta)|_{\infty} \geq c)	
			\leq \E \left \{\sup_{\tau, \theta,\lambda \in \Theta_{\vps}}| \frac{1}{\sqrt{NT}} \sum_i \sum_t \tpsi_{it}(\tau,\delta)|_{\infty} \right\}^2/ c^2 
			\leq (NT)^{-1/4}/c^2  
		\end{equation*}
		where $\Theta_{\vps}$ is the discretized function set $A_{\vps}$. 
		For example, We can pick $(NT)^{-1/8}/c= o(1)$ such that $K_{NT} = o(1)$. Also $\vps/ (NT)^{-1/4} = o(1)$.
	\end{proof}

	\begin{lemma}\label{lemmaB2} 
		Define $\eta_{it}(\tau, \delta) \defeq \sup_{\tau, \theta,\lambda \in \Theta_{\vps}} \{ \check{s}_{it}(\theta , \lambda, \tau) - \check{s}_{it}(\theta^{0} , 0, \tau) \}$, where $\Theta_{\vps}$ is the discretized function set $A_{\vps}$. For each $\tau$ and $\| \delta \|_2 \leq M < \infty$, if $c_{w} > (1 + \frac{1}{1-q})d$, then
		\begin{align*}
			\Var\left [ \frac{1}{\sqrt{NT}} \displaystyle \sum_{i = 1}^{N}\sum_{t = 1}^{T} \eta_{it}(\tau, \delta) \right ]
			&  = O(\{NT\}^{-\frac{1}{4}}).
		\end{align*}
	\end{lemma}

	
	\begin{proof} \;
		
		For simplicity, denote $ \sup_{\theta,\lambda,\tau \in {\Theta}_{\vps}}(\psi_{\tau}( u_{it}(\gamma_1, \phi,\lambda, \tau))- \tilde{\psi}_{\tau}( u^*_{it})) \Psi_{it}$ as $\eta_{it}$. Then, it is easily seen that $\eta_{it}$ is NED.
		For any $i \in 1, \cdots, N$,  $t \in 1, \cdots, T$ and any $s > 0$, define $R_{it}^{s} = \E(\eta_{it}|\mathscr{F}_{it}(s))$ and $T_{it}^{s} =  \eta_{it} - R_{it}^{s}$. 
		By the Jenson and Lyapunov inequalities, we have for all $i \in 1, \cdots, N$, $t \in 1, \cdots, T$ and any $1\leq q \leq 2+\delta$,
		\begin{equation*}
			\E |R_{it}^{s}|_1^{q} = \E\{ |\E( \eta_{it}|\mathscr{F}_{it}(s))|_1^{q} \} \leq \E\{ \E(| \eta_{it}|_1^{q}|\mathscr{F}_{it}(s)) \} =  \E | \eta_{it}|_1^{q}.
		\end{equation*}
		Hence, 
		\begin{align*}
			\|  R_{it}^{s} \|_{q} & \leq \|  \eta_{it}\|_{q} \leq \|  \eta_{it} \|_{2+\delta} \\
			\|  T_{it}^{s} \|_{q} & \leq \| \eta_{it}\|_{q} + \| R_{it}^{s} \|_{q} \leq 2\| \eta_{it}\|_{q} \leq 2 \|  \eta_{it}\|_{2+\delta}.
		\end{align*}
		Therefore, both $R_{it}^{s}$ and $T_{it}^{s}$ are uniformly $L_{2+\delta}$ bounded. As $\{\tilde{\psi}_{\tau}\}_{i,t}$ is uniformly $L_{2}$-NED on $\{ U_{it} \}_{i,t}$ and the NED-scaling factors can be chosen as one, we obtain:
		\begin{equation*}
			\displaystyle \sup_{i,t} \|  T_{it}^{s} \|_{2} \leq \varphi(s),
		\end{equation*}
		Further, let $\sigma(R_{it}^{s})$ denote the $\sigma-$field generated by $R_{it}^{s}$. Since $\sigma(R_{it}^{s}) \subseteq \mathscr{F}_{it}(s)$, the mixing coefficients of $R_{it}^{s}$ satisfy:
		\begin{equation*}
			\overline{\alpha}_{R}(1, 1, h) \leq \left\{\begin{array}{cc}
				1,                                          & h \leq 2s, \\
				\overline{\alpha}(Ms^{d}, Ms^{d}, h - 2s),  & h > 2s,
			\end{array}\right.
		\end{equation*}
		where $\overline{\alpha}(u, v, h)$ are the mixing coefficients of the input process $\{  U_{it} \} $, because the $s-$neighborhood of any point on $D$ contains at most $Ms^{2}$ points of $D$ for some $M$ that does not depend on $s$, see Lemma A.1 of \citet{Jenish2009}.
		
		We decompose $\eta_{it}$ and $\eta_{i',t'}$ as follows:
		\begin{equation*}
			\eta_{it} = R_{it}^{[s/3]} + T_{it}^{[s/3]}, \quad \eta_{i',t'} = R_{i',t'}^{[s/3]} + T_{i',t'}^{[s/3]}.
		\end{equation*}
		where $s = \rho((i,t), (i',t'))$. Then,
		\begin{align*}
			&|\Cov(\eta_{it}, \eta_{i',t'})| =  |\Cov(R_{it}^{[s/3]} + T_{it}^{[s/3]}, R_{i',t'}^{[s/3]} + T_{i',t'}^{[s/3]})| \\
			& \leq |\Cov(R_{it}^{[s/3]}, R_{i',t'}^{[s/3]})| + |\Cov(R_{it}^{[s/3]}, T_{i',t'}^{[s/3]})| + |\Cov(T_{it}^{[s/3]}, R_{i',t'}^{[s/3]})| + |\Cov(T_{it}^{[s/3]}, T_{i',t'}^{[s/3]})|.
		\end{align*}
		
		We then bound each term in the last inequality.
		First, using Lemma \ref{Lem6} with $p = 2+\delta, q = 2$, and $h = 2(2+\delta)/\delta$, we obtain the following bound on the first term:
		\begin{align*}
			|\Cov(R_{it}^{[s/3]}, R_{i',t'}^{[s/3]})| & \leq 4 \|R_{it}^{[s/3]}\|_{2+\delta} \|R_{i',t'}^{[s/3]}\|_{2} \overline{\alpha}_{R}^{\delta/(4+2\delta)}(1, 1, [s/3])  \\
			& \leq 4 \| \eta_{it} \|_{2+\delta} \| \eta_{it} \|_{2} \overline{\alpha}^{\delta/(4+2\delta)}(M[s/3]^{d}, M[s/3]^{d}, s - 2[s/3]) \\
			& \leq C_{1} \| \eta_{it} \|_{2+\delta} \| \eta_{it} \|_{2} [s/3]^{d\varsigma_{0}} \widehat{\alpha}^{\delta/(4 + 2\delta)}([s/3]),
		\end{align*}
		where $\varsigma_{0} = \delta \varsigma/(4+2\delta)$.
		By the Cauchy-Schwartz inequality, the second and third terms are bounded by
		\begin{equation*}
			|\Cov(R_{it}^{[s/3]}, T_{i',t'}^{[s/3]})| \leq 4 \|R_{it}^{[s/3]}\|_{2} \|T_{i',t'}^{[s/3]}\|_{2} \leq 4 \| \eta_{it} \|_{2} \varphi([s/3]).
		\end{equation*}
		Similarly, the fourth term is bounded by
		\begin{equation*}
			|\Cov(T_{it}^{[s/3]}, T_{i',t'}^{[s/3]})| \leq 4 \|T_{it}^{[s/3]}\|_{2} \|T_{i',t'}^{[s/3]}\|_{2} \leq 8 \| \eta_{it} \|_{2} \varphi([s/3]).
		\end{equation*}
		Collecting these results, we have:
		\begin{equation} \label{appcov}
			|\Cov(R_{it}^{[s/3]}, R_{i',t'}^{[s/3]})|  \leq \| \eta_{it} \|_{2} \left \{ C_{1} \| \eta_{it} \|_{2+\delta} [s/3]^{d\tau_{0}} \widehat{\alpha}^{\delta/(4 + 2\delta)}([s/3]) +  C_{2}\varphi([s/3])  \right \}.
		\end{equation}
		
		Using the inequality in (\ref{appcov}), and  the definition of random fields, we have:
		\begin{align*}
			&\Var\left [ \frac{1}{\sqrt{NT}} \displaystyle \sum_{i = 1}^{N}\sum_{t = 1}^{T} \eta_{it} \right ]\\
			\leq &  \frac{1}{NT}  \left \{ \displaystyle \sum_{i = 1}^{N}\sum_{t = 1}^{T} \Var(\eta_{it}) + \displaystyle \sum_{i = 1}^{N}\sum_{t = 1}^{T} \displaystyle \sum_{(i',t') \neq (i, t)} |\Cov(R_{it}^{[s/3]}, R_{i',t'}^{[s/3]})| \right \}\\
			\leq &  4 \| \eta_{it} \|_{2}^{2} + C_{1} \frac{1}{NT} \displaystyle \sum_{i = 1}^{N}\sum_{t = 1}^{T} \displaystyle \sum_{(i',t') \neq (i, t)} \| \eta_{it} \|_{2} \| \eta_{it} \|_{2+\delta} [s/3]^{d\varsigma_{0}} \widehat{\alpha}^{\delta/(4 + 2\delta)}([s/3]) \\
			&  + C_{2} \frac{1}{NT} \displaystyle \sum_{i = 1}^{N}\sum_{t = 1}^{T} \displaystyle \sum_{(i',t') \neq (i, t)} \| \eta_{it} \|_{2}\varphi([s/3]) \\
			\leq&   4 \| \eta_{it} \|_{2}^{2} + C_{1} \frac{1}{NT} \displaystyle \sum_{i = 1}^{N}\sum_{t = 1}^{T} \displaystyle \sum_{h = 1}^{\infty} \sum_{(i',t'): h\leq \rho((i,t), (i',t'))/3 < h + 1} \\
			&\| \eta_{it} \|_{2} \| \eta_{it} \|_{2+\delta} [\rho((i,t), (i',t'))/3]^{d\varsigma_{0}} \widehat{\alpha}^{\delta/(4 + 2\delta)}([\rho((i,t), (i',t'))/3]) \\
			& +  C_{2} \frac{1}{NT} \displaystyle \sum_{i = 1}^{N}\sum_{t = 1}^{T} \displaystyle \sum_{h = 1}^{\infty} \sum_{(i',t'): h\leq \rho((i,t), (i',t'))/3 < h + 1} \| \eta_{it} \|_{2}\varphi([\rho((i,t), (i',t'))/3]) \\
			\leq &  4 \| \eta_{it} \|_{2}^{2} + C_{3} \| \eta_{it} \|_{2} \left\{ \displaystyle \sum_{h = 1}^{\infty}h^{d(\varsigma_{0} + 1) -1}\widehat{\alpha}^{\delta/(4 + 2\delta)}(h) + \displaystyle \sum_{h = 1}^{\infty} h^{d-1} \varphi(h) \right\},
		\end{align*}
		where the second inequality is obtained by substituting \eqref{appcov}, the third inequality by using the properties of random field, and the last inequality by Lemma \ref{Lem2} and the $L_{2+\delta}$-bound property of $\{\eta_{it}\}$.
		
		We discuss $\displaystyle \sum_{h = 1}^{\infty} h^{d-1}\varphi(h)$ under the two cases of the NED coefficients of $\{ \eta_{it} \}_{i=1}^{n}$: (1) Under Assumptions \ref{Ass1}-\ref{Ass3}(1) and  \ref{Ass_stationray}(1), the NED coefficients become $\varphi(s) = s^{-(1-q)(c_{w} - d)}$. If $c_{w} > (1 + \frac{1}{1-q})d$, then $\displaystyle \sum_{h = 1}^{\infty} h^{d-1}\varphi(h) = \displaystyle \sum_{h = 1}^{\infty} h^{d - (1-q)(c_{w} - d)-1} < \infty$. (2) Under Assumptions \ref{Ass1}-\ref{Ass3}(2) and  \ref{Ass_stationray}(1), we have: $\varphi(s) = \Upsilon^{(1-q)s/\bar{\rho}_{0}}$. Then, $\displaystyle \sum_{h = 1}^{\infty} h^{d-1}\varphi(h) = \displaystyle \sum_{h = 1}^{\infty} h^{d-1} \Upsilon^{(1-q)h/\bar{\rho}_{0}} < \infty$, due to $\Upsilon < 1$. Therefore, $\displaystyle \sum_{h = 1}^{\infty} h^{d-1}\varphi(h)< \infty$.
		
		Further, under assumption  \ref{Ass_stationray}(1), $\displaystyle \sum_{h = 1}^{\infty}h^{d(\varsigma_{0} + 1) -1}\widehat{\alpha}^{\delta/(4 + 2\delta)}(h) < \infty$. Combining this with $L_{2+\delta}$-bound of $\{\eta_{it}\}$, we obtain: for some $C < \infty$,
		\begin{equation*}
			\Var\left [ \frac{1}{\sqrt{NT}} \displaystyle \sum_{i = 1}^{N}\sum_{t = 1}^{T} \eta_{it}\right ]  \leq C \max_{i,t} \| \eta_{it} \|_{2}.
		\end{equation*}
		
		
		Next, we analyze $\| \eta_{it} \|_{2}$. Note that
		\begin{align*}
			\left| \E\left( \eta_{it}^{\top} \eta_{it}\right) \right|  \leq 
			& 2\left| \E \left\{ \left [ \II({-(NT)}^{-\frac{1}{2}}\delta^{\top}|\xi_{it}|_a\leq u_{it}^* < {(NT)}^{-\frac{1}{2}}\delta^{\top}|\xi_{it}|_a) \right ] \{\Psi_{it}^{\top} \Psi_{it}\vee\E|\Psi_{it}|_a^{\top}\E |\Psi_{it}|_a\} \right\} \right|\\
			\leq & 2\left|\E \displaystyle \int_{-{NT}^{-\frac{1}{2}}\delta^{\top}|\xi_{it}|_a}^{{NT}^{-\frac{1}{2}}\delta^{\top}|\xi_{it}|_a} \{\Psi_{it}^{\top} \Psi_{it}\vee\E|\Psi_{it}|_a^{\top}\E |\Psi_{it}|_a\} f(u)du\right| \\
			= &4\left| \E\{NT\}^{-\frac{1}{2}}\delta^{\top}|\xi_{it}|_a f(u) \{\Psi_{it}^{\top} \Psi_{it}\vee\E|\Psi_{it}|_a^{\top}\E |\Psi_{it}|_a\} \right|,
		\end{align*}
		where {{$\xi_{it} = (\overline{Y}_{it},X_{it}^{\top})^{\top}$}},   $u\in (0, \{NT\}^{-\frac{1}{2}}\delta^{\top}|\xi_{it}|_a)$ and $f(u)\leq \pi_5$ (by Assumption \ref{assumption_estimation}(2)) is the density function of $u^*_{it}$ conditioning on $\mathcal{C}$ and $\xi_{it}$. Therefore,
		\begin{align*} 
			\| \eta_{it} \|_{2}
			= & \left[\E\left( \eta_{it}^{\top} \eta_{it}\right) \right]^{1/2} \leq \left| \{NT\}^{-\frac{1}{2}}\E \left\{ \delta^{\top}\xi_{it} f(u) \{\Psi_{it}^{\top} \Psi_{it} \vee\E|\Psi_{it}|_a^{\top}\E |\Psi_{it}|_a\} \right\} \right| ^{1/2} 
			\\ \leq & \pi_{5}\{NT\}^{-\frac{1}{4}} \E \left \{ \left|\delta^{\top}\xi_{it}\{\Psi_{it}^{\top} \Psi_{it}\vee\E|\Psi_{it}|_a^{\top}\E |\Psi_{it}|_a\} \right| \right\}^{1/2}.
		\end{align*}
		By Assumption  \ref{Ass_stationray}(1), the last term  $\E \left \{ \left|\delta^{\top}\xi_{it}\{\Psi_{it}^{\top} \Psi_{it}\vee\E|\Psi_{it}|_a^{\top}\E |\Psi_{it}|_a\} \right| \right\}^{1/2}$ is bounded. 
		Hence, we obtain: for some $C < \infty$,
		\begin{align*}
			\Var\left [ \frac{1}{\sqrt{NT}} \displaystyle \sum_{i = 1}^{N}\sum_{t = 1}^{T} \eta_{it} \right ]
			& \leq C\mbox{max}_{i,t} \| \eta_{it} \|_{2} = O_{p}(\{NT\}^{-\frac{1}{4}}).
		\end{align*}
		
	\end{proof}

	\newpage
	
	\section{Additional Monte Carlo Simulations}\label{appendixb}
	\setcounter{table}{0}
	\renewcommand{\thetable}{B\arabic{table}} 
	\label{simulappd}
	
	\subsection{The Simulation Results under Alternative Networks} \label{appendixb_b1}
	To check the robustness of the finite sample performance of the IVQR estimator, we consider the two alternative network matrices, which we rewrite here for convenience:
	
	\textsc{Type 2.} (Stochastic Block Model) We first consider the Stochastic Block Model 
	with an important application in community detection by \cite{zhao2012consistency}. We
	follow \cite{nowicki2001estimation} and randomly assign each node a block label index from 1 to $L$, where $L\in \{5,10,20\}$. We then set $%
	\P(a_{ij}=1)=0.3N^{-0.3}$ if $i$ and $j$ are in the same block, and $\P(a_{ij}=1)=0.3N^{-1}$ otherwise. Thus, the nodes within the same block have higher probability of connecting with each other than the nodes between blocks.

	\textsc{Type 3.} (Power-law Distribution Network) In practice, the majority of nodes in the network have a small number of links while a small
	number of nodes have a large number of links, see  \cite{barabasi1999emergence}. In this case the degrees of nodes can be characterized by the power-law distribution. We simulate the adjacency matrix as follows: For each node, we generate the in-degree, $d_{i}=\sum_{j}a_{ji}$ according to the discrete power-law distribution such as $\P(d_{i}=\check{k})=c\check{k}^{-\beta }$, where $c$ is a
	normalizing constant and the exponent parameter $\beta $ is set at $2.5$ as in \cite{clauset2009power}. Finally, for the $i$-th node, we randomly select $d_{i}$ nodes as its followers.
	
	Under the same DGP setup specified in Sectioon \ref{mcsetup}, the simulation results for RMSEs and coverage probabilities, are presented in Tables \ref{tab_simu_RMSE_W2}-\ref{tab_simu_RMSE_W3}. Similar to those for \textsc{Type 1} network in Table \ref{tab_simu_RMSE_W1}, we find that the RMSEs of all the parameters decrease as the size of nodes or the time periods increases across the different quantiles ($\tau =0.1,0.5,0.9$) and the different distributions of $\mathfrak{u}_{it}$ under both \textsc{Type 2} (Table \ref{tab_simu_RMSE_W2}) and \textsc{Type 3} (Table \ref{tab_simu_RMSE_W3}) networks, which is also in line with the asymptotic theory. RMSEs of $\gamma_{1}$ are slightly larger compared with other parameters, especially in a small sample ($N=100$), which may reflect uncertainty associated with the selection of the IV variables. 
	The biases of the IVQR estimators are mostly negligible across the different quantiles, for the different distributions of $\mathfrak{u}_{it}$ and for all the sample pairs of $(N, T)$, see Tables \ref{tab_simu_Bias_W1}--\ref{tab_simu_Bias_W3}.
	
	The results for coverage probabilities, reported in Tables \ref{tab_simu_CP_W2}-\ref{tab_simu_CP_W3}, are also similar to those for \textsc{Type 1} network presented in Table \ref{tab_simu_CP_W1}. In sum, That is, coverage probabilities are mostly close to the nominal 95\% level.
	
	\subsection{The Simulation Results for the Ordinary QR Estimator}\label{appendixb_b2}
	
	We report the additional simulation results by applying the ordinary QR estimator instead of the IVQR estimator in Tables \ref{tab_simu_RMSE_W1_NonIV}--\ref{tab_simu_Bias_W3_NonIV}. We observe that RMSEs are relatively larger for all the sample sizes, and they decrease very slightly even as the sample size increases, especially for the endogenous coefficient, $\hat{\gamma}_{1}$. The coverage probability is far below the nominal 95\% level across all quantiles, $\tau = 0.1,0.5,0.9$ under the three different network structures and for the all sample sizes. The biases are large especially for $\gamma_{1}$, which remain substantial as the sample size increases. This clearly highlights the importance of explicitly dealing with simultaneous network endogeneity by the IVQR estimator.
	
	\begin{table}[!htbp]
		\begin{center}
			\caption{RMSE ($\times 100$) for  \textsc{Type 2} Network}
			\label{tab_simu_RMSE_W2}
			{\scriptsize 
				\begin{tabular}{ccc|ccccccccccccc}
					\hline
					\hline
					&Dist. & $\tau$ &$\gamma_0 $  & $\gamma_1$&  $\gamma_2$ &  $\gamma_3$ & $\alpha_1$ & $\alpha_2$ &$\alpha_3$ & $\alpha_4$ & $\alpha_5$ & $\beta_1$ & $\beta_2$ & $\beta_3$ & $\beta_4$  \\
					
					\hline
					&&& \multicolumn{13}{c}{$ N = 100 $} \\
					\multirow{6}*{$T=100$} &$N(0,1)$ &  0.1 & 1.74 & 5.14 & 1.76 & 2.99 & 1.78 & 1.75 & 1.79 & 1.82 & 1.60 & 1.37 & 1.51 & 1.38 & 1.34 \\ 
					&  &  0.5 & 1.43 & 4.49 & 1.55 & 2.70 & 1.29 & 1.56 & 1.39 & 1.34 & 1.26 & 1.05 & 1.16 & 0.99 & 0.97 \\ 
					&  &  0.9 & 1.72 & 4.97 & 1.80 & 3.01 & 1.82 & 1.75 & 1.71 & 1.72 & 1.60 & 1.31 & 1.46 & 1.23 & 1.26 \\ 
					&$t(5)$ &  0.1 & 1.98 & 4.94 & 1.59 & 2.86 & 2.10 & 2.21 & 2.14 & 2.05 & 1.87 & 1.62 & 1.74 & 1.61 & 1.59 \\ 
					&  &  0.5 & 1.47 & 3.87 & 1.22 & 2.15 & 1.28 & 1.62 & 1.39 & 1.39 & 1.26 & 1.08 & 1.18 & 1.04 & 1.05 \\ 
					&  &  0.9 & 2.00 & 4.85 & 1.56 & 2.88 & 2.04 & 2.10 & 2.03 & 2.05 & 1.84 & 1.51 & 1.70 & 1.49 & 1.50 \\ 
					\cline{2-16}
					\multirow{6}*{$T=200$} &$N(0,1)$ &  0.1 & 1.22 & 3.89 & 1.34 & 2.26 & 1.43 & 1.33 & 1.26 & 1.31 & 1.19 & 0.98 & 1.07 & 0.93 & 0.92 \\ 
					&  &  0.5 & 0.98 & 3.13 & 1.11 & 1.89 & 0.93 & 1.10 & 0.96 & 0.95 & 0.85 & 0.70 & 0.77 & 0.64 & 0.66 \\ 
					&  &  0.9 & 1.24 & 3.74 & 1.31 & 2.21 & 1.50 & 1.37 & 1.30 & 1.28 & 1.14 & 0.92 & 1.09 & 0.86 & 0.87 \\ 
					&$t(5)$ &  0.1 & 1.38 & 3.55 & 1.10 & 2.00 & 1.45 & 1.52 & 1.52 & 1.48 & 1.36 & 1.14 & 1.19 & 1.17 & 1.08 \\ 
					&  &  0.5 & 1.06 & 2.61 & 0.81 & 1.43 & 0.95 & 1.18 & 0.98 & 0.98 & 0.89 & 0.74 & 0.82 & 0.70 & 0.72 \\ 
					&  &  0.9 & 1.39 & 3.42 & 1.06 & 2.07 & 1.52 & 1.49 & 1.46 & 1.44 & 1.30 & 1.06 & 1.21 & 1.06 & 1.06 \\ 
					\cline{2-16}
					\multirow{6}*{$T=500$} &$N(0,1)$ &  0.1 & 1.00 & 3.30 & 1.09 & 1.92 & 1.26 & 1.11 & 1.09 & 1.06 & 1.01 & 0.79 & 0.91 & 0.73 & 0.74 \\ 
					&     &  0.5 & 0.80 & 2.57 & 0.88 & 1.58 & 0.75 & 0.95 & 0.79 & 0.79 & 0.73 & 0.55 & 0.62 & 0.53 & 0.53 \\ 
					&     &  0.9 & 1.04 & 3.28 & 1.17 & 1.89 & 1.44 & 1.16 & 1.15 & 1.09 & 1.02 & 0.72 & 0.86 & 0.64 & 0.72 \\ 
					&$t(5)$ &  0.1 & 1.14 & 2.97 & 1.01 & 1.64 & 1.37 & 1.23 & 1.22 & 1.25 & 1.10 & 0.87 & 0.98 & 0.84 & 0.90 \\ 
					&     &  0.5 & 0.83 & 2.06 & 0.73 & 1.13 & 0.74 & 1.04 & 0.80 & 0.76 & 0.68 & 0.55 & 0.59 & 0.52 & 0.53 \\ 
					&     &  0.9 & 1.21 & 2.89 & 0.99 & 1.77 & 1.55 & 1.32 & 1.24 & 1.27 & 1.15 & 0.86 & 1.06 & 0.80 & 0.85 \\ 
					\cline{2-16}
					&&& \multicolumn{13}{c}{$ N = 200 $} \\
					\multirow{6}*{$T=100$} &$N(0,1)$ &  0.1 & 1.16 & 3.78 & 1.27 & 2.20 & 1.30 & 1.24 & 1.27 & 1.23 & 1.13 & 1.01 & 1.02 & 0.91 & 0.99 \\ 
					&  &  0.5 & 1.00 & 3.19 & 1.10 & 1.94 & 0.90 & 1.13 & 0.93 & 0.93 & 0.85 & 0.77 & 0.83 & 0.71 & 0.70 \\ 
					&  &  0.9 & 1.21 & 3.65 & 1.24 & 2.20 & 1.31 & 1.25 & 1.28 & 1.22 & 1.09 & 0.94 & 1.05 & 0.86 & 0.89 \\ 
					&$t(5)$ &  0.1 & 1.40 & 3.46 & 1.18 & 1.96 & 1.53 & 1.49 & 1.49 & 1.48 & 1.31 & 1.17 & 1.25 & 1.11 & 1.14 \\ 
					&  &  0.5 & 1.08 & 2.41 & 0.85 & 1.36 & 0.88 & 1.10 & 0.90 & 0.92 & 0.83 & 0.70 & 0.74 & 0.69 & 0.70 \\ 
					&  &  0.9 & 1.39 & 3.29 & 1.08 & 2.00 & 1.63 & 1.48 & 1.40 & 1.39 & 1.29 & 1.06 & 1.22 & 1.04 & 1.08 \\ 
					\cline{2-16}
					\multirow{6}*{$T=200$} &$N(0,1)$ &  0.1 & 0.86 & 2.83 & 0.88 & 1.66 & 1.07 & 0.92 & 0.91 & 0.94 & 0.79 & 0.67 & 0.76 & 0.66 & 0.67 \\ 
					&  &  0.5 & 0.72 & 2.12 & 0.80 & 1.31 & 0.60 & 0.82 & 0.64 & 0.68 & 0.59 & 0.50 & 0.53 & 0.47 & 0.47 \\ 
					&  &  0.9 & 0.88 & 2.78 & 0.93 & 1.58 & 1.17 & 0.93 & 0.92 & 0.91 & 0.81 & 0.64 & 0.77 & 0.59 & 0.62 \\ 
					&$t(5)$ &  0.1 & 0.91 & 2.55 & 0.84 & 1.42 & 1.15 & 1.07 & 1.06 & 1.09 & 0.93 & 0.79 & 0.86 & 0.79 & 0.83 \\ 
					&  &  0.5 & 0.75 & 1.77 & 0.55 & 1.02 & 0.63 & 0.91 & 0.63 & 0.65 & 0.61 & 0.48 & 0.55 & 0.48 & 0.46 \\ 
					&  &  0.9 & 0.95 & 2.46 & 0.79 & 1.40 & 1.21 & 1.07 & 0.99 & 1.02 & 0.92 & 0.73 & 0.89 & 0.74 & 0.73 \\ 
					\cline{2-16}
					\multirow{6}*{$T=500$} &$N(0,1)$ &  0.1 & 0.67 & 2.31 & 0.73 & 1.29 & 0.97 & 0.89 & 0.82 & 0.83 & 0.72 & 0.55 & 0.60 & 0.54 & 0.54 \\ 
					&     &  0.5 & 0.54 & 1.87 & 0.67 & 1.16 & 0.55 & 0.72 & 0.53 & 0.51 & 0.48 & 0.43 & 0.40 & 0.39 & 0.38 \\ 
					&     &  0.9 & 0.76 & 2.47 & 0.73 & 1.29 & 1.15 & 0.83 & 0.72 & 0.83 & 0.71 & 0.48 & 0.59 & 0.51 & 0.45 \\ 
					&$t(5)$ &  0.1 & 0.76 & 2.43 & 0.65 & 1.26 & 0.86 & 0.86 & 0.88 & 0.88 & 0.79 & 0.68 & 0.81 & 0.66 & 0.63 \\ 
					&     &  0.5 & 0.60 & 1.72 & 0.50 & 0.95 & 0.56 & 0.76 & 0.59 & 0.63 & 0.55 & 0.45 & 0.46 & 0.38 & 0.44 \\ 
					&     &  0.9 & 0.79 & 2.35 & 0.69 & 1.26 & 0.93 & 0.88 & 0.86 & 0.84 & 0.79 & 0.67 & 0.69 & 0.56 & 0.61 \\ 
					\hline
					&&& \multicolumn{13}{c}{$ N = 500 $} \\
					\multirow{6}*{$T=100$} &$N(0,1)$ &  0.1 & 0.97 & 3.00 & 1.08 & 1.74 & 1.15 & 1.06 & 1.11 & 1.02 & 0.92 & 0.76 & 0.85 & 0.78 & 0.81 \\ 
					&  &  0.5 & 0.77 & 2.35 & 0.88 & 1.49 & 0.67 & 0.90 & 0.73 & 0.73 & 0.65 & 0.56 & 0.63 & 0.55 & 0.55 \\ 
					&  &  0.9 & 0.93 & 2.96 & 0.96 & 1.81 & 1.18 & 1.00 & 1.00 & 1.01 & 0.86 & 0.75 & 0.87 & 0.72 & 0.71 \\ 
					&$t(5)$ &  0.1 & 1.09 & 2.86 & 0.90 & 1.59 & 1.22 & 1.19 & 1.24 & 1.18 & 1.04 & 0.96 & 1.02 & 0.86 & 0.90 \\ 
					&  &  0.5 & 0.85 & 1.94 & 0.65 & 1.09 & 0.69 & 0.95 & 0.77 & 0.73 & 0.67 & 0.60 & 0.64 & 0.59 & 0.56 \\ 
					&  &  0.9 & 1.08 & 2.71 & 0.91 & 1.63 & 1.39 & 1.16 & 1.16 & 1.16 & 1.10 & 0.91 & 0.98 & 0.86 & 0.85 \\ 
					\cline{2-16}
					\multirow{6}*{$T=200$} &$N(0,1)$ &  0.1 & 0.68 & 2.17 & 0.76 & 1.31 & 0.99 & 0.76 & 0.76 & 0.74 & 0.61 & 0.60 & 0.60 & 0.56 & 0.51 \\ 
					&     &  0.5 & 0.58 & 1.73 & 0.67 & 1.06 & 0.46 & 0.75 & 0.50 & 0.53 & 0.49 & 0.40 & 0.43 & 0.34 & 0.37 \\ 
					&     &  0.9 & 0.78 & 2.34 & 0.78 & 1.30 & 1.11 & 0.81 & 0.72 & 0.72 & 0.63 & 0.50 & 0.59 & 0.47 & 0.51 \\ 
					&$t(5)$ &  0.1 & 0.84 & 2.19 & 0.68 & 1.28 & 0.95 & 0.88 & 0.86 & 0.93 & 0.84 & 0.68 & 0.70 & 0.65 & 0.62 \\ 
					&     &  0.5 & 0.59 & 1.30 & 0.46 & 0.73 & 0.52 & 0.72 & 0.51 & 0.51 & 0.50 & 0.38 & 0.43 & 0.41 & 0.38 \\ 
					&     &  0.9 & 0.87 & 2.04 & 0.63 & 1.23 & 1.20 & 0.89 & 0.84 & 0.91 & 0.81 & 0.63 & 0.70 & 0.62 & 0.60 \\ 
					\cline{2-16}
					\multirow{6}*{$T=500$} &$N(0,1)$ &  0.1 & 0.56 & 2.08 & 0.63 & 1.07 & 0.92 & 0.70 & 0.66 & 0.62 & 0.62 & 0.41 & 0.48 & 0.46 & 0.46 \\ 
					&  &  0.5 & 0.46 & 1.38 & 0.51 & 0.90 & 0.44 & 0.75 & 0.45 & 0.46 & 0.45 & 0.29 & 0.40 & 0.29 & 0.32 \\ 
					&  &  0.9 & 0.54 & 2.09 & 0.68 & 0.98 & 1.13 & 0.80 & 0.67 & 0.68 & 0.63 & 0.43 & 0.51 & 0.40 & 0.40 \\ 
					&$t(5)$ &  0.1 & 0.70 & 1.68 & 0.54 & 0.87 & 1.12 & 0.80 & 0.72 & 0.80 & 0.65 & 0.55 & 0.58 & 0.55 & 0.54 \\ 
					&  &  0.5 & 0.52 & 0.95 & 0.36 & 0.55 & 0.47 & 0.68 & 0.43 & 0.46 & 0.40 & 0.33 & 0.35 & 0.27 & 0.34 \\ 
					&  &  0.9 & 0.71 & 1.92 & 0.52 & 1.10 & 1.21 & 0.76 & 0.73 & 0.92 & 0.72 & 0.53 & 0.58 & 0.51 & 0.53 \\ 
					\hline
					\hline
				\end{tabular}
			}
		\end{center}
		{\footnotesize{Notes: The simulation results are based on the DGP in Section \ref{mcsetup} with 1000 replications and reported across the three different quantiles, $\tau = (0.1,0.5,0.9$) for the sample pairs, $(N,T) = 100, 200, 500$, where we generate $\mathfrak{u}_{it}$ from either a standard normal distribution, $N(0,1)$ or a $t$-distribution with 5 degrees of freedom, $t(5)$.}}
	\end{table}
	\normalsize

	\begin{table}[!htbp]
		\begin{center}
			\caption{RMSE ($\times 100$) for \textsc{Type 3} Network}
			\label{tab_simu_RMSE_W3}
			{\scriptsize 
\begin{tabular}{ccc|ccccccccccccc}
					\hline
					\hline
					&Dist. & $\tau$ &$\gamma_0 $  & $\gamma_1$&  $\gamma_2$ &  $\gamma_3$ & $\alpha_1$ & $\alpha_2$ &$\alpha_3$ & $\alpha_4$ & $\alpha_5$ & $\beta_1$ & $\beta_2$ & $\beta_3$ & $\beta_4$  \\
					
					\hline
					&&& \multicolumn{13}{c}{$ N = 100 $} \\
					\multirow{6}*{$T=100$} &$N(0,1)$ &  0.1 & 1.74 & 4.39 & 1.28 & 2.58 & 1.71 & 1.88 & 1.85 & 1.80 & 1.59 & 1.44 & 1.57 & 1.42 & 1.39 \\ 
					&  &  0.5 & 1.50 & 3.77 & 1.08 & 2.19 & 1.31 & 1.59 & 1.47 & 1.51 & 1.32 & 1.14 & 1.29 & 1.13 & 1.08 \\ 
					&  &  0.9 & 1.73 & 4.25 & 1.24 & 2.45 & 1.61 & 1.74 & 1.75 & 1.76 & 1.57 & 1.38 & 1.48 & 1.34 & 1.32 \\ 
					&$t(5)$ &  0.1 & 1.94 & 4.28 & 1.18 & 2.30 & 1.94 & 2.15 & 2.17 & 2.19 & 1.88 & 1.65 & 1.78 & 1.65 & 1.70 \\ 
					&  &  0.5 & 1.52 & 3.14 & 0.85 & 1.57 & 1.38 & 1.61 & 1.53 & 1.51 & 1.35 & 1.18 & 1.25 & 1.15 & 1.12 \\ 
					&  &  0.9 & 1.96 & 4.02 & 1.11 & 2.16 & 1.87 & 2.07 & 2.06 & 2.07 & 1.84 & 1.58 & 1.65 & 1.58 & 1.59 \\
\hline
					\hline
				\end{tabular}
			}
		\end{center}
		{\footnotesize{Notes: See the notes to Table \ref{tab_simu_RMSE_W2}.}}
	\end{table}
	\normalsize

	\begin{table}[!htbp]
		\begin{center}
			\caption{Coverage Probability ($\times 100$) for \textsc{Type 2} Network}
			\label{tab_simu_CP_W2}
			{\scriptsize 
				\begin{tabular}{ccc|ccccccccccccc}
					\hline
					\hline
					&Dist. & $\tau$ &$\gamma_0 $  & $\gamma_1$&  $\gamma_2$ &  $\gamma_3$ & $\alpha_1$ & $\alpha_2$ &$\alpha_3$ & $\alpha_4$ & $\alpha_5$ & $\beta_1$ & $\beta_2$ & $\beta_3$ & $\beta_4$  \\
					
					\hline
					&&& \multicolumn{13}{c}{$ N = 100 $} \\
					\multirow{6}*{$T=100$} &$N(0,1)$ &  0.1 & 94.6 &96.0 &95.0 &95.4 &95.2 &95.1 &95.2 &94.9 &96.0 &95.6 &95.8 &95.4 &95.6 \\ 
					&  &  0.5 & 93.2 & 93.1 & 93.6 & 92.5 & 96.2 & 94.6 & 96.1 & 94.0 & 94.8 & 95.4 & 95.6 & 96.1 & 96.2 \\ 
					&  & 0.9 &92.0 &94.4 &91.8 &94.7 &93.7 &94.6 &96.5 &95.0 &93.7 &96.3 &93.8 &96.0 &95.2 \\ 
					&$t(5)$ & 0.1 &92.8 &95.0 &94.5 &97.5 &96.2 &95.3 &94.2 &94.5 &94.8 &96.0 &96.2 &96.5 &95.1 \\ 
					&  & 0.5 &93.5 &92.7 &93.6 &92.8 &96.6 &94.2 &96.4 &96.3 &97.7 &96.0 &95.6 &96.0 &96.2 \\ 
					&  & 0.9 &93.7 &95.2 &92.5 &94.4 &94.6 &95.4 &96.4 &96.4 &95.3 &97.3 &95.1 &96.4 &96.3 \\
\hline
					\hline
				\end{tabular}
			}
		\end{center}
		{\footnotesize{Notes: See the notes to Table \ref{tab_simu_RMSE_W2}.}}
	\end{table}
	\normalsize

	\begin{table}[!htbp]
		\begin{center}
			\caption{Coverage Probability ($\times 100$) for \textsc{Type 3} Network}
			\label{tab_simu_CP_W3}
			{\scriptsize 
				\begin{tabular}{ccc|ccccccccccccc}
					\hline
					\hline
					&Dist. & $\tau$ &$\gamma_0 $  & $\gamma_1$&  $\gamma_2$ &  $\gamma_3$ & $\alpha_1$ & $\alpha_2$ &$\alpha_3$ & $\alpha_4$ & $\alpha_5$ & $\beta_1$ & $\beta_2$ & $\beta_3$ & $\beta_4$  \\
					\hline
					
					&&& \multicolumn{13}{c}{$ N = 100 $} \\
					\multirow{6}*{$T=100$} &$N(0,1)$ &  0.1 & 93.7 & 97.3 & 93.9 & 95.2 & 95.3 & 95.3 & 95.5 & 95.9 & 96.6 & 95.3 & 95.6 & 95.2 & 96.0 \\ 
					&  &  0.5 & 92.0 & 92.8 & 92.8 & 94.5 & 96.3 & 94.1 & 96.0 & 95.3 & 95.1 & 96.3 & 95.2 & 95.7 & 96.9 \\ 
					&  &  0.9 & 92.8 & 96.2 & 93.6 & 94.8 & 95.1 & 96.1 & 94.8 & 95.3 & 95.0 & 95.8 & 95.6 & 95.6 & 95.8 \\ 
					&$t(5)$ & 0.1 & 94.9 & 96.1 & 94.8 & 95.8 & 96.3 & 96.1 & 95.3 & 95.4 & 96.0 & 96.3 & 96.3 & 95.8 & 95.8 \\ 
					&  &  0.5 & 92.6 & 93.5 & 94.8 & 94.7 & 96.0 & 94.9 & 95.5 & 94.9 & 95.8 & 96.2 & 94.9 & 95.6 & 95.6 \\ 
					&  &  0.9 & 93.6 & 96.1 & 94.6 & 95.9 & 95.8 & 95.8 & 96.1 & 96.0 & 96.1 & 96.4 & 96.6 & 96.3 & 96.1 \\
\hline
					\hline
				\end{tabular}
			}
		\end{center}
		{\footnotesize{Notes: See the notes to Table \ref{tab_simu_RMSE_W2}.}}
	\end{table}
	\normalsize

	\begin{table}[!htbp]
		\begin{center}
			\caption{Bias ($\times 100$) for \textsc{Type 1} Network}
			\label{tab_simu_Bias_W1}
			{\scriptsize  
				\begin{tabular}{ccc|ccccccccccccc}
					\hline
					\hline
					&Dist. & $\tau$ &$\gamma_0 $  & $\gamma_1$&  $\gamma_2$ &  $\gamma_3$ & $\alpha_1$ & $\alpha_2$ &$\alpha_3$ & $\alpha_4$ & $\alpha_5$ & $\beta_1$ & $\beta_2$ & $\beta_3$ & $\beta_4$  \\
					
					\hline
					&& \multicolumn{13}{c}{$ N = 100 $} \\
					\multirow{6}*{$T=100$} &$N(0,1)$ & 0.1 & -0.01 & 0.07 & -0.11 & 0.12 & 0.13 & 0.09 & 0.01 & -0.05 & 0.08 & 0.15 & 0.10 & -0.06 & 0.01 \\ 
					&     &  0.5 & -0.07 & -0.05 & -0.03 & 0.10 & 0.11 & 0.07 & 0.02 & -0.01 & 0.10 & 0.07 & 0.15 & 0.02 & 0.02 \\ 
					&     &  0.9 & 0.03 & -0.06 & -0.08 & -0.15 & -0.05 & -0.06 & -0.10 & 0.01 & -0.07 & -0.04 & -0.11 & -0.04 & 0.01 \\ 
					&$t(5)$ &   0.1 & -0.13 & -0.08 & -0.02 & 0.07 & 0.07 & 0.03 & 0.02 & -0.07 & 0.06 & 0.03 & 0.19 & 0.01 & -0.05 \\ 
					&     &  0.5 & -0.03 & -0.05 & -0.01 & 0.14 & 0.13 & 0.04 & -0.01 & 0.02 & 0.06 & 0.04 & 0.14 & -0.02 & 0.07 \\ 
					&     &  0.9 & 0.08 & -0.09 & -0.07 & -0.13 & -0.10 & -0.14 & -0.07 & 0.01 & -0.10 & -0.14 & -0.09 & -0.02 & 0.00 \\
\hline
					\hline
				\end{tabular}
			}
		\end{center}
		{\footnotesize{Notes: See the notes to Table \ref{tab_simu_RMSE_W2}.}}
	\end{table}
	\normalsize

	\begin{table}[!htbp]
		\begin{center}
			\caption{Bias ($\times 100$) for \textsc{Type 2} Network}
			\label{tab_simu_Bias_W2}
			{\scriptsize  
				\begin{tabular}{ccc|ccccccccccccc}
					\hline
					\hline
					&Dist. & $\tau$ &$\gamma_0 $  & $\gamma_1$&  $\gamma_2$ &  $\gamma_3$ & $\alpha_1$ & $\alpha_2$ &$\alpha_3$ & $\alpha_4$ & $\alpha_5$ & $\beta_1$ & $\beta_2$ & $\beta_3$ & $\beta_4$  \\
					
					\hline
					&& \multicolumn{13}{c}{$ N = 100 $} \\
					\multirow{6}*{$T=100$} &$N(0,1)$ &  0.1 & -0.07 & 0.04 & -0.06 & 0.13 & 0.05 & 0.05 & 0.06 & 0.01 & 0.04 & 0.07 & 0.22 & 0.03 & 0.04 \\ 
					&     &  0.5 & -0.13 & -0.08 & -0.03 & 0.14 & 0.11 & 0.08 & 0.08 & -0.02 & 0.12 & 0.05 & 0.11 & 0.03 & 0.06 \\ 
					&     &  0.9 & 0.08 & -0.09 & 0.01 & -0.12 & -0.06 & -0.04 & -0.08 & 0.02 & -0.06 & -0.13 & -0.08 & 0.01 & -0.07 \\ 
					&$t(5)$ &   0.1 & -0.01 & -0.05 & -0.05 & 0.06 & 0.07 & 0.09 & 0.04 & 0.06 & 0.01 & 0.10 & 0.10 & 0.05 & 0.06 \\ 
					&     &  0.5 & -0.11 & -0.02 & -0.02 & 0.04 & 0.10 & 0.12 & 0.10 & 0.02 & 0.05 & 0.05 & 0.03 & 0.06 & 0.09 \\ 
					&     &  0.9 & 0.10 & -0.03 & -0.05 & -0.06 & -0.02 & -0.01 & -0.03 & 0.02 & -0.04 & -0.11 & -0.04 & -0.03 & -0.11 \\
\hline
					\hline
				\end{tabular}
			}
		\end{center}
		{\footnotesize{Notes: See the notes to Table \ref{tab_simu_RMSE_W2}.}}
	\end{table}
	\normalsize

	\begin{table}[!htbp]
		\begin{center}
			\caption{Bias ($\times 100$) for \textsc{Type 3} Network}
			\label{tab_simu_Bias_W3}
			{\scriptsize  
				\begin{tabular}{ccc|ccccccccccccc}
					\hline
					\hline
					&Dist. & $\tau$ &$\gamma_0 $  & $\gamma_1$&  $\gamma_2$ &  $\gamma_3$ & $\alpha_1$ & $\alpha_2$ &$\alpha_3$ & $\alpha_4$ & $\alpha_5$ & $\beta_1$ & $\beta_2$ & $\beta_3$ & $\beta_4$  \\
					
					\hline
					&& \multicolumn{13}{c}{$ N = 100 $} \\
					\multirow{6}*{$T=100$} &$N(0,1)$ &  0.1 & -0.02 & 0.03 & -0.08 & 0.05 & 0.14 & 0.00 & -0.01 & 0.03 & 0.03 & 0.01 & 0.08 & 0.05 & 0.02 \\ 
					&     &  0.5 & -0.07 & -0.01 & -0.05 & 0.05 & 0.07 & 0.01 & 0.00 & 0.01 & 0.07 & 0.03 & 0.05 & 0.00 & 0.01 \\ 
					&     &  0.9 & 0.04 & -0.02 & -0.08 & -0.08 & -0.04 & -0.14 & 0.03 & 0.00 & -0.04 & -0.02 & -0.05 & 0.02 & -0.04 \\ 
					&$t(5)$ &   0.1 & -0.06 & -0.02 & -0.08 & 0.04 & 0.02 & 0.05 & 0.08 & -0.04 & 0.00 & 0.06 & 0.11 & -0.05 & -0.02 \\ 
					&     &  0.5 & 0.00 & -0.08 & -0.02 & 0.12 & 0.08 & 0.02 & 0.03 & -0.02 & 0.07 & 0.10 & 0.09 & 0.01 & 0.01 \\ 
					&     &  0.9 & 0.01 & 0.02 & -0.09 & -0.07 & -0.06 & -0.04 & -0.09 & 0.07 & -0.05 & -0.03 & -0.04 & -0.04 & -0.05 \\
\hline
					\hline
				\end{tabular}
			}
		\end{center}
		{\footnotesize{Notes: See the notes to Table \ref{tab_simu_RMSE_W2}.}}
	\end{table}
	\normalsize

	\begin{table}[!htbp]
		\begin{center}
			\caption{RMSE ($\times 100$) for \textsc{Type 1} Network by Ordinary QR Estimator}
			\label{tab_simu_RMSE_W1_NonIV}
			{\scriptsize 
				\begin{tabular}{ccc|ccccccccccccc}
					\hline
					\hline
					&Dist. & $\tau$ &$\gamma_0 $  & $\gamma_1$&  $\gamma_2$ &  $\gamma_3$ & $\alpha_1$ & $\alpha_2$ &$\alpha_3$ & $\alpha_4$ & $\alpha_5$ & $\beta_1$ & $\beta_2$ & $\beta_3$ & $\beta_4$  \\
					
					\hline
					&& \multicolumn{13}{c}{$ N = 100 $} \\
					\multirow{6}*{$T=100$} &$N(0,1)$ &  0.1 & 1.55 & 6.57 & 1.45 & 3.52 & 1.78 & 1.88 & 1.77 & 1.85 & 1.63 & 1.41 & 1.62 & 1.37 & 1.36 \\ 
					&  &  0.5 & 1.41 & 6.04 & 1.28 & 3.18 & 1.40 & 1.61 & 1.48 & 1.53 & 1.33 & 1.13 & 1.33 & 1.07 & 1.09 \\ 
					&  &  0.9 & 1.55 & 6.10 & 1.45 & 3.61 & 1.65 & 1.75 & 1.74 & 1.74 & 1.52 & 1.32 & 1.64 & 1.25 & 1.26 \\ 
					&$t(5)$ &  0.1 & 1.88 & 5.86 & 1.31 & 2.87 & 1.96 & 2.16 & 2.16 & 2.22 & 1.94 & 1.63 & 1.75 & 1.60 & 1.64 \\ 
					&  &  0.5 & 1.49 & 4.55 & 0.98 & 2.39 & 1.38 & 1.68 & 1.59 & 1.53 & 1.41 & 1.12 & 1.24 & 1.10 & 1.10 \\ 
					&  &  0.9 & 1.84 & 5.30 & 1.29 & 3.40 & 1.95 & 2.08 & 2.03 & 2.04 & 1.88 & 1.54 & 1.78 & 1.48 & 1.47 \\
\hline
					\hline
				\end{tabular}
			}
		\end{center}
		{\footnotesize{Notes: The simulation results are based on the DGP in Section \ref{mcsetup} with 1000 replications and reported across the three different quantiles, $\tau = (0.1,0.5,0.9$) for the sample pairs, $(N,T) = 100, 200, 500$, where we generate $\mathfrak{u}_{it}$ from either a standard normal distribution, $N(0,1)$ or a $t$-distribution with 5 degrees of freedom, $t(5)$. Here we apply the ordinary QR estimator.}}
	\end{table}
	\normalsize

	\begin{table}[!htbp]
		\begin{center}
			\caption{RMSE ($\times 100$) for \textsc{Type 2} Network by Ordinary QR Estimator}
			\label{tab_simu_RMSE_W2_NonIV}
			{\scriptsize 
				\begin{tabular}{ccc|ccccccccccccc}
					\hline
					\hline
					&Dist. & $\tau$ &$\gamma_0 $  & $\gamma_1$&  $\gamma_2$ &  $\gamma_3$ & $\alpha_1$ & $\alpha_2$ &$\alpha_3$ & $\alpha_4$ & $\alpha_5$ & $\beta_1$ & $\beta_2$ & $\beta_3$ & $\beta_4$  \\
					
					\hline
					&& \multicolumn{13}{c}{$ N = 100 $} \\
					\multirow{6}*{$T=100$} &$N(0,1)$ &  0.1 & 1.96 & 14.60 & 3.69 & 7.41 & 2.04 & 1.80 & 1.85 & 1.83 & 1.62 & 1.49 & 2.03 & 1.25 & 1.24 \\ 
					&  &  0.5 & 1.43 & 13.52 & 3.43 & 7.28 & 1.40 & 1.87 & 1.49 & 1.42 & 1.35 & 1.26 & 1.98 & 0.91 & 0.90 \\ 
					&  &  0.9 & 1.58 & 13.25 & 3.47 & 7.92 & 2.13 & 1.87 & 1.79 & 1.78 & 1.65 & 1.58 & 2.51 & 1.13 & 1.17 \\ 
					&$t(5)$ &  0.1 & 1.88 & 12.23 & 3.04 & 5.72 & 2.22 & 2.19 & 2.16 & 2.08 & 1.93 & 1.62 & 1.98 & 1.50 & 1.50 \\ 
					&  &  0.5 & 1.45 & 9.78 & 2.40 & 5.07 & 1.36 & 1.86 & 1.47 & 1.46 & 1.33 & 1.14 & 1.60 & 0.98 & 0.99 \\ 
					&  &  0.9 & 1.83 & 11.02 & 2.78 & 6.69 & 2.36 & 2.22 & 2.05 & 2.04 & 1.85 & 1.71 & 2.41 & 1.40 & 1.42 \\ 
					\cline{2-16}
					\multirow{6}*{$T=200$} &$N(0,1)$ &  0.1 & 1.66 & 13.76 & 3.64 & 6.94 & 1.74 & 1.38 & 1.30 & 1.31 & 1.21 & 1.12 & 1.73 & 0.88 & 0.89 \\ 
					&  &  0.5 & 1.08 & 12.74 & 3.44 & 6.88 & 1.02 & 1.52 & 1.07 & 1.08 & 0.98 & 1.03 & 1.75 & 0.60 & 0.64 \\ 
					&  &  0.9 & 1.17 & 12.46 & 3.44 & 7.51 & 1.91 & 1.57 & 1.39 & 1.39 & 1.26 & 1.35 & 2.32 & 0.79 & 0.84 \\ 
					&$t(5)$ &  0.1 & 1.48 & 11.21 & 2.31 & 5.27 & 1.61 & 1.56 & 1.60 & 1.53 & 1.44 & 1.23 & 1.68 & 1.10 & 1.03 \\ 
					&  &  0.5 & 1.14 & 8.80 & 1.87 & 4.51 & 1.03 & 1.43 & 1.06 & 1.08 & 0.97 & 0.88 & 1.36 & 0.69 & 0.68 \\ 
					&  &  0.9 & 1.30 & 10.08 & 2.19 & 6.09 & 1.80 & 1.59 & 1.50 & 1.49 & 1.33 & 1.31 & 2.04 & 0.98 & 1.04 \\ 
					\cline{2-16}
					\multirow{6}*{$T=500$} &$N(0,1)$ &  0.1 & 1.49 & 14.61 & 3.69 & 7.63 & 1.62 & 1.16 & 1.18 & 1.16 & 1.07 & 1.04 & 1.69 & 0.69 & 0.70 \\ 
					&  &  0.5 & 0.97 & 13.48 & 3.52 & 7.51 & 0.91 & 1.49 & 0.91 & 0.93 & 0.88 & 0.99 & 1.76 & 0.50 & 0.50 \\ 
					&  &  0.9 & 1.02 & 13.25 & 3.54 & 8.18 & 1.95 & 1.37 & 1.23 & 1.23 & 1.15 & 1.31 & 2.36 & 0.60 & 0.71 \\ 
					&$t(5)$ &  0.1 & 1.40 & 13.03 & 3.37 & 5.97 & 1.63 & 1.27 & 1.29 & 1.31 & 1.17 & 1.04 & 1.60 & 0.79 & 0.87 \\ 
					&  &  0.5 & 0.98 & 10.35 & 2.71 & 5.32 & 0.84 & 1.48 & 0.94 & 0.84 & 0.81 & 0.82 & 1.42 & 0.52 & 0.51 \\ 
					&  &  0.9 & 1.11 & 11.58 & 3.09 & 6.98 & 2.13 & 1.55 & 1.31 & 1.35 & 1.27 & 1.30 & 2.28 & 0.76 & 0.83 \\ 
					\hline
					&& \multicolumn{13}{c}{$ N = 200 $} \\
					\multirow{6}*{$T=100$} &$N(0,1)$ &  0.1 & 1.37 & 13.52 & 3.12 & 6.92 & 1.49 & 1.28 & 1.29 & 1.23 & 1.16 & 1.21 & 1.76 & 0.83 & 0.90 \\ 
					&  &  0.5 & 0.99 & 12.48 & 2.93 & 6.81 & 0.94 & 1.47 & 1.01 & 1.04 & 0.94 & 1.01 & 1.74 & 0.67 & 0.66 \\ 
					&  &  0.9 & 1.14 & 12.26 & 2.93 & 7.50 & 1.67 & 1.39 & 1.36 & 1.26 & 1.14 & 1.25 & 2.21 & 0.80 & 0.85 \\ 
					&$t(5)$ &  0.1 & 1.40 & 12.62 & 3.04 & 5.87 & 1.69 & 1.50 & 1.52 & 1.52 & 1.34 & 1.23 & 1.72 & 1.04 & 1.08 \\ 
					&  &  0.5 & 1.04 & 9.91 & 2.45 & 5.15 & 0.94 & 1.47 & 1.02 & 1.02 & 0.91 & 0.90 & 1.47 & 0.67 & 0.68 \\ 
					&  &  0.9 & 1.25 & 11.17 & 2.81 & 6.81 & 2.13 & 1.62 & 1.47 & 1.43 & 1.36 & 1.37 & 2.30 & 0.97 & 1.01 \\ 
					\cline{2-16}
					\multirow{6}*{$T=200$} &$N(0,1)$ &  0.1 & 1.28 & 14.23 & 3.39 & 7.26 & 1.35 & 0.95 & 0.95 & 0.96 & 0.83 & 0.97 & 1.73 & 0.60 & 0.65 \\ 
					&  &  0.5 & 0.82 & 13.23 & 3.19 & 7.29 & 0.71 & 1.33 & 0.73 & 0.77 & 0.69 & 0.96 & 1.74 & 0.44 & 0.46 \\ 
					&  &  0.9 & 0.84 & 12.91 & 3.15 & 7.87 & 1.68 & 1.13 & 1.06 & 0.97 & 0.91 & 1.20 & 2.21 & 0.56 & 0.62 \\ 
					&$t(5)$ &  0.1 & 1.14 & 12.41 & 2.82 & 5.71 & 1.36 & 1.13 & 1.13 & 1.10 & 0.93 & 0.99 & 1.60 & 0.73 & 0.79 \\ 
					&  &  0.5 & 0.86 & 9.78 & 2.30 & 5.02 & 0.69 & 1.33 & 0.73 & 0.72 & 0.72 & 0.76 & 1.36 & 0.47 & 0.47 \\ 
					&  &  0.9 & 0.91 & 11.09 & 2.63 & 6.72 & 1.78 & 1.23 & 1.05 & 1.08 & 0.97 & 1.16 & 2.15 & 0.68 & 0.72 \\ 
					\cline{2-16}
					\multirow{6}*{$T=500$} &$N(0,1)$ &  0.1 & 1.07 & 13.29 & 3.02 & 6.97 & 1.24 & 0.88 & 0.84 & 0.84 & 0.77 & 0.85 & 1.55 & 0.51 & 0.53 \\ 
					&  &  0.5 & 0.67 & 12.32 & 2.92 & 6.95 & 0.62 & 1.20 & 0.62 & 0.60 & 0.59 & 0.84 & 1.60 & 0.37 & 0.38 \\ 
					&  &  0.9 & 0.69 & 11.93 & 2.85 & 7.55 & 1.60 & 1.09 & 0.81 & 0.94 & 0.82 & 1.11 & 2.07 & 0.45 & 0.48 \\ 
					&$t(5)$ &  0.1 & 1.07 & 13.65 & 2.53 & 6.37 & 1.10 & 0.93 & 0.91 & 0.92 & 0.76 & 1.00 & 1.76 & 0.58 & 0.64 \\ 
					&  &  0.5 & 0.73 & 10.70 & 2.09 & 5.32 & 0.64 & 1.22 & 0.65 & 0.68 & 0.63 & 0.77 & 1.49 & 0.36 & 0.43 \\ 
					&  &  0.9 & 0.78 & 12.25 & 2.41 & 6.98 & 1.47 & 1.06 & 0.92 & 0.88 & 0.85 & 1.14 & 2.13 & 0.52 & 0.62 \\ 
					\hline
					&& \multicolumn{13}{c}{$ N = 500 $} \\
					\multirow{6}*{$T=100$} &$N(0,1)$ &  0.1 & 1.30 & 14.17 & 3.37 & 7.49 & 1.45 & 1.08 & 1.10 & 1.01 & 0.96 & 1.05 & 1.71 & 0.72 & 0.78 \\ 
					&  &  0.5 & 0.80 & 13.19 & 3.27 & 7.37 & 0.76 & 1.40 & 0.84 & 0.82 & 0.74 & 0.99 & 1.77 & 0.53 & 0.53 \\ 
					&  &  0.9 & 0.92 & 12.81 & 3.23 & 8.04 & 1.70 & 1.16 & 1.06 & 1.09 & 0.96 & 1.27 & 2.30 & 0.66 & 0.75 \\ 
					&$t(5)$ &  0.1 & 1.14 & 11.81 & 2.57 & 5.49 & 1.38 & 1.24 & 1.23 & 1.21 & 1.08 & 1.11 & 1.57 & 0.81 & 0.87 \\ 
					&  &  0.5 & 0.83 & 9.32 & 2.13 & 4.78 & 0.73 & 1.29 & 0.84 & 0.78 & 0.74 & 0.79 & 1.36 & 0.57 & 0.57 \\ 
					&  &  0.9 & 0.98 & 10.51 & 2.43 & 6.38 & 1.86 & 1.35 & 1.20 & 1.19 & 1.13 & 1.25 & 2.05 & 0.78 & 0.84 \\ 
					\cline{2-16}
					\multirow{6}*{$T=200$} &$N(0,1)$ &  0.1 & 1.23 & 14.40 & 3.68 & 7.31 & 1.36 & 0.85 & 0.80 & 0.77 & 0.63 & 0.93 & 1.58 & 0.52 & 0.50 \\ 
					&  &  0.5 & 0.72 & 13.38 & 3.44 & 7.28 & 0.53 & 1.30 & 0.58 & 0.59 & 0.60 & 0.97 & 1.77 & 0.34 & 0.35 \\ 
					&  &  0.9 & 0.76 & 13.07 & 3.42 & 7.94 & 1.70 & 1.07 & 0.70 & 0.77 & 0.70 & 1.17 & 2.23 & 0.46 & 0.55 \\ 
					&$t(5)$ &  0.1 & 1.04 & 11.68 & 2.66 & 5.33 & 1.20 & 0.93 & 0.92 & 0.96 & 0.86 & 0.87 & 1.43 & 0.60 & 0.61 \\ 
					&  &  0.5 & 0.69 & 9.13 & 2.11 & 4.66 & 0.60 & 1.15 & 0.57 & 0.58 & 0.57 & 0.67 & 1.23 & 0.41 & 0.39 \\ 
					&  &  0.9 & 0.84 & 10.39 & 2.43 & 6.27 & 1.72 & 1.06 & 0.92 & 1.01 & 0.91 & 1.04 & 1.92 & 0.55 & 0.60 \\ 
					\cline{2-16}
					\multirow{6}*{$T=500$} &$N(0,1)$ &  0.1 & 0.98 & 12.90 & 2.94 & 6.94 & 1.21 & 0.70 & 0.66 & 0.56 & 0.58 & 0.78 & 1.42 & 0.42 & 0.42 \\ 
					&  &  0.5 & 0.54 & 11.77 & 2.86 & 6.80 & 0.50 & 1.21 & 0.51 & 0.55 & 0.51 & 0.82 & 1.47 & 0.28 & 0.32 \\ 
					&  &  0.9 & 0.60 & 11.46 & 2.82 & 7.51 & 1.63 & 1.09 & 0.74 & 0.80 & 0.77 & 1.04 & 1.97 & 0.36 & 0.45 \\ 
					&$t(5)$ &  0.1 & 0.95 & 11.80 & 2.63 & 5.63 & 1.35 & 0.83 & 0.76 & 0.80 & 0.63 & 0.79 & 1.33 & 0.50 & 0.54 \\ 
					&  &  0.5 & 0.66 & 9.26 & 2.20 & 4.93 & 0.49 & 1.14 & 0.56 & 0.56 & 0.52 & 0.68 & 1.20 & 0.27 & 0.33 \\ 
					&  &  0.9 & 0.67 & 10.44 & 2.53 & 6.71 & 1.85 & 1.00 & 0.81 & 0.95 & 0.80 & 1.10 & 1.99 & 0.48 & 0.55 \\ 
					\hline
					\hline
				\end{tabular}
			}
		\end{center}
		{\footnotesize{Notes: See the notes to Table \ref{tab_simu_RMSE_W1_NonIV}.}}
	\end{table}
	\normalsize

	\begin{table}[!htbp]
		\begin{center}
			\caption{RMSE ($\times 100$) for \textsc{Type 3} Network by Ordinary QR Estimator}
			\label{tab_simu_RMSE_W3_NonIV}
			{\scriptsize 
				\begin{tabular}{ccc|ccccccccccccc}
					\hline
					\hline
					&Dist. & $\tau$ &$\gamma_0 $  & $\gamma_1$&  $\gamma_2$ &  $\gamma_3$ & $\alpha_1$ & $\alpha_2$ &$\alpha_3$ & $\alpha_4$ & $\alpha_5$ & $\beta_1$ & $\beta_2$ & $\beta_3$ & $\beta_4$  \\
					
					\hline
					&& \multicolumn{13}{c}{$ N = 100 $} \\
					\multirow{6}*{$T=100$} &$N(0,1)$ &  0.1 & 1.61 & 1.58 & 1.26 & 1.71 & 1.72 & 1.87 & 1.86 & 1.80 & 1.60 & 1.41 & 1.43 & 1.43 & 1.39 \\ 
					&  &  0.5 & 1.41 & 1.31 & 1.07 & 1.39 & 1.32 & 1.58 & 1.46 & 1.51 & 1.32 & 1.10 & 1.15 & 1.13 & 1.07 \\ 
					&  &  0.9 & 1.64 & 1.49 & 1.23 & 1.60 & 1.60 & 1.74 & 1.74 & 1.74 & 1.56 & 1.35 & 1.35 & 1.34 & 1.31 \\ 
					&$t(5)$ &  0.1 & 1.82 & 1.53 & 1.17 & 1.49 & 1.93 & 2.15 & 2.18 & 2.21 & 1.90 & 1.60 & 1.65 & 1.64 & 1.68 \\ 
					&  &  0.5 & 1.43 & 1.10 & 0.84 & 1.02 & 1.36 & 1.59 & 1.51 & 1.51 & 1.35 & 1.14 & 1.16 & 1.14 & 1.11 \\ 
					&  &  0.9 & 1.86 & 1.39 & 1.11 & 1.51 & 1.85 & 2.06 & 2.05 & 2.06 & 1.83 & 1.58 & 1.54 & 1.57 & 1.57 \\ 
					\cline{2-16}
					\multirow{6}*{$T=200$} &$N(0,1)$ &  0.1 & 1.10 & 1.21 & 0.94 & 1.26 & 1.19 & 1.36 & 1.33 & 1.30 & 1.15 & 0.98 & 1.01 & 0.99 & 0.98 \\ 
					&  &  0.5 & 1.01 & 0.97 & 0.78 & 0.99 & 0.94 & 1.08 & 1.10 & 1.06 & 0.93 & 0.77 & 0.79 & 0.79 & 0.75 \\ 
					&  &  0.9 & 1.14 & 1.11 & 0.90 & 1.19 & 1.18 & 1.20 & 1.24 & 1.26 & 1.11 & 0.92 & 0.94 & 0.90 & 0.93 \\ 
					&$t(5)$ &  0.1 & 1.21 & 1.09 & 0.83 & 1.06 & 1.39 & 1.52 & 1.53 & 1.51 & 1.36 & 1.12 & 1.18 & 1.18 & 1.17 \\ 
					&  &  0.5 & 1.03 & 0.76 & 0.60 & 0.72 & 0.98 & 1.12 & 1.10 & 1.09 & 0.97 & 0.76 & 0.76 & 0.81 & 0.79 \\ 
					&  &  0.9 & 1.33 & 1.06 & 0.80 & 1.17 & 1.40 & 1.52 & 1.50 & 1.53 & 1.27 & 1.12 & 1.09 & 1.08 & 1.10 \\ 
					\cline{2-16}
					\multirow{6}*{$T=500$} &$N(0,1)$ &  0.1 & 0.89 & 1.09 & 0.76 & 1.07 & 0.95 & 1.06 & 1.09 & 1.11 & 0.98 & 0.82 & 0.81 & 0.79 & 0.79 \\ 
					&  &  0.5 & 0.84 & 0.90 & 0.64 & 0.81 & 0.79 & 0.95 & 0.83 & 0.86 & 0.78 & 0.61 & 0.63 & 0.63 & 0.64 \\ 
					&  &  0.9 & 0.93 & 0.99 & 0.74 & 1.06 & 0.95 & 0.97 & 0.95 & 1.02 & 0.91 & 0.74 & 0.80 & 0.72 & 0.72 \\ 
					&$t(5)$ &  0.1 & 1.04 & 0.90 & 0.67 & 0.85 & 1.15 & 1.25 & 1.27 & 1.22 & 1.10 & 0.89 & 0.92 & 0.98 & 0.97 \\ 
					&  &  0.5 & 0.80 & 0.61 & 0.48 & 0.55 & 0.81 & 0.91 & 0.90 & 0.86 & 0.79 & 0.62 & 0.64 & 0.65 & 0.61 \\ 
					&  &  0.9 & 1.04 & 0.78 & 0.65 & 0.93 & 1.15 & 1.20 & 1.20 & 1.18 & 1.09 & 0.89 & 0.92 & 0.89 & 0.87 \\ 
					\hline
					&& \multicolumn{13}{c}{$ N = 200 $} \\
					\multirow{6}*{$T=100$} &$N(0,1)$ &  0.1 & 1.09 & 1.14 & 0.94 & 1.21 & 1.19 & 1.28 & 1.26 & 1.31 & 1.19 & 0.99 & 0.98 & 0.98 & 1.02 \\ 
					&  &  0.5 & 0.99 & 0.94 & 0.74 & 0.99 & 0.97 & 1.10 & 1.02 & 1.03 & 0.95 & 0.82 & 0.80 & 0.79 & 0.78 \\ 
					&  &  0.9 & 1.10 & 1.14 & 0.89 & 1.20 & 1.07 & 1.14 & 1.23 & 1.20 & 1.11 & 0.91 & 0.94 & 0.94 & 0.94 \\ 
					&$t(5)$ &  0.1 & 1.27 & 0.99 & 0.84 & 1.04 & 1.32 & 1.47 & 1.46 & 1.51 & 1.29 & 1.16 & 1.19 & 1.20 & 1.16 \\ 
					&  &  0.5 & 1.01 & 0.70 & 0.61 & 0.69 & 0.95 & 1.10 & 1.02 & 1.01 & 0.92 & 0.79 & 0.80 & 0.81 & 0.78 \\ 
					&  &  0.9 & 1.31 & 0.95 & 0.77 & 1.04 & 1.34 & 1.38 & 1.46 & 1.38 & 1.18 & 1.11 & 1.13 & 1.08 & 1.13 \\  
					\cline{2-16}
					\multirow{6}*{$T=200$} &$N(0,1)$ &  0.1 & 0.77 & 0.84 & 0.63 & 0.86 & 0.83 & 0.88 & 0.93 & 0.90 & 0.84 & 0.69 & 0.71 & 0.73 & 0.72 \\ 
					&  &  0.5 & 0.69 & 0.70 & 0.58 & 0.68 & 0.66 & 0.80 & 0.72 & 0.70 & 0.64 & 0.56 & 0.55 & 0.55 & 0.52 \\ 
					&  &  0.9 & 0.84 & 0.73 & 0.63 & 0.83 & 0.81 & 0.91 & 0.92 & 0.84 & 0.75 & 0.61 & 0.67 & 0.65 & 0.67 \\ 
					&$t(5)$ &  0.1 & 0.93 & 0.83 & 0.57 & 0.81 & 0.93 & 1.09 & 1.04 & 1.03 & 0.97 & 0.81 & 0.84 & 0.81 & 0.85 \\ 
					&  &  0.5 & 0.71 & 0.58 & 0.44 & 0.55 & 0.64 & 0.83 & 0.72 & 0.76 & 0.68 & 0.57 & 0.55 & 0.57 & 0.53 \\ 
					&  &  0.9 & 0.88 & 0.77 & 0.53 & 0.86 & 0.88 & 1.04 & 1.00 & 0.98 & 0.87 & 0.76 & 0.80 & 0.74 & 0.75 \\ 
					\cline{2-16}
					\multirow{6}*{$T=500$} &$N(0,1)$ &  0.1 & 0.63 & 0.75 & 0.49 & 0.71 & 0.73 & 0.73 & 0.69 & 0.72 & 0.68 & 0.56 & 0.58 & 0.55 & 0.53 \\ 
					&  &  0.5 & 0.53 & 0.58 & 0.41 & 0.54 & 0.60 & 0.62 & 0.56 & 0.57 & 0.50 & 0.43 & 0.44 & 0.42 & 0.44 \\ 
					&  &  0.9 & 0.67 & 0.67 & 0.54 & 0.70 & 0.71 & 0.71 & 0.72 & 0.70 & 0.58 & 0.52 & 0.56 & 0.52 & 0.52 \\ 
					&$t(5)$ &  0.1 & 0.77 & 0.61 & 0.48 & 0.65 & 0.83 & 0.86 & 0.91 & 0.80 & 0.73 & 0.65 & 0.66 & 0.65 & 0.66 \\ 
					&  &  0.5 & 0.61 & 0.43 & 0.35 & 0.40 & 0.55 & 0.68 & 0.63 & 0.60 & 0.50 & 0.46 & 0.43 & 0.49 & 0.43 \\ 
					&  &  0.9 & 0.72 & 0.55 & 0.48 & 0.68 & 0.78 & 0.79 & 0.85 & 0.87 & 0.81 & 0.64 & 0.62 & 0.60 & 0.61 \\ \hline
					&& \multicolumn{13}{c}{$ N = 500 $} \\
					\multirow{6}*{$T=100$} &$N(0,1)$ &  0.1 & 1.06 & 0.99 & 0.75 & 1.02 & 0.92 & 1.00 & 1.09 & 1.09 & 0.99 & 0.80 & 0.89 & 0.80 & 0.82 \\ 
					&  &  0.5 & 0.84 & 0.82 & 0.63 & 0.87 & 0.74 & 0.91 & 0.80 & 0.81 & 0.72 & 0.63 & 0.69 & 0.61 & 0.65 \\ 
					&  &  0.9 & 1.07 & 0.93 & 0.70 & 1.02 & 0.96 & 0.90 & 0.96 & 0.99 & 0.89 & 0.75 & 0.74 & 0.74 & 0.76 \\ 
					&$t(5)$ &  0.1 & 1.02 & 0.86 & 0.66 & 0.88 & 1.08 & 1.26 & 1.20 & 1.19 & 1.08 & 0.97 & 1.01 & 0.94 & 0.94 \\ 
					&  &  0.5 & 0.82 & 0.61 & 0.46 & 0.60 & 0.75 & 0.93 & 0.90 & 0.86 & 0.82 & 0.66 & 0.68 & 0.67 & 0.64 \\ 
					&  &  0.9 & 1.04 & 0.81 & 0.63 & 0.88 & 1.10 & 1.17 & 1.13 & 1.12 & 1.07 & 0.91 & 0.91 & 0.94 & 0.92 \\ 
					\cline{2-16}
					\multirow{6}*{$T=200$} &$N(0,1)$ &  0.1 & 0.67 & 0.63 & 0.51 & 0.64 & 0.70 & 0.61 & 0.66 & 0.70 & 0.69 & 0.57 & 0.58 & 0.57 & 0.54 \\ 
					&  &  0.5 & 0.61 & 0.52 & 0.44 & 0.56 & 0.55 & 0.69 & 0.62 & 0.58 & 0.53 & 0.43 & 0.46 & 0.42 & 0.46 \\ 
					&  &  0.9 & 0.64 & 0.60 & 0.49 & 0.62 & 0.67 & 0.69 & 0.65 & 0.70 & 0.63 & 0.49 & 0.54 & 0.49 & 0.58 \\ 
					&$t(5)$ &  0.1 & 0.80 & 0.66 & 0.50 & 0.71 & 0.80 & 0.83 & 0.82 & 0.81 & 0.76 & 0.70 & 0.63 & 0.63 & 0.61 \\ 
					&  &  0.5 & 0.60 & 0.46 & 0.32 & 0.52 & 0.57 & 0.66 & 0.60 & 0.60 & 0.50 & 0.46 & 0.41 & 0.42 & 0.45 \\ 
					&  &  0.9 & 0.76 & 0.63 & 0.44 & 0.74 & 0.79 & 0.80 & 0.79 & 0.76 & 0.66 & 0.66 & 0.68 & 0.66 & 0.59 \\ 
					\cline{2-16}
					\multirow{6}*{$T=500$} &$N(0,1)$ &  0.1 & 0.64 & 0.57 & 0.47 & 0.62 & 0.54 & 0.57 & 0.57 & 0.69 & 0.52 & 0.47 & 0.43 & 0.52 & 0.43 \\ 
					&  &  0.5 & 0.48 & 0.46 & 0.40 & 0.46 & 0.45 & 0.57 & 0.45 & 0.39 & 0.38 & 0.38 & 0.35 & 0.36 & 0.31 \\ 
					&  &  0.9 & 0.60 & 0.58 & 0.41 & 0.77 & 0.63 & 0.54 & 0.49 & 0.56 & 0.55 & 0.37 & 0.41 & 0.40 & 0.43 \\ 
					&$t(5)$ &  0.1 & 0.61 & 0.57 & 0.38 & 0.54 & 0.70 & 0.76 & 0.74 & 0.60 & 0.62 & 0.52 & 0.56 & 0.53 & 0.58 \\ 
					&  &  0.5 & 0.48 & 0.34 & 0.33 & 0.35 & 0.46 & 0.62 & 0.46 & 0.47 & 0.44 & 0.37 & 0.35 & 0.38 & 0.35 \\ 
					&  &  0.9 & 0.60 & 0.49 & 0.39 & 0.60 & 0.66 & 0.70 & 0.79 & 0.70 & 0.54 & 0.54 & 0.56 & 0.44 & 0.57 \\ 
					\hline
					\hline
				\end{tabular}
			}
		\end{center}
		{\footnotesize{Notes: See the notes to Table \ref{tab_simu_RMSE_W1_NonIV}.}}
	\end{table}
	\normalsize

	\begin{table}[!htbp]
		\begin{center}
			\caption{Coverage Probability ($\times 100$) for \textsc{Type 1} Network by Ordinary QR Estimator}
			\label{tab_simu_CP_W1_NonIV}
			{\scriptsize 
\begin{tabular}{ccc|ccccccccccccc}
					\hline
					\hline
					&Dist. & $\tau$ &$\gamma_0 $  & $\gamma_1$&  $\gamma_2$ &  $\gamma_3$ & $\alpha_1$ & $\alpha_2$ &$\alpha_3$ & $\alpha_4$ & $\alpha_5$ & $\beta_1$ & $\beta_2$ & $\beta_3$ & $\beta_4$  \\
					
					\hline
					&& \multicolumn{13}{c}{$ N = 100 $} \\
					\multirow{6}*{$T=100$} &$N(0,1)$ &  0.1 & 91.6 & 37.5 & 91.6 & 74.3 & 93.6 & 93.0 & 94.9 & 94.9 & 95.3 & 93.9 & 91.0 & 94.8 & 94.6 \\ 
					&  &  0.5 & 87.2 & 38.3 & 88.2 & 69.5 & 94.0 & 92.8 & 94.4 & 93.8 & 94.4 & 94.1 & 90.6 & 94.6 & 94.2 \\ 
					&  &  0.9 & 87.8 & 38.1 & 88.6 & 70.8 & 93.8 & 93.2 & 94.1 & 94.3 & 94.7 & 94.8 & 90.0 & 94.7 & 94.8 \\ 
					&$t(5)$ &  0.1 & 92.0 & 39.8 & 91.2 & 74.5 & 94.5 & 94.8 & 94.5 & 94.7 & 95.0 & 95.1 & 92.6 & 95.0 & 94.6 \\ 
					&  &  0.5 & 87.0 & 43.0 & 89.0 & 69.8 & 94.8 & 91.6 & 93.8 & 94.3 & 93.6 & 94.6 & 92.2 & 94.8 & 94.3 \\ 
					&  &  0.9 & 88.6 & 40.2 & 90.2 & 72.9 & 93.0 & 94.0 & 93.8 & 93.5 & 92.2 & 94.8 & 91.6 & 94.8 & 94.4 \\
\hline
					\hline
				\end{tabular}
			}
		\end{center}
		{\footnotesize{Notes: See the notes to Table \ref{tab_simu_RMSE_W1_NonIV}.}}
	\end{table}
	\normalsize

	\begin{table}[!htbp]
		\begin{center}
			\caption{Coverage Probability ($\times 100$) for \textsc{Type 2} Network by Ordinary QR Estimator}
			\label{tab_simu_CP_W2_NonIV}
			{\scriptsize 
				\begin{tabular}{ccc|ccccccccccccc}
					\hline
					\hline
					&Dist. & $\tau$ &$\gamma_0 $  & $\gamma_1$&  $\gamma_2$ &  $\gamma_3$ & $\alpha_1$ & $\alpha_2$ &$\alpha_3$ & $\alpha_4$ & $\alpha_5$ & $\beta_1$ & $\beta_2$ & $\beta_3$ & $\beta_4$  \\
					
					\hline
					&& \multicolumn{13}{c}{$ N = 100 $} \\
					\multirow{6}*{$T=100$} &$N(0,1)$ &  0.1 & 87.4 & 33.6 & 81.5 & 61.5 & 92.6 & 94.4 & 94.2 & 93.6 & 93.8 & 94.9 & 91.8 & 93.9 & 94.7 \\ 
					&  &  0.5 & 85.6 & 33.3 & 73.1 & 53.5 & 93.8 & 90.0 & 93.2 & 94.6 & 93.8 & 94.6 & 91.6 & 95.5 & 96.2 \\ 
					&  &  0.9 & 84.3 & 30.5 & 74.6 & 55.4 & 91.6 & 91.6 & 92.9 & 93.0 & 91.8 & 93.0 & 90.5 & 94.6 & 94.0 \\ 
					&$t(5)$ &  0.1 & 88.6 & 33.0 & 81.0 & 60.6 & 91.7 & 93.2 & 94.1 & 94.8 & 94.0 & 94.4 & 92.5 & 94.3 & 94.6 \\ 
					&  &  0.5 & 83.9 & 33.6 & 74.6 & 58.9 & 94.6 & 89.6 & 94.6 & 95.3 & 94.4 & 94.5 & 91.8 & 95.2 & 94.6 \\ 
					&  &  0.9 & 85.3 & 30.6 & 77.5 & 57.0 & 90.0 & 91.2 & 92.3 & 92.5 & 92.3 & 94.1 & 90.3 & 94.7 & 94.6 \\
\hline
					\hline
				\end{tabular}
			}
		\end{center}
		{\footnotesize{Notes: See the notes to Table \ref{tab_simu_RMSE_W1_NonIV}.}}
	\end{table}
	\normalsize

	\begin{table}[!htbp]
		\begin{center}
			\caption{Coverage Probability ($\times 100$) for \textsc{Type 3} Network by Ordinary QR Estimator}
			\label{tab_simu_CP_W3_NonIV}
			{\scriptsize 
				\begin{tabular}{ccc|ccccccccccccc}
					\hline
					\hline
					&Dist. & $\tau$ &$\gamma_0 $  & $\gamma_1$&  $\gamma_2$ &  $\gamma_3$ & $\alpha_1$ & $\alpha_2$ &$\alpha_3$ & $\alpha_4$ & $\alpha_5$ & $\beta_1$ & $\beta_2$ & $\beta_3$ & $\beta_4$  \\
					
					\hline
					&& \multicolumn{13}{c}{$ N = 100 $} \\
					\multirow{6}*{$T=100$} &$N(0,1)$ &  0.1 & 91.2 & 47.2 & 92.6 & 77.6 & 94.4 & 94.3 & 94.3 & 94.6 & 95.7 & 94.3 & 91.9 & 94.3 & 94.9 \\ 
					&  &  0.5 & 87.3 & 44.8 & 89.8 & 73.3 & 94.5 & 92.0 & 94.0 & 93.0 & 93.3 & 94.3 & 90.8 & 94.0 & 95.6 \\ 
					&  &  0.9 & 88.5 & 47.2 & 91.3 & 76.2 & 93.6 & 94.5 & 93.6 & 93.7 & 93.5 & 94.2 & 91.8 & 93.8 & 94.6 \\ 
					&$t(5)$ &  0.1 & 91.5 & 47.1 & 92.9 & 79.0 & 95.1 & 94.4 & 94.1 & 94.3 & 94.8 & 94.8 & 92.9 & 94.3 & 94.1 \\ 
					&  &  0.5 & 86.3 & 49.0 & 91.6 & 80.6 & 94.2 & 93.0 & 94.0 & 94.7 & 94.6 & 94.1 & 93.0 & 94.5 & 95.3 \\ 
					&  &  0.9 & 88.4 & 47.8 & 92.6 & 79.8 & 93.8 & 94.1 & 94.2 & 94.3 & 94.6 & 94.4 & 93.2 & 94.6 & 94.4 \\
\hline
					\hline
				\end{tabular}
			}
		\end{center}
		{\footnotesize{Notes: See the notes to Table \ref{tab_simu_RMSE_W1_NonIV}.}}
	\end{table}
	\normalsize
	
	\begin{table}[!htbp]
		\begin{center}
			\caption{Bias ($\times 100$) for \textsc{Type 1} Network by Ordinary QR Estimator}
			\label{tab_simu_Bias_W1_NonIV}
			{\scriptsize  
				\begin{tabular}{ccc|ccccccccccccc}
					\hline
					\hline
					&Dist. & $\tau$ &$\gamma_0 $  & $\gamma_1$&  $\gamma_2$ &  $\gamma_3$ & $\alpha_1$ & $\alpha_2$ &$\alpha_3$ & $\alpha_4$ & $\alpha_5$ & $\beta_1$ & $\beta_2$ & $\beta_3$ & $\beta_4$  \\
					
					\hline
					&& \multicolumn{13}{c}{$ N = 100 $} \\
					\multirow{6}*{$T=100$} &$N(0,1)$ &  0.1 & 0.14 & 6.27 & -0.59 & -2.95 & 0.30 & 0.03 & 0.08 & 0.02 & -0.01 & -0.33 & -0.81 & -0.03 & -0.09 \\ 
					&     &  0.5 & 0.01 & 5.80 & -0.60 & -2.74 & 0.20 & 0.59 & 0.01 & -0.05 & 0.10 & -0.35 & -0.77 & 0.00 & -0.08 \\ 
					&     &  0.9 & -0.09 & 5.82 & -0.63 & -3.11 & -0.27 & -0.09 & -0.11 & 0.01 & -0.06 & -0.45 & -1.00 & -0.06 & -0.09 \\ 
					&$t(5)$ &   0.1 & 0.14 & 5.57 & -0.53 & -2.26 & 0.34 & 0.05 & 0.04 & -0.05 & 0.06 & -0.34 & -0.66 & -0.03 & -0.18 \\ 
					&     &  0.5 & 0.09 & 4.33 & -0.41 & -2.00 & 0.12 & 0.65 & -0.01 & 0.03 & 0.06 & -0.25 & -0.54 & -0.04 & -0.01 \\ 
					&     &  0.9 & 0.04 & 5.01 & -0.55 & -2.93 & -0.35 & -0.16 & -0.07 & 0.00 & -0.11 & -0.48 & -0.88 & -0.04 & -0.12 \\
\hline
					\hline
				\end{tabular}
			}
		\end{center}
		{\footnotesize{Notes: See the notes to Table \ref{tab_simu_RMSE_W1_NonIV}.}}
	\end{table}
	\normalsize
	
	\begin{table}[!htbp]
		\begin{center}
			\caption{Bias ($\times 100$) for \textsc{Type 2} Network by Ordinary QR Estimator}
			\label{tab_simu_Bias_W2_NonIV}
			{\scriptsize  
				\begin{tabular}{ccc|ccccccccccccc}
					\hline
					\hline
					&Dist. & $\tau$ &$\gamma_0 $  & $\gamma_1$&  $\gamma_2$ &  $\gamma_3$ & $\alpha_1$ & $\alpha_2$ &$\alpha_3$ & $\alpha_4$ & $\alpha_5$ & $\beta_1$ & $\beta_2$ & $\beta_3$ & $\beta_4$  \\
					
					\hline
					&& \multicolumn{13}{c}{$ N = 100 $} \\
					\multirow{6}*{$T=100$} &$N(0,1)$ &  0.1 & 0.98 & 14.50 & -3.42 & -7.22 & 1.02 & 0.11 & 0.03 & 0.05 & 0.07 & -0.82 & -1.61 & -0.05 & -0.19 \\ 
					&     &  0.5 & 0.16 & 13.41 & -3.22 & -7.12 & 0.20 & 1.14 & 0.10 & 0.00 & 0.23 & -0.86 & -1.75 & -0.01 & -0.08 \\ 
					&     &  0.9 & -0.41 & 13.14 & -3.18 & -7.73 & -1.20 & -0.58 & -0.12 & 0.05 & -0.20 & -1.08 & -2.23 & -0.04 & -0.28 \\ 
					&$t(5)$ &   0.1 & 0.49 & 12.12 & -2.76 & -5.51 & 0.91 & 0.17 & 0.05 & 0.08 & 0.02 & -0.67 & -1.28 & -0.02 & -0.22 \\ 
					&     &  0.5 & 0.25 & 9.68 & -2.21 & -4.93 & 0.05 & 1.06 & 0.12 & 0.02 & 0.21 & -0.61 & -1.25 & 0.02 & -0.06 \\ 
					&     &  0.9 & 0.17 & 10.91 & -2.51 & -6.52 & -1.32 & -0.71 & -0.08 & 0.04 & -0.28 & -0.99 & -1.96 & -0.11 & -0.33 \\
\hline
					\hline
				\end{tabular}
			}
		\end{center}
		{\footnotesize{Notes: See the notes to Table \ref{tab_simu_RMSE_W1_NonIV}.}}
	\end{table}
	\normalsize

	\begin{table}[!htbp]
		\begin{center}
			\caption{Bias ($\times 100$) for \textsc{Type 3} Network by Ordinary QR Estimator}
			\label{tab_simu_Bias_W3_NonIV}
			{\scriptsize  
				\begin{tabular}{ccc|ccccccccccccc}
					\hline
					\hline
					&Dist. & $\tau$ &$\gamma_0 $  & $\gamma_1$&  $\gamma_2$ &  $\gamma_3$ & $\alpha_1$ & $\alpha_2$ &$\alpha_3$ & $\alpha_4$ & $\alpha_5$ & $\beta_1$ & $\beta_2$ & $\beta_3$ & $\beta_4$  \\
					
					\hline
					&& \multicolumn{13}{c}{$ N = 100 $} \\
					
					\multirow{6}*{$T=100$} &$N(0,1)$ &  0.1 & 0.06 & 1.06 & -0.13 & -0.71 & 0.13 & -0.02 & -0.01 & 0.09 & -0.06 & -0.07 & -0.11 & 0.01 & -0.04 \\ 
					&     &  0.5 & -0.01 & 0.95 & -0.05 & -0.63 & 0.11 & 0.43 & 0.01 & -0.05 & 0.15 & -0.05 & -0.03 & 0.03 & 0.04 \\ 
					&     &  0.9 & -0.04 & 0.91 & -0.10 & -0.81 & -0.20 & -0.03 & -0.05 & 0.03 & -0.05 & -0.15 & -0.21 & -0.06 & -0.02 \\ 
					&$t(5)$ &   0.1 & -0.04 & 0.58 & -0.06 & -0.14 & 0.20 & 0.00 & 0.03 & -0.07 & 0.04 & 0.00 & 0.17 & 0.06 & -0.09 \\ 
					&     &  0.5 & -0.11 & 0.46 & -0.09 & -0.06 & 0.20 & 0.42 & 0.06 & 0.00 & -0.03 & 0.04 & 0.06 & 0.03 & 0.06 \\ 
					&     &  0.9 & -0.04 & 0.57 & -0.11 & -0.40 & -0.36 & -0.14 & -0.01 & 0.08 & -0.06 & -0.06 & -0.20 & -0.07 & 0.00 \\
\hline
					\hline
				\end{tabular}
			}
		\end{center}
		{\footnotesize{Notes: See the notes to Table \ref{tab_simu_RMSE_W1_NonIV}.}}
	\end{table}
	\normalsize

	\newpage
	\section{Additional Empirical Results}\label{appendixc}
	\setcounter{table}{0}
	\renewcommand{\thetable}{C\arabic{table}} 
	\setcounter{figure}{0}
	\renewcommand{\thefigure}{C\arabic{figure}} 
	
	As the robustness check, we now provide the two additional estimation results using alternative network structures.

	\subsection{The Common Shareholder Networks $W_{CS3}$ and $W_{CS7}$} \label{appendixc1}
	
\begin{figure}[!htbp]
		\centering
		\includegraphics[scale=0.7]{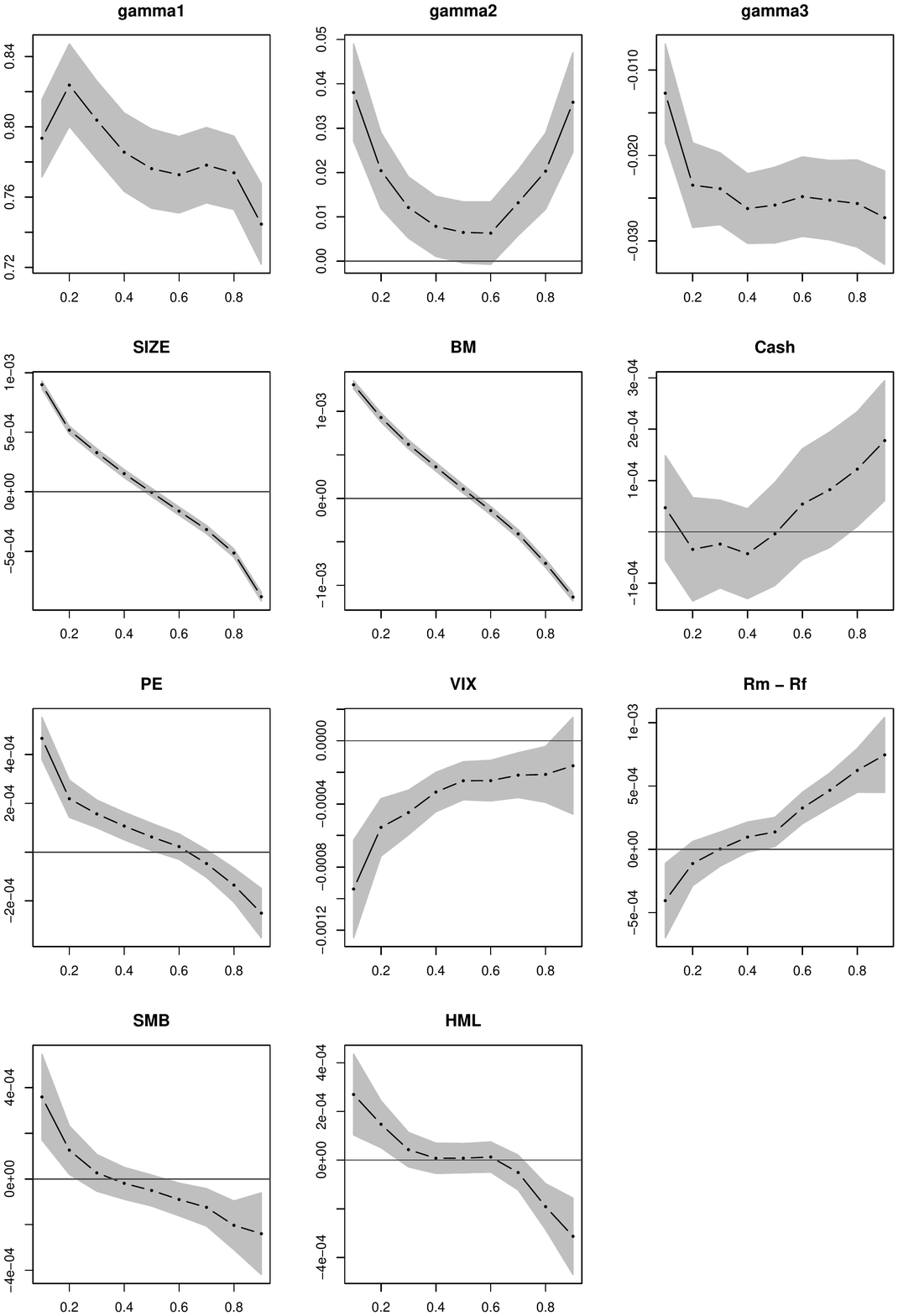}
		\caption{Quantile-specific Coefficients for the Network $W_{CS3}$}\label{AppParW3}
		\caption*{\footnotesize Notes: The dashed line is the QR coefficient while the grey area indicates a kernel density based confidence band advanced by \cite{powell1991}. They are displayed across quantiles, $\tau = 0.1, 0.2, \cdots, 0.9$.}
		
	\end{figure}

\begin{figure}[!htbp]
		\centering
		\includegraphics[scale=0.7]{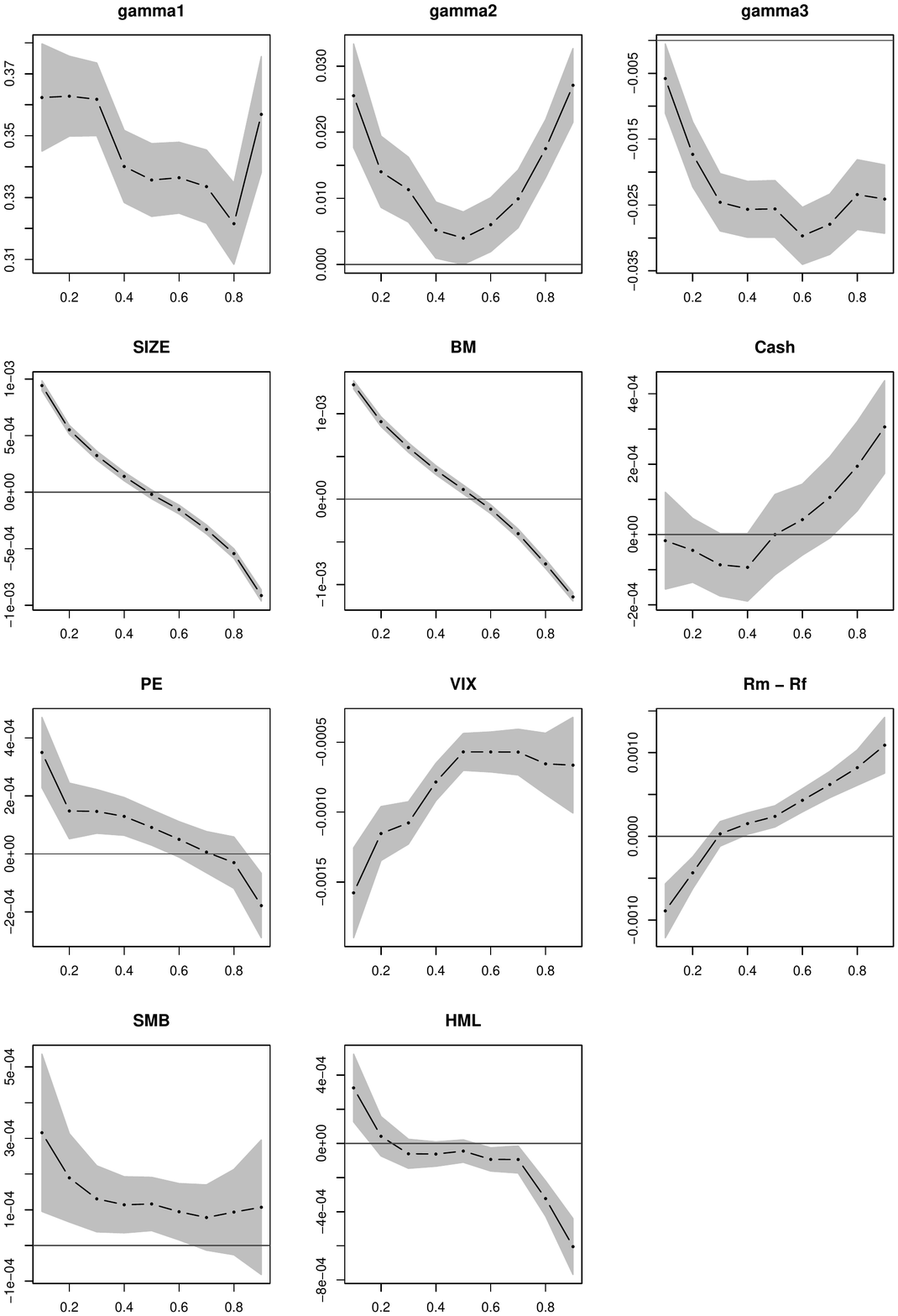}
		\caption{Quantile-specific Coefficients for the Network $W_{CS7}$}\label{AppParW7}
		\caption*{\footnotesize Notes: The dashed line is the QR coefficient while the grey area indicates a kernel density based confidence band advanced by \cite{powell1991}. They are displayed across quantiles, $\tau = 0.1, 0.2, \cdots, 0.9$.}
	\end{figure}

	First, we reconstruct the network matrix by changing the number of common shareholders to three $(W_{CS3})$ and seven $(W_{CS7})$, and report the respective results in Tables \ref{TabAppW3}--\ref{TabAppW7} and Figures \ref{AppParW3}--\ref{AppParW7}. Notice that the network density is dense at 25.25\% for $W_{CS3}$ and relatively sparse at 0.41\% for $W_{CS7}$. 
	
	Overall, we find qualitatively similar results to those reported for $W_{CS5}$. 
	One notable observation is that the contemporaneous network effects measured by $\gamma_1$ tend to decrease monotonically as the network becomes more sparse. For example, at $\tau = 0.1$,  $\hat{\gamma}_1$ is estimated at 0.69 for $W_{CS3}$, 0.54 for $W_{CS5}$, and 0.35 $W_{CS7}$, respectively. Still, we find that the patterns of the quantile specific coefficients reported in Figures \ref{AppParW3}--\ref{AppParW7} are qualitatively similar to those displayed in Figure \ref{AppParW1}.
	
	
	\begin{table}[!htbp]
		\begin{center}
			\scriptsize 
			\caption{Estimation Results for the Network $W_{CS3}$}\label{TabAppW3}
			\begin{tabular}{c|ccc|ccc|ccc}
				\hline
				\hline			
				& \multicolumn{3}{c}{DNQM } & \multicolumn{3}{c}{NQARF} & \multicolumn{3}{c}{NQAR } \\ \hline
				& $ \tau=0.1$  & $ \tau=0.5$  & $ \tau=0.9$ & $ \tau=0.1$  & $ \tau=0.5$  & $ \tau=0.9$ & $ \tau=0.1$  & $ \tau=0.5$  & $ \tau=0.9$ \\ 
				$ \hat\gamma_0 $  &   $-2.18^{***}$ &  $0.01^{***}$ &  $2.22^{***}$ &  $-2.55^{***}$ &  $0.05^{***}$ &  $2.64^{***}$ &  $-2.55^{***}$ &  $0.04^{***}$ &  $2.65^{***}$ \\ 
				&   (0.01)  &  (0.00)  &  (0.01)  &  (0.01)  &  (0.00)  &  (0.01)  &  (0.01)  &  (0.00)  &  (0.01)  \\ 
				$ \hat\gamma_1 $  &   $79.75^{***}$ &  $77.98^{***}$ &  $75.12^{***}$ & - & - & -  & - & - & -  \\ 
				&   (1.34)  &  (1.40)  &  (1.39)  & && & && \\ 
				$ \hat\gamma_2 $  &   $4.06^{***}$ &  $0.78^{***}$ &  $3.88^{***}$ &  $5.64^{***}$ &  $-1.10^{***}$ &  $6.53^{***}$ &  $7.03^{***}$ &  $-1.10^{***}$ &  $6.08^{***}$ \\ 
				&   (0.69)  &  (0.42)  &  (0.70)  &  (0.83)  &  (0.27)  &  (0.82)  &  (0.85)  &  (0.27)  &  (0.83)  \\ 
				$ \hat\gamma_3 $  &   $-1.27^{***}$ &  $-2.51^{***}$ &  $-2.69^{***}$ &  $-0.51$ &  $-1.96$ &  $-2.74$ &  $-0.44$ &  $-1.47$ &  $-2.80$ \\ 
				&   (0.35)  &  (0.27)  &  (0.33)  &  (0.42)  &  (0.10)  &  (0.40)  &  (0.42)  &  (0.13)  &  (0.39)  \\ 
				&&&&&&&&& \\ 
				SIZE  &   $0.09^{***}$ &  $0.00^{***}$ &  $-0.09^{***}$ &  $0.09^{***}$ &  $0.00^{***}$ &  $-0.09^{***}$ &  $0.10^{***}$ &  $0.00^{***}$ &  $-0.09^{***}$ \\ 
				&   (0.00)  &  (0.00)  &  (0.00)  &  (0.00)  &  (0.00)  &  (0.00)  &  (0.00)  &  (0.00)  &  (0.00)  \\ 
				BM  &   $0.13^{***}$ &  $0.01^{***}$ &  $-0.11^{***}$ &  $0.14^{***}$ &  $0.02^{***}$ &  $-0.12^{***}$ &  $0.14^{***}$ &  $0.02^{***}$ &  $-0.12^{***}$ \\ 
				&   (0.00)  &  (0.00)  &  (0.00)  &  (0.00)  &  (0.00)  &  (0.00)  &  (0.00)  &  (0.00)  &  (0.00)  \\ 
				Cash  &   $0.00$ &  $0.00$ &  $0.02$ &  $0.00$ &  $0.01$ &  $0.02$ &  $0.00$ &  $0.01$ &  $0.02$ \\ 
				&   (0.01)  &  (0.01)  &  (0.01)  &  (0.01)  &  (0.01)  &  (0.01)  &  (0.01)  &  (0.01)  &  (0.01)  \\ 
				PE  &   $0.05^{***}$ &  $0.01^{***}$ &  $-0.02^{***}$ &  $0.03^{***}$ &  $0.01^{***}$ &  $-0.01^{***}$ &  $0.03^{***}$ &  $0.01^{***}$ &  $-0.01^{***}$ \\ 
				&   (0.01)  &  (0.00)  &  (0.01)  &  (0.01)  &  (0.00)  &  (0.01)  &  (0.01)  &  (0.00)  &  (0.01)  \\ 
				VIX  &   $-0.09^{***}$ &  $-0.03^{***}$ &  $-0.01^{***}$ &  $-0.23^{***}$ &  $-0.08^{***}$ &  $-0.10^{***}$ & - & - & -  \\ 
				&   (0.02)  &  (0.01)  &  (0.02)  &  (0.02)  &  (0.01)  &  (0.02)  & && \\ 
				Rm - Rf  &   $-0.04^{**}$ &  $0.01^{**}$ &  $0.07^{**}$ &  $-0.13^{***}$ &  $0.04^{***}$ &  $0.15^{***}$ & - & - & -  \\ 
				&   (0.02)  &  (0.01)  &  (0.02)  &  (0.02)  &  (0.01)  &  (0.02)  & && \\ 
				SMB  &   $0.04^{***}$ &  $-0.01^{***}$ &  $-0.02^{***}$ &  $-0.01$ &  $0.01$ &  $0.04$ & - & - & -  \\ 
				&   (0.01)  &  (0.00)  &  (0.01)  &  (0.01)  &  (0.00)  &  (0.01)  & && \\ 
				HML  &   $0.03^{**}$ &  $0.00^{**}$ &  $-0.03^{**}$ &  $0.01$ &  $-0.01$ &  $-0.10$ & - & - & -  \\ 
				&   (0.01)  &  (0.00)  &  (0.01)  &  (0.01)  &  (0.00)  &  (0.01)  & && \\ 
				\hline &&&&&&&&& \\  
				Goodn.fit. &  - & - & -  & 10.79 & 12.52 & 9.77 & 11.48 & 12.57 & 9.89 \\ 
				
				\hline
				\hline
			\end{tabular}
		\end{center}    
		{\footnotesize Notes: The dataset consists of $N = 943$ stocks with $T = 252$ time periods. The network matrix, $W_{CS3}$ is constructed by checking if the stocks are invested in by at least three common shareholders with the network density, 25.25\%. The estimates ($\times 10^{2}$) are reported across different quantiles $\tau = 0.1, 0.5, 0.9$, and the value in parentheses is the standard error ($\times 10^{2}$). DNQR denotes the proposed model, NQAR is the model without contemporaneous network effects and common factors, and NQARF is the factor-augmented NQAR model. Goodn.fit. ($\times 10^{2}$) represents the goodness of fit of DNQR model with respect to the other models. The 1\%, 5\% and 10\% significance levels are denoted by ***, **, *, respectively.} 
	\end{table}
	
	\begin{table}[!htbp]
		\begin{center}
			\scriptsize 
			\caption{Estimation Results for the Network $W_{CS7}$}\label{TabAppW7}
			\begin{tabular}{c|ccc|ccc|ccc}
				\hline
				\hline			
				& \multicolumn{3}{c}{DNQM } & \multicolumn{3}{c}{NQARF} & \multicolumn{3}{c}{NQAR}\\ \hline
				& $ \tau=0.1$  & $ \tau=0.5$  & $ \tau=0.9$ & $ \tau=0.1$  & $ \tau=0.5$  & $ \tau=0.9$ & $ \tau=0.1$  & $ \tau=0.5$  & $ \tau=0.9$ \\ 
				$ \hat\gamma_0 $  &   $-2.40^{***}$ &  $0.04^{***}$ &  $2.46^{***}$ &  $-2.54^{***}$ &  $0.05^{***}$ &  $2.65^{***}$ &  $-2.55^{***}$ &  $0.04^{***}$ &  $2.65^{***}$ \\ 
				&   (0.01)  &  (0.00)  &  (0.01)  &  (0.01)  &  (0.00)  &  (0.01)  &  (0.01)  &  (0.00)  &  (0.01)  \\ 
				$ \hat\gamma_1 $  &   $34.76^{***}$ &  $33.62^{***}$ &  $35.76^{***}$ & - & - & -  & - & - & -  \\ 
				&   (1.06)  &  (0.70)  &  (1.12)  & && & && \\ 
				$ \hat\gamma_2 $  &   $2.15^{***}$ &  $0.67^{***}$ &  $2.51^{***}$ &  $1.63^{***}$ &  $-0.64^{***}$ &  $2.35^{***}$ &  $2.97^{***}$ &  $-0.60^{***}$ &  $2.41^{***}$ \\ 
				&   (0.49)  &  (0.26)  &  (0.41)  &  (0.58)  &  (0.18)  &  (0.55)  &  (0.59)  &  (0.18)  &  (0.54)  \\ 
				$ \hat\gamma_3 $  &   $-0.43$ &  $-2.68$ &  $-2.41$ &  $0.22$ &  $-2.08$ &  $-1.90$ &  $0.32$ &  $-1.57$ &  $-2.13$ \\ 
				&   (0.30)  &  (0.27)  &  (0.34)  &  (0.40)  &  (0.12)  &  (0.38)  &  (0.40)  &  (0.12)  &  (0.37)  \\ 
				&&&&&&&&& \\ 
				SIZE  &   $0.09^{***}$ &  $0.00^{***}$ &  $-0.09^{***}$ &  $0.09^{***}$ &  $0.00^{***}$ &  $-0.09^{***}$ &  $0.10^{***}$ &  $0.00^{***}$ &  $-0.09^{***}$ \\ 
				&   (0.00)  &  (0.00)  &  (0.00)  &  (0.00)  &  (0.00)  &  (0.00)  &  (0.00)  &  (0.00)  &  (0.00)  \\ 
				BM  &   $0.13^{***}$ &  $0.01^{***}$ &  $-0.12^{***}$ &  $0.14^{***}$ &  $0.02^{***}$ &  $-0.12^{***}$ &  $0.14^{***}$ &  $0.02^{***}$ &  $-0.12^{***}$ \\ 
				&   (0.00)  &  (0.00)  &  (0.00)  &  (0.00)  &  (0.00)  &  (0.01)  &  (0.00)  &  (0.00)  &  (0.00)  \\ 
				Cash  &   $0.00$ &  $0.00$ &  $0.02$ &  $0.00$ &  $0.01$ &  $0.02$ &  $0.00$ &  $0.01$ &  $0.03$ \\ 
				&   (0.01)  &  (0.01)  &  (0.01)  &  (0.01)  &  (0.01)  &  (0.01)  &  (0.01)  &  (0.01)  &  (0.01)  \\ 
				PE  &   $0.04^{***}$ &  $0.01^{***}$ &  $-0.02^{***}$ &  $0.03^{***}$ &  $0.01^{***}$ &  $-0.01^{***}$ &  $0.03^{***}$ &  $0.01^{***}$ &  $-0.01^{***}$ \\ 
				&   (0.01)  &  (0.00)  &  (0.01)  &  (0.01)  &  (0.00)  &  (0.01)  &  (0.01)  &  (0.00)  &  (0.01)  \\ 
				VIX  &   $-0.18^{***}$ &  $-0.07^{***}$ &  $-0.07^{***}$ &  $-0.23^{***}$ &  $-0.08^{***}$ &  $-0.12^{***}$ & - & - & -  \\ 
				&   (0.02)  &  (0.01)  &  (0.02)  &  (0.02)  &  (0.01)  &  (0.02)  & && \\ 
				Rm - Rf  &   $-0.08^{***}$ &  $0.03^{***}$ &  $0.10^{***}$ &  $-0.13^{***}$ &  $0.03^{***}$ &  $0.15^{***}$ & - & - & -  \\ 
				&   (0.02)  &  (0.01)  &  (0.02)  &  (0.02)  &  (0.01)  &  (0.02)  & && \\ 
				SMB  &   $0.02^{*}$ &  $0.01^{*}$ &  $0.00^{*}$ &  $-0.01$ &  $0.01$ &  $0.04$ & - & - & -  \\ 
				&   (0.01)  &  (0.00)  &  (0.01)  &  (0.01)  &  (0.00)  &  (0.01)  & && \\ 
				HML  &   $0.03^{***}$ &  $-0.01^{***}$ &  $-0.06^{***}$ &  $0.01$ &  $-0.01$ &  $-0.09$ & - & - & -  \\ 
				&   (0.01)  &  (0.00)  &  (0.01)  &  (0.01)  &  (0.00)  &  (0.01)  & && \\ 
				\hline &&&&&&&&& \\  
				Goodn.fit. &  - & - & -  & 4.66 & 4.86 & 3.63 & 5.41 & 4.92 & 3.75 \\ 
				\hline
				\hline
			\end{tabular}
		\end{center}
		{\footnotesize Notes: The dataset consists of $N = 943$ stocks with $T = 252$ time periods. The network matrix, $W_{CS3}$ is constructed by checking if the stocks are invested in by at least seven common shareholders with the network density, 0.41\%. See also notes to Table \ref{TabAppW3}.} 
	\end{table}
	
	\subsection{The Uniform Headquarter Location Network $W_{HQ}$} \label{appendixc2}
	
	Next, we use the uniform headquarter location network, $W_{HQ}$, and redo the application.  
	From Table \ref{TabAppW2}, we find that the contemporaneous network effects from the DNQR model are smaller than the counterparts in Table \ref{TabAppW1}, mainly reflecting that this network is more sparse than the network with $(W_{CS5})$. Still, the contemporaneous network effects are significant and dominate all other effects across all quantiles. The overall loss function of the DNQR model drops 3.5\%--5.2\% relative to NQAR and 3.4\%--4.5\% relative to NQARF, highlighting the importance of explicitly the contemporaneous network effects. 
	Overall, we have obtained qualitatively similar results.
	
\begin{figure}[!htbp]
		\centering
		\includegraphics[scale=0.7]{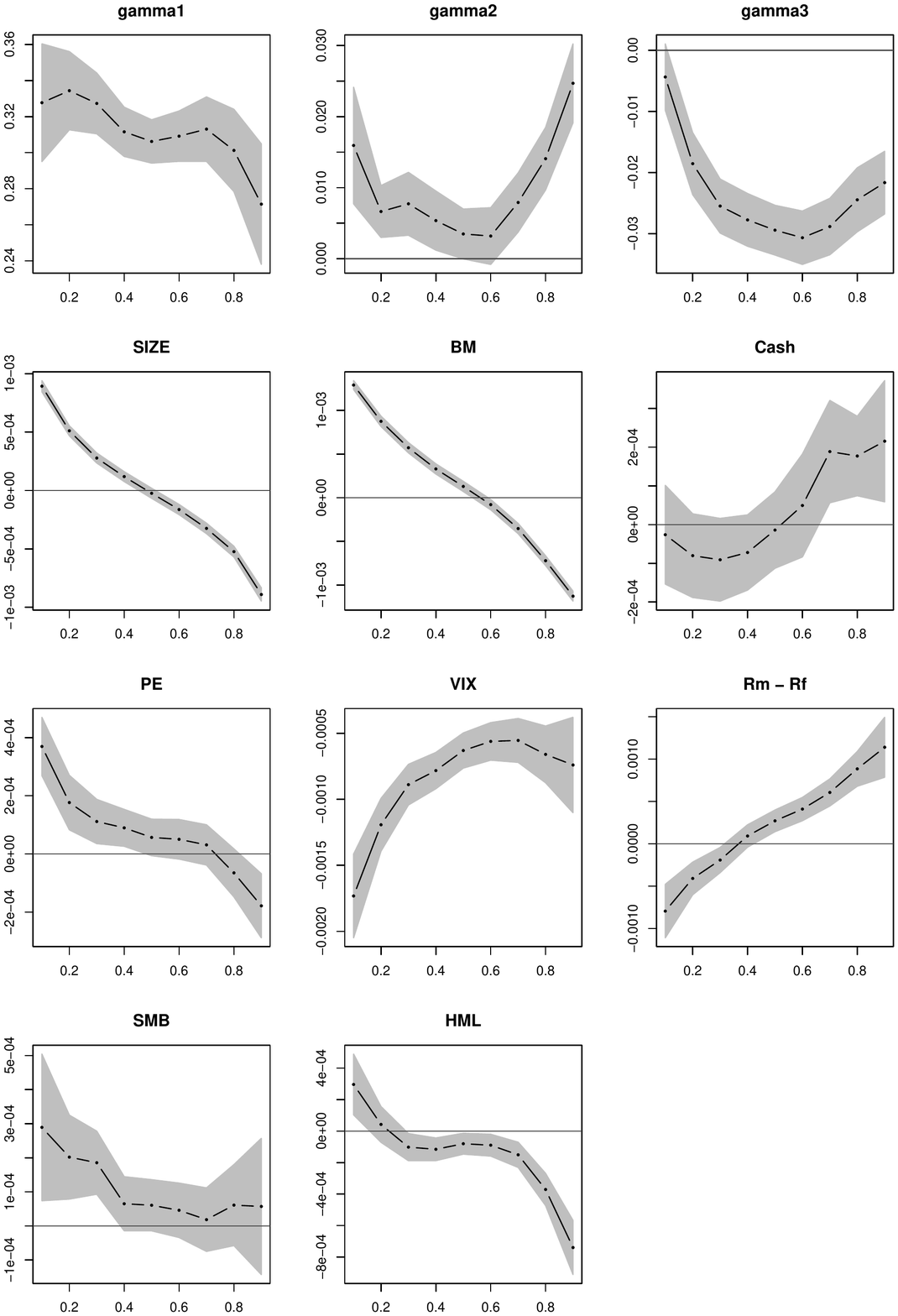}
		\caption{Quantile-specific Coefficients for the Network $W_{HQ}$}\label{AppParW2}
		\caption*{\footnotesize Notes: The dashed line is the QR coefficient while the grey area indicates a kernel density based confidence band advanced by \cite{powell1991}. They are displayed across quantiles, $\tau = 0.1, 0.2, \cdots, 0.9$.}
	\end{figure}

	Furthermore, the patterns of the QR-specific coefficients displayed in Figure \ref{AppParW2} are also qualitatively similar to those in Figure \ref{AppParW1}.


	\newpage
	
	
	\begin{table}[!htbp]
		\begin{center}
			\scriptsize 
			\caption{Estimation Results for the Network $W_{HQ}$} \label{TabAppW2}
			\begin{tabular}{c|ccc|ccc|ccc}
				\hline
				\hline
				
				& \multicolumn{3}{c}{DNQM} & \multicolumn{3}{c}{NQARF} & \multicolumn{3}{c}{NQAR} \\ \hline
				
				& $ \tau=0.1$  & $ \tau=0.5$  & $ \tau=0.9$ & $ \tau=0.1$  & $ \tau=0.5$  & $ \tau=0.9$ & $ \tau=0.1$  & $ \tau=0.5$  & $ \tau=0.9$ \\ 
				$ \hat\gamma_0 $  &   $-2.40^{***}$ &  $0.05^{***}$ &  $2.49^{***}$ &  $-2.54^{***}$ &  $0.05^{***}$ &  $2.65^{***}$ &  $-2.55^{***}$ &  $0.04^{***}$ &  $2.65^{***}$ \\ 
				&   (0.01)  &  (0.00)  &  (0.01)  &  (0.01)  &  (0.00)  &  (0.01)  &  (0.01)  &  (0.00)  &  (0.01)  \\ 
				$ \hat\gamma_1 $  &   $32.77^{***}$ &  $30.62^{***}$ &  $27.14^{***}$ & - & - & -  & - & - & -  \\ 
				&   (1.99)  &  (0.73)  &  (2.03)  & && & && \\ 
				$ \hat\gamma_2 $  &   $1.59^{***}$ &  $0.35^{***}$ &  $2.47^{***}$ &  $2.09^{***}$ &  $-0.21^{***}$ &  $2.40^{***}$ &  $3.23^{***}$ &  $-0.25^{***}$ &  $2.33^{***}$ \\ 
				&   (0.50)  &  (0.21)  &  (0.34)  &  (0.57)  &  (0.03)  &  (0.46)  &  (0.57)  &  (0.16)  &  (0.48)  \\ 
				$ \hat\gamma_3 $  &   $-0.44$ &  $-2.94$ &  $-2.17$ &  $0.15$ &  $-2.18$ &  $-1.88$ &  $0.16$ &  $-1.66$ &  $-2.08$ \\ 
				&   (0.33)  &  (0.25)  &  (0.31)  &  (0.40)  &  (0.10)  &  (0.38)  &  (0.40)  &  (0.12)  &  (0.37)  \\ 
				&&&&&&&&& \\ 
				SIZE  &   $0.09^{***}$ &  $0.00^{***}$ &  $-0.09^{***}$ &  $0.09^{***}$ &  $0.00^{***}$ &  $-0.09^{***}$ &  $0.10^{***}$ &  $0.00^{***}$ &  $-0.09^{***}$ \\ 
				&   (0.00)  &  (0.00)  &  (0.00)  &  (0.00)  &  (0.00)  &  (0.00)  &  (0.00)  &  (0.00)  &  (0.00)  \\ 
				BM  &   $0.13^{***}$ &  $0.01^{***}$ &  $-0.11^{***}$ &  $0.14^{***}$ &  $0.02^{***}$ &  $-0.12^{***}$ &  $0.14^{***}$ &  $0.02^{***}$ &  $-0.12^{***}$ \\ 
				&   (0.00)  &  (0.00)  &  (0.00)  &  (0.00)  &  (0.00)  &  (0.00)  &  (0.00)  &  (0.00)  &  (0.00)  \\ 
				Cash  &   $0.00$ &  $0.00$ &  $0.02$ &  $0.01$ &  $0.01$ &  $0.02$ &  $0.00$ &  $0.01$ &  $0.02$ \\ 
				&   (0.01)  &  (0.01)  &  (0.01)  &  (0.01)  &  (0.01)  &  (0.01)  &  (0.01)  &  (0.01)  &  (0.01)  \\ 
				PE  &   $0.04^{***}$ &  $0.01^{***}$ &  $-0.02^{***}$ &  $0.03^{***}$ &  $0.01^{***}$ &  $-0.01^{***}$ &  $0.03^{***}$ &  $0.01^{***}$ &  $-0.01^{***}$ \\ 
				&   (0.01)  &  (0.00)  &  (0.01)  &  (0.01)  &  (0.00)  &  (0.01)  &  (0.01)  &  (0.00)  &  (0.01)  \\ 
				VIX  &   $-0.17^{***}$ &  $-0.06^{***}$ &  $-0.07^{***}$ &  $-0.23^{***}$ &  $-0.08^{***}$ &  $-0.12^{***}$ & - & - & -  \\ 
				&   (0.02)  &  (0.01)  &  (0.02)  &  (0.02)  &  (0.01)  &  (0.02)  & && \\ 
				Rm - Rf  &   $-0.08^{***}$ &  $0.03^{***}$ &  $0.11^{***}$ &  $-0.14^{***}$ &  $0.04^{***}$ &  $0.16^{***}$ & - & - & -  \\ 
				&   (0.02)  &  (0.01)  &  (0.02)  &  (0.02)  &  (0.01)  &  (0.02)  & && \\ 
				SMB  &   $0.03^{**}$ &  $0.01^{**}$ &  $0.01^{**}$ &  $-0.01$ &  $0.01$ &  $0.03$ & - & - & -  \\ 
				&   (0.01)  &  (0.00)  &  (0.01)  &  (0.01)  &  (0.00)  &  (0.01)  & && \\ 
				HML  &   $0.03^{**}$ &  $-0.01^{**}$ &  $-0.07^{**}$ &  $0.01$ &  $-0.01$ &  $-0.09$ & - & - & -  \\ 
				&   (0.01)  &  (0.00)  &  (0.01)  &  (0.01)  &  (0.00)  &  (0.01)  & && \\ 
				\hline &&&&&&&&& \\  
				Goodn.fit. &  - & - & -  & 4.49 & 4.50 & 3.42 & 5.24 & 4.56 & 3.54 \\ 
				
				\hline
				\hline
			\end{tabular}	
		\end{center}
		{\footnotesize Notes: The dataset consists of $N = 943$ stocks with $T = 252$ time periods. The network $W_{HQ}$ is constructed by by checking if the headquarters of companies are located in the same city with the network density, 0.63\%. See also notes to Table \ref{TabAppW3}.}  
	\end{table}
	
\end{appendices}

\end{document}